\documentclass[twocolumn]{aastex631}
\newcommand{\Lsun}{$\rm L_{\odot}$}

\newcommand{\Av}{$A_V$}

\newcommand{\mic}{$\rm \mu$m}

\usepackage{amsmath}
\usepackage{booktabs}
\usepackage{multirow}

\shorttitle{Embedded Protostars Viewed in CO Emission}
\shortauthors{Rubinstein et al.}

\begin{document}

\title{IPA. Class 0 Protostars Viewed in CO Emission Using JWST}

\correspondingauthor{Adam E. Rubinstein}
\email{arubinst@ur.rochester.edu}

\author[0000-0001-8790-9484]{Adam E. Rubinstein}
\affiliation{Department of Physics and Astronomy, Bausch \& Lomb Hall, University of Rochester, Rochester, NY 14627, USA}

\author[0000-0001-5175-1777]{Neal J. Evans II}
\affiliation{Department of Astronomy, University of Texas at Austin, 1 University Station C1400, Austin, TX 78712, USA}

\author[0000-0002-9497-8856]{Himanshu Tyagi}
\affiliation{Department of Astronomy \& Astrophysics Tata Institute of Fundamental Research, Homi Bhabha Rd, Colaba, Mumbai, Maharashtra, IN}

\author[0000-0002-0554-1151]{Mayank Narang}
\affiliation{Academia Sinica Institute of Astronomy \& Astrophysics, No. 1 Sec. 4 Roosevelt Rd., Taipei 10617, TW, R.O.C.}

\author[0000-0002-4448-3871]{Pooneh Nazari}
\affiliation{Leiden Observatory, Leiden University, PO Box 9513, 2300 RA Leiden, South Holland, NL}
\affiliation{European Southern Observatory, Karl-Schwarzschild-Strasse 2, 85748 Garching, DE}

\author[0000-0002-6447-899X]{Robert Gutermuth}
\affiliation{Department of Astronomy, University of Massachusetts Amherst, 710 North Pleasant Street, Amherst, MA 01003, USA}

\author[0000-0002-6136-5578]{Samuel Federman}
\affiliation{Ritter Astrophysical Research Center, Department of Physics and Astronomy, University of Toledo, Toledo, OH 43606, USA}

\author[0000-0002-3530-304X]{P. Manoj}
\affiliation{Department of Astronomy \& Astrophysics Tata Institute of Fundamental Research, Homi Bhabha Rd, Colaba, Mumbai, Maharashtra, IN}

\author[0000-0003-1665-5709]{Joel D. Green}
\affiliation{Space Telescope Science Institute, 3700 San Martin Drive, Baltimore, MD 21218, USA}

\author[0000-0001-8302-0530]{Dan M. Watson}
\affiliation{Department of Physics and Astronomy, Bausch \& Lomb Hall, University of Rochester, Rochester, NY 14627, USA}

\author[0000-0001-7629-3573]{S. Thomas Megeath}
\affiliation{Ritter Astrophysical Research Center, Department of Physics and Astronomy, University of Toledo, Toledo, OH 43606, USA}
\author[0000-0001-6144-4113]{Will R. M. Rocha}
\affiliation{Leiden Observatory, Leiden University, PO Box 9513, 2300 RA Leiden, South Holland, NL}

\author[0000-0001-7826-7934]{Nashanty G. C. Brunken}
\affiliation{Leiden Observatory, Leiden University, PO Box 9513, 2300 RA Leiden, South Holland, NL}

\author[0000-0002-7433-1035]{Katerina Slavicinska}
\affiliation{Leiden Observatory, Leiden University, PO Box 9513, 2300 RA Leiden, South Holland, NL}

\author[0000-0001-7591-1907]{Ewine F. van Dishoeck}
\affiliation{Leiden Observatory, Leiden University, PO Box 9513, 2300 RA Leiden, South Holland, NL}


\author[0000-0002-1700-090X]{Henrik Beuther}
\affiliation{Max Planck Institute for Astronomy, Heidelberg, Baden-Württemberg, DE}

\author[0000-0001-7491-0048]{Tyler L. Bourke}
\affiliation{SKA Observatory, Jodrell Bank, Lower Withington, Macclesfield SK11 9FT, UK}

\author[0000-0001-8876-6614]{Alessio Caratti o Garatti}
\affiliation{Dublin Institute for Advanced Studies, 31 Fitzwilliam Place, Dublin, IE}

\author[0000-0003-1430-8519]{Lee Hartmann}
\affiliation{Department of Astronomy, University of Michigan -- Ann Arbor, 1085 S. University Ave, Ann Arbor, MI 48109, USA}

\author[0000-0001-9443-0463]{Pamela Klaassen}
\affiliation{United Kingdom Astronomy Technology Centre, Royal Observatory Edinburgh, Blackford Hill, Edinburgh EH9 3HJ, GB}

\author[0000-0002-5943-1222]{Hendrik Linz}
\affiliation{Max Planck Institute for Astronomy, Heidelberg, Baden-Württemberg, DE}
\affiliation{Friedrich Schiller University, Jena, Thuringia, DE}

\author[0000-0002-4540-6587]{Leslie W. Looney}
\affiliation{Department of Astronomy, University of Illinois, 1002 West Green St, Urbana, IL 61801, USA}
\affiliation{National Radio Astronomy Observatory, 520 Edgemont Rd., Charlottesville, VA 22903, USA}

\author[0000-0002-5943-1222]{James Muzerolle Page}
\affiliation{Space Telescope Science Institute, 3700 San Martin Drive, Baltimore, MD 21218, USA}

\author[0000-0002-5812-9232]{Thomas Stanke}
\affiliation{Max Planck Institute for Extraterrestrial Physics, Gie\ss{}enbachstraße 185748 Garching, DE}

\author[0000-0002-6195-0152]{John J. Tobin}
\affiliation{National Radio Astronomy Observatory, 520 Edgemont Rd., Charlottesville, VA 22903, USA}

\author[0000-0002-0826-9261]{Scott J. Wolk}
\affiliation{Harvard-Smithsonian Center for Astrophysics, 60 Garden St., Cambridge, MA 02138, USA}

\author[0000-0001-8227-2816]{Yao-Lun Yang}
\affiliation{RIKEN Cluster for Pioneering Research, Wako-shi, Saitama, 351-0106, JP}

\begin{abstract}

    We investigate the bright CO fundamental emission in the central regions of five protostars in their primary mass assembly phase using new observations from JWST's Near-Infrared Spectrograph (NIRSpec) and Mid-Infrared Instrument (MIRI). 
    CO line emission images and fluxes are extracted for a forest of $\sim$150 ro-vibrational transitions from two vibrational bands, $v=1-0$ and $v=2-1$. 
    However, ${}^{13}$CO is undetected, indicating that ${}^{12}$CO emission is optically thin. 
    We use H$_2$ emission lines to correct fluxes for extinction and then construct rotation diagrams for the CO lines with the highest spectral resolution and sensitivity to estimate rotational temperatures and numbers of CO molecules. Two distinct rotational temperature components are required for $v=1$ ($\sim600$ to 1000~K and 2000 to $\sim 10^4$~K), while one hotter component is required for $v=2$ ($\gtrsim 3500$~K). ${}^{13}$CO is depleted compared to the abundances found in the ISM, indicating selective UV photodissociation of ${}^{13}$CO; therefore, UV radiative pumping may explain the higher rotational temperatures in $v=2$.  The average vibrational temperature is $\sim 1000$~K for our sources and is similar to the lowest rotational temperature components.
    Using the measured rotational and vibrational temperatures to infer a total number of CO molecules, we find that the total gas masses 
    range 
    from lower limits of $\sim10^{22}$~g for the lowest mass protostars to $\sim 10^{26}$~g for the highest mass protostars.
    Our gas mass lower limits are compatible with those in more evolved systems, which  suggest the lowest rotational temperature component comes from the inner disk, scattered into our line of sight, but we also cannot exclude the contribution to the CO emission from disk winds for higher mass targets.
    

\end{abstract}


\section{Introduction} 
    \label{intro}    Protostellar evolution is an interplay of an infalling envelope, accretion through a disk onto a central mass, and feedback by outflows and jets. 
For all protostellar processes, CO emission at infrared and longer wavelengths probe different excitation levels and can be used to estimate gas column densities and kinetic temperatures  \citep[e.g.][]{Scoville_1979, Watson_1985, Mitchell_1990, Evans_1991, Hayashi_1994, Blake_1995, Bontemps_1996, Jorgensen_2002, Fuller_2002, Tachihara_2002, Hatchell_2005}. At sub-mm wavelengths, one finds gas kinematics of infalling and outflowing material \citep[e.g. using the Submillimeter Array, ][]{Bjerkeli_2016} and molecular cavity walls \citep[e.g. with the Atacama Large Millimeter/submillimeter Array, ALMA, ][]{Bjerkeli_2019, Hsieh_2023}. At far-infrared wavelengths, the \textit{Herschel} Space Telescope frequently observed and spectrally resolved hot shocked gas associated with protostellar sources \citep[e.g.][]{vanKempen_2010A, vanKempen_2010B, Bjerkeli_2011, Herczeg_2012, Goicoechea_2012, Bjerkeli_2013, Karska_2013, Manoj_2013, Tafalla_2013, Bjerkeli_2014, Manoj_2016, Kristensen_2017, Karska_2018}. In such conditions, CO becomes rotationally excited (denoted by the quantum number $J$) and undergoes transitions from upper levels as high as $J_{u}$=50.
Spatially and spectrally resolved observations of rotationally excited CO gas emission in the ground vibrational state (denoted by quantum number $v$) reveal regions near the protostar itself \citep[e.g.][]{Yildiz_2010, Green_2013, Manoj_2013, Yildiz_2013, Yildiz_2015}.

At shorter wavelengths ($\lambda<5$ \mic), ground-based and space-based telescopes observe vibrationally excited CO lines probing the inner gaseous disk, disk surfaces, and the base of outflows through disk winds in both low- and high-mass young stellar objects, also known as YSOs \citep[e.g.][]{Herczeg_2011, Ilee_2013, Salyk_2022}. Recent NIR interferometry of more evolved intermediate-mass YSOs (Herbig Ae/Be stars) and more massive protostars has pinpointed the origin of the emitting region down to 1 au \citep[][]{Caratti_2020, Koutoulaki_2021}. 
NIR CO emission in overtones also correlate with accretion rates, providing lessons about inner disk and protostellar conditions  \citep{Ilee_2014, Poorta_2023, LeGoulec_2024}. 
Classifying by spectral energy distributions (SEDs), more evolved YSOs are called Class II \citep{Lada_1987}. M-band (4.7--5 \mic) emission and absorption lines from warm water and CO gas are inner disk tracers for Class II sources as well as Class I (less-evolved) YSOs \citep{Mitchell_1990, Najita_2003, Pontoppidan_2003, Blake_2004, Brown_2013, Podio_2020, Smith_2021, Banzatti_2022}. The column densities and temperatures expected from CO emission 
can be difficult to interpret for Class I disks when due to a mixture of the protostellar disk, colder molecular outflows, and the outflow cavity's walls \citep[e.g.][]{Jorgensen_2005, Herczeg_2011}.  



Class 0 YSOs are the least evolved. The many interacting components of the protostar makes simulations complex \citep[e.g.][]{Dunham_2014, Federrath_2014}, while the embedded nature of the sources makes observations difficult as well \citep[e.g.][]{Beltran_2016}.
Observatories using mm/sub-mm wavelengths, like NOEMA, SMA, \citep{Anderl_2020} and ALMA \citep{Kristensen_2013_1, Oya_2020, Hsieh_2023}, can probe Class 0 disks down to 10 to 100 au scales but are restricted to longer ($>$350 \mic) wavelengths and therefore cannot reveal the hottest gas components. 
Far-IR gas-phase CO lines have also been observed from Class 0 YSOs and interpreted as being due to winds \citep[e.g.][]{Manoj_2013, Manoj_2016, Yang_2018}. Complete surveys of M-band CO lines have not, to our knowledge, been reported for Class 0 sources. Only one source, IRAS 15398-3359, has partial coverage of the the M-band using JWST's MIRI \citep{Yang_2022, Salyk_2024}. The higher spectral resolution of ground-based observations is also limited in the coverage of $J_u$. 

Our project, Investigating Protostellar Accretion (IPA), has obtained novel JWST observations of CO rotational-vibrational (hereafter ro-vibrational) emission from the innermost regions of five Class 0 protostars (\citealt{Federman_2023}). 
In past work with Class I and II sources, the exact spatial location that produces the NIR to MIR CO emission can be uncertain between the inner gaseous disk, dust-rich disk, and winds without sufficient physical modeling \citep[e.g.][]{Herczeg_2011, Bosman_2019, Banzatti_2022}. 
We do not have spectrally resolved line profiles as found in past ground-based observations (e.g. for Class II protoplanetary disks see \citealt{Najita_2003}, \citealt{Banzatti_2022}; for Class I see \citealt{Herczeg_2011}). Instead, we uniquely detect a full suite of CO fundamental ro-vibrational lines for 2 vibrationally excited bands of CO at relatively high spatial resolution ($\sim$0.1\arcsec).
As in prior work using high-resolution observations \citep[e.g. with ALMA, ][]{Yen_2017A, Yen_2017B, Okoda_2018}, we infer physical properties (e.g. gas mass, temperature) of CO emission to investigate its origins around the central protostar.





\section{Observations and Data Reduction}  
    \label{obs}

    \subsection{JWST IPA Sample}
    \label{jwst_obs}    
    IPA is a Cycle 1 medium General Observers proposal (ID 1802, PI: S. T. Megeath), which includes JWST's NIRSpec and MIRI/MRS Integral Field Unit (IFU) observations for a sample of 5 Class 0 protostars obtained from 22 July 2022 to 16 October 2022 (see \citealt{Federman_2023} for IPA details and initial data release, and see \citealt{Rigby_2023, Boker_2022a, Boker_2023, Wright_2023} for JWST instrument and launch details). These protostars cover a few orders of magnitude in protostellar mass, dust disk properties, and bolometric luminosity (see Table \ref{tab:source_props}). 
One similarity among our sources is that their disks are closer to edge-on than face-on (with 0$^{\circ}$ as face-on, $i > \pm 65^{\circ}$, see \citealt{Federman_2023} and references therein), which enables studying scattered light above and below the plane of the disk, isolating jets or outflows from disks, and ultimately assessing envelopes and accretion for our sources. For general details about the NIRSpec and MIRI/MRS data reduction from the JWST pipeline, see \citet{Federman_2023} and \citet{Narang_2024}, which used a custom noise masking and alignment procedure instead of the Mikulski Archive for Space Telescopes (MAST) products that are output by automated pipeline reduction routines. We discuss observations, reductions, and basic results, first for NIRSpec (Section \ref{jwst_nirspec}), then for MIRI (Section \ref{jwst_miri}).

\begin{deluxetable*}{ccccccc}
    \tablewidth{0pt}
    \tablecaption{IPA Sample of Class 0 Protostars}
    \tablehead{
        \colhead{Identification} & \colhead{Dust Continuum Coordinates} & \colhead{Distance} & \colhead{Luminosity} & \colhead{Dust Disk Radius} & \colhead{Major Axis P.A.} & \colhead{Ref.\tablenotemark{a}}  \\ 
        \colhead{--} & \colhead{R.A., Decl.} & \colhead{$d_{star}$} & \colhead{$L_{bol}$} & \colhead{$R_{disk,dust}$} & \colhead{$PA$} & \colhead{--} \\
        \colhead{(--)} & \colhead{(hh:mm:ss, dd:mm:ss)} & \colhead{(pc)} & \colhead{(\Lsun)} & \colhead{(au)} & \colhead{($\circ$)}  & \colhead{(--)}
        }
    \startdata
        IRAS 16253-2429 &   16:28:21.62,-24:36:24.33    & 140   & 0.16                      & 16  &   -23      & \textit{1,6,1,12,12}   \\ 
        B335            &   19:37:00.9,+7:34:09.4      & 165   & 1.4\tablenotemark{*}      & $<$10  &   5  & \textit{2,7,10,13,14}  \\ 
        HOPS 153        &   5:37:57.021,-7:06:56.25     & 390   & 3.8                     & 150  &   -33    & \textit{3,3,3,3,3}   \\ 
        HOPS 370        &   5:35:27.635,-5:09:34.44     & 390   & 315.7                    & 100  &   109.7    & \textit{4,3,4,3,3}  \\ 
        IRAS 20126+4104 &   20:14:26.036,+41:13:32.52   & 1550  & ${10}^{4}$             & 860  &   56    & \textit{5,8,11,5,5}  \\ 
    \enddata
    \tablenotetext{a}{References for protostellar distance, bolometric luminosity, disk radius probing mm/sub-mm continuum from dust, the protostar positions from mm/sub-mm photocenters (see \citealt{Federman_2023} for details about coordinate alignment), and the position angle (P.A.) along the major axis of the mm continuum source. The only exception is B335, which has moved between 2017 (e.g. \citealt{Maury_2018}) and 2023, and will discussed in future work by Kim, C. et al. in prep.} 
    \tablenotetext{*}{Variable.}
    \tablenotetext{}{ \textit{1}.  \citealt{Aso_2023}, \textit{2}. Kim et al. in prep, \textit{3}. \citealt{Tobin_2020A}, \textit{4}. \citealt{Tobin_2020B}, \textit{5}. \citealt{Cesaroni_2014}, \textit{6}. \citealt{Zucker_2020}, \textit{7}. \citealt{Watson_2020}, \textit{8}. \citealt{Reid_2019}, \textit{10}. \citealt{Evans_2023},  \textit{11}. \citealt{Cesaroni_2023}, \textit{12}. \citealt{TH-Hsieh_2019}, \textit{13}. \citealt{Yen_2015}. \textit{14} \citealt{Bjerkeli_2023}.} 
     \label{tab:source_props}
\end{deluxetable*}

    \subsection{NIRSpec Spectral Cubes}
    \label{jwst_nirspec}
    The NIRSpec IFU observations for IPA used the G395M medium-resolution grating with wavelength coverage from 2.87 to 5.27 \mic. The data result in an image taken at each wavelength limited in spacing by the spectral resolution (wavelength spacing per resolution element of 0.00156 \mic\ to 0.00155 \mic), which for G395M varies with wavelength as $R = \frac{\lambda}{\Delta\lambda} \sim 700-1300$ or velocity as $\Delta v \sim 200-400 \ \rm km \ {sec}^{-1} $. The spatial resolution is approximately $\Delta\theta=$0\farcs17 to 0\farcs21. The Calibration Data Reference System (CRDS) used for all NIRSpec cubes was version 11.16.20 with pipeline mapping (pmap) 1069.

\begin{figure*}
    \centering
        \includegraphics[width=1\linewidth, trim = 2in 0.2in 2in 0.3in,clip]{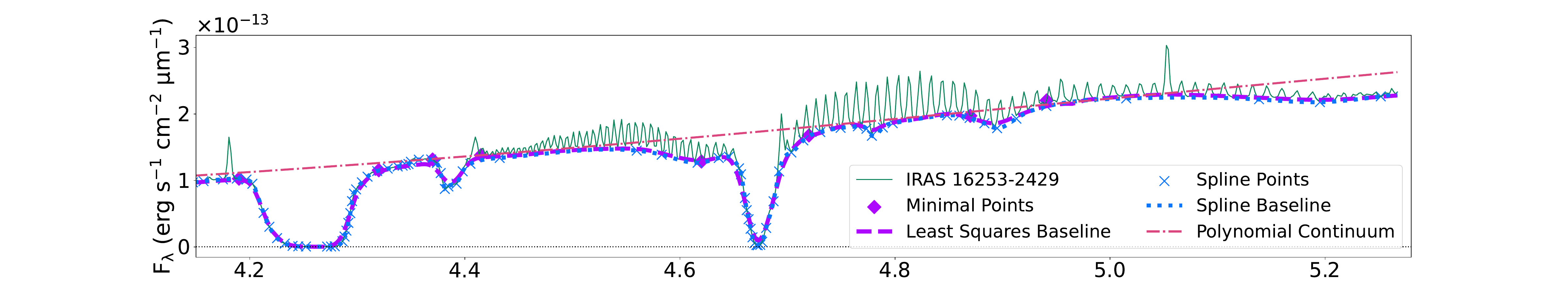}
        \includegraphics[width=1\linewidth, trim = 2in 0.2in 2in 0.3in,clip]{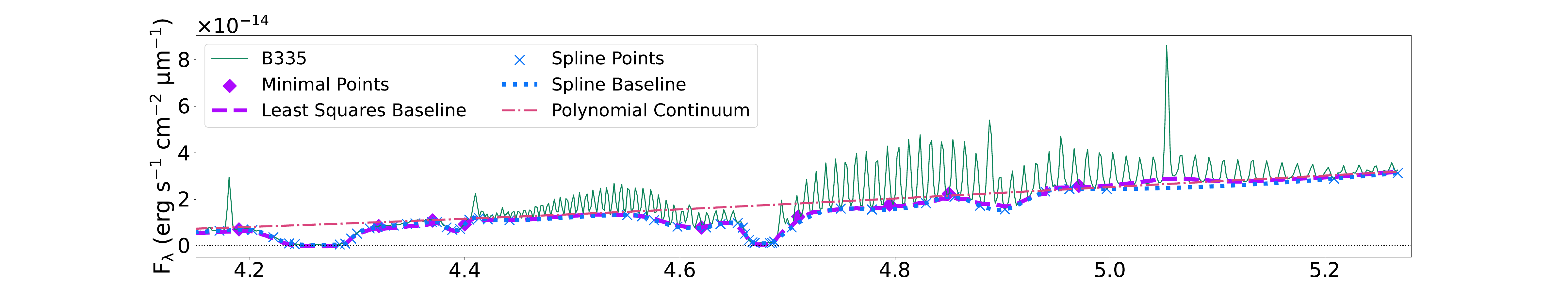}
        \includegraphics[width=1\linewidth, trim = 2in 0.2in 2in 0.3in,clip]{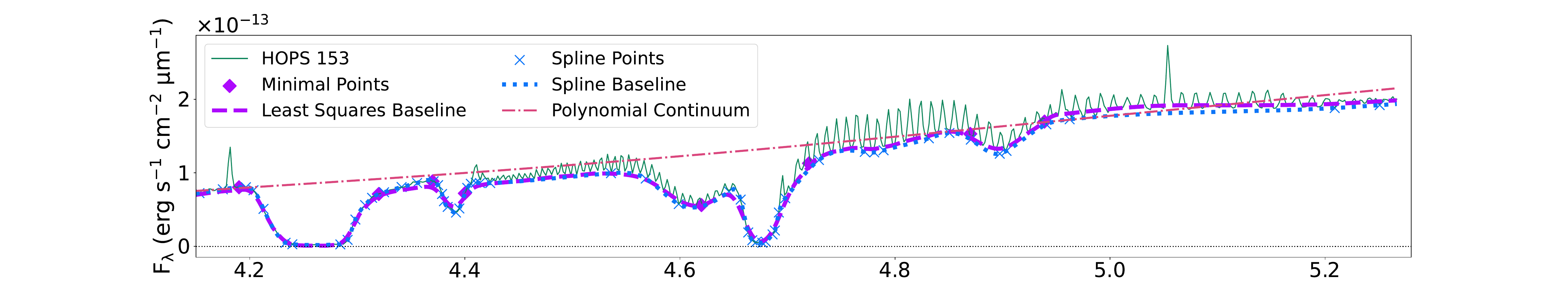}
        \includegraphics[width=1\linewidth, trim = 2in 0.2in 2in 0.3in,clip]{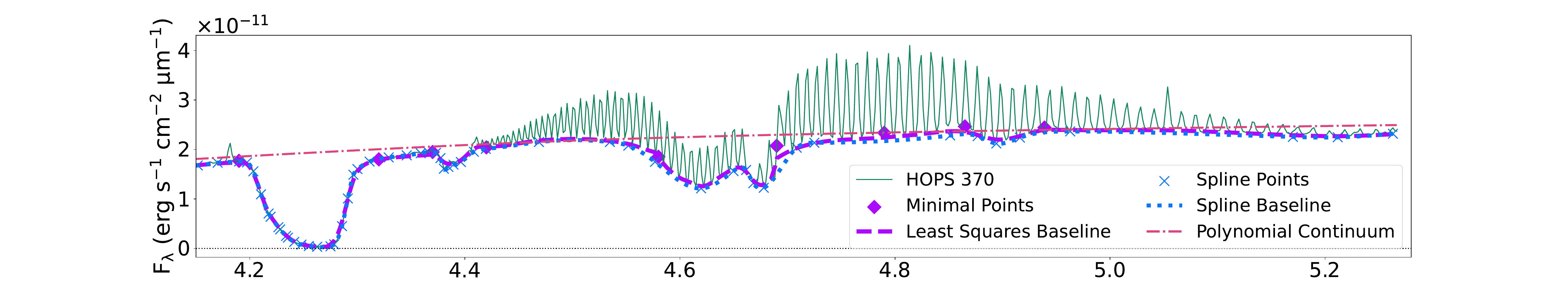}
        \includegraphics[width=1\linewidth, trim = 2in 0.2in 2in 0.3in,clip]{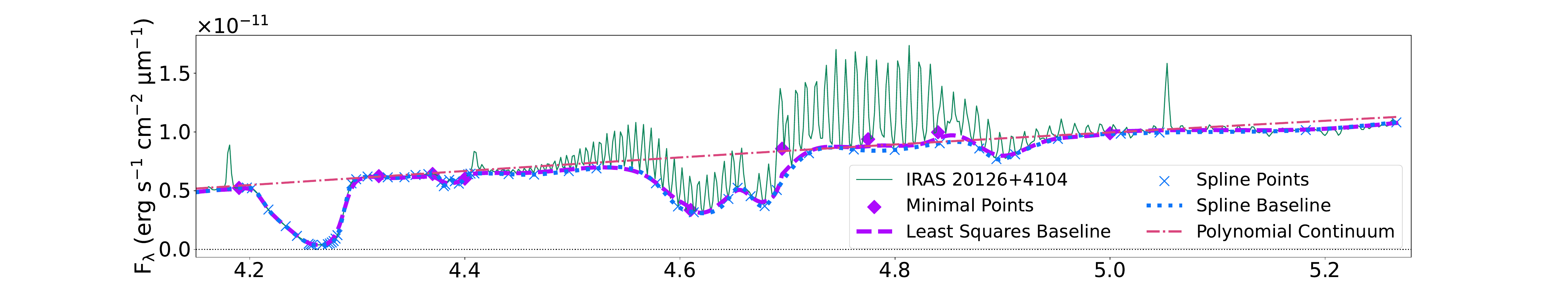}
        \includegraphics[width=1\linewidth, trim = 2in 0in 2in 0.9in,clip]{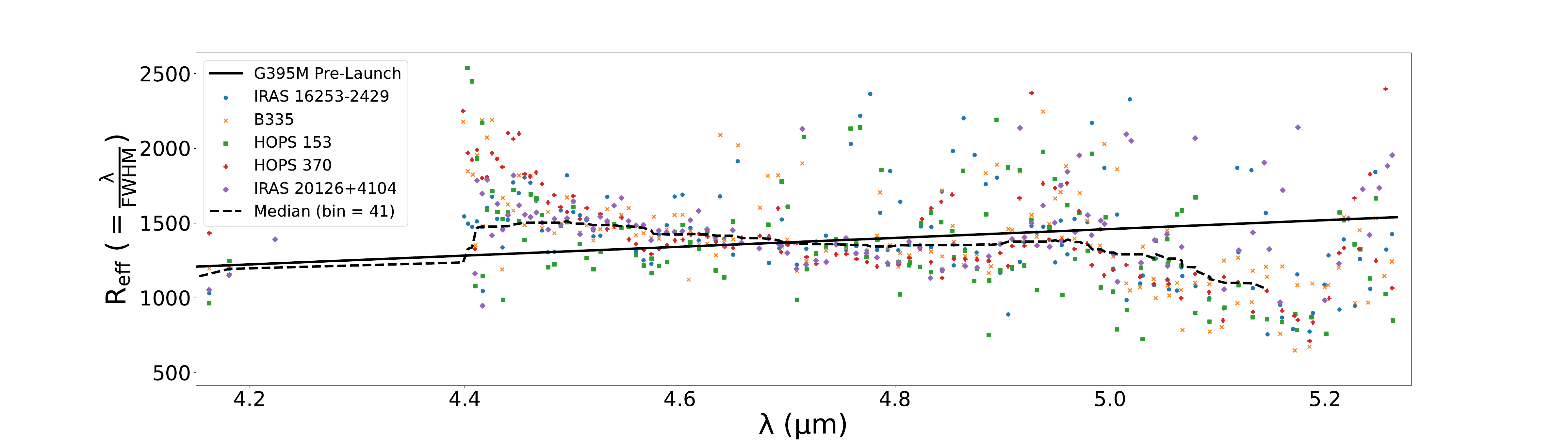}
    \caption{Extracted spectra and systematic effects (see Appendix \ref{sect:fitting_methods} for details). \underline{In the upper five panels} (observations in green), baselines for line images (dashed purple curve) are smoothed between ice features (purple diamonds), the splined baseline is used in our high S/N fits (blue dotted line and x's). A polynomial continuum is shown by the pink dash-dot line. \underline{The bottom panel} shows R$_{eff}$ = $\lambda$/FWHM of line profiles for each protostar and line fit in Figure \ref{fig:example_spaxels}. Our median is the dashed curve and the pre-launch behavior is the solid curve (assumed for making images in Figure \ref{fig:co_imgs}).}\label{fig:systematics}
\end{figure*}

\begin{figure*}
    \centering
    \includegraphics[width=\linewidth, trim = 5in 0.1in 4in 1in,clip]{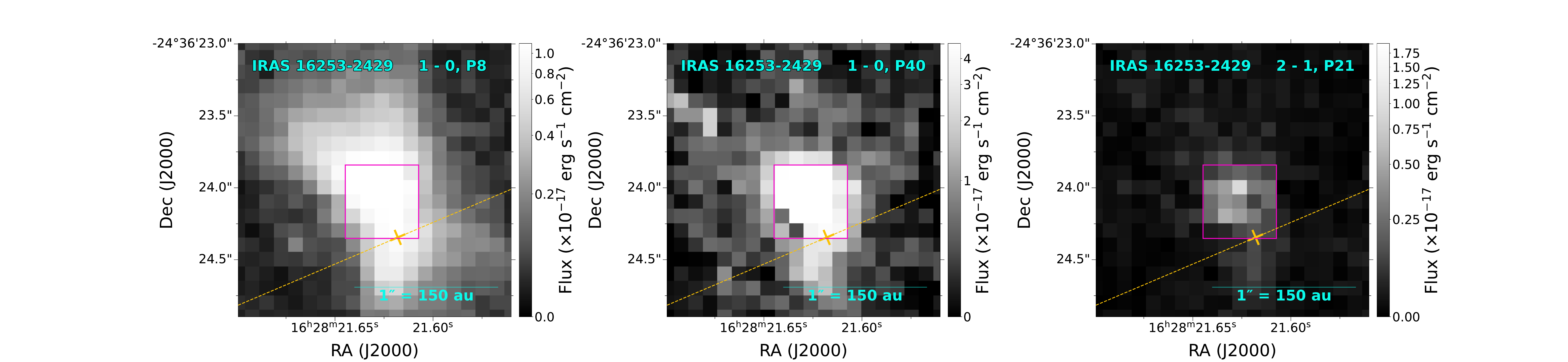}
    \includegraphics[width=\linewidth, trim = 5in 0.1in 4in 1in,clip]{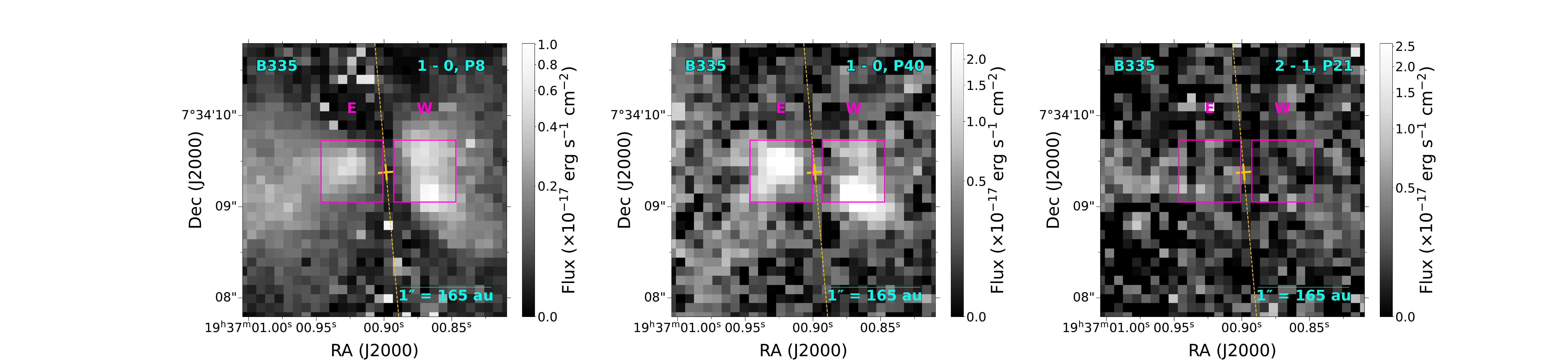}
    \includegraphics[width=\linewidth, trim = 5in 0.1in 4in 1in,clip]{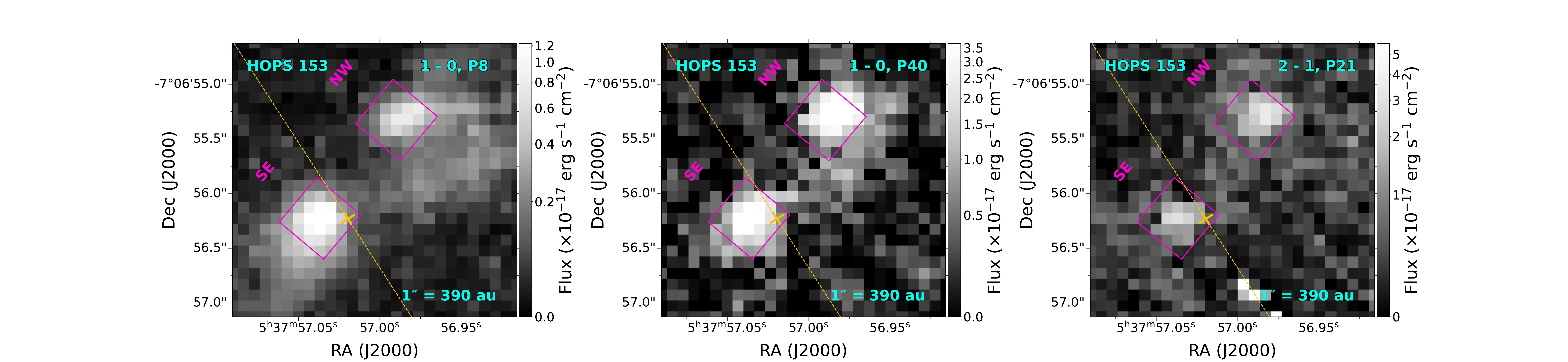}
    \includegraphics[width=\linewidth, trim = 5in 0.1in 4in 1in,clip]{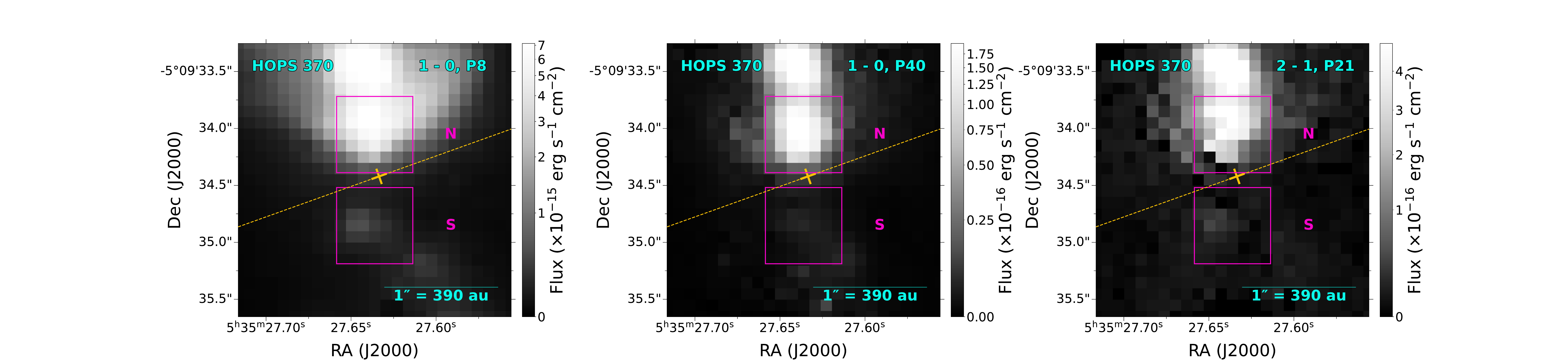}
    \includegraphics[width=\linewidth, trim = 5in 0.1in 4in 1in,clip]{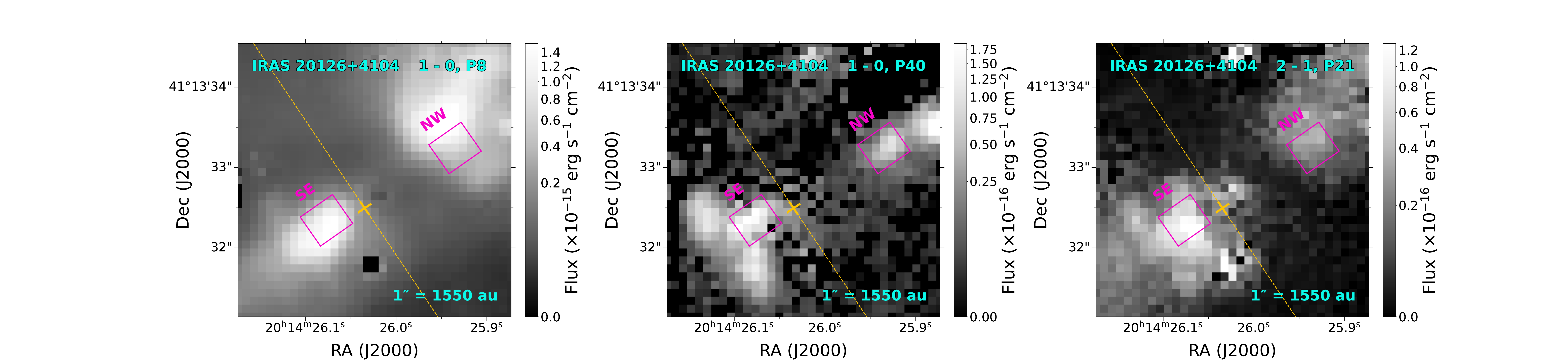}
    \caption{CO line emission images (see Appendix \ref{sect:fitting_methods} for details). Each image has the dust disk's continuum coordinates and disk's major-axis P.A. (see Table \ref{tab:source_props}) marked with a yellow plus sign and dashed line. 
    Rows correspond to a protostar in our sample (upper left of each panel). The images are used to define apertures that maximize the S/N of CO line emission (magenta boxes, from Table \ref{tab:apertures}). Different CO spectral lines are shown (see upper right of each panel) to highlight variations in $J$ (low-$J$ on the left images and high-$J$ in the middle images) and $v$ (left and middle images compared to the right one). 
    }\label{fig:co_imgs}
\end{figure*}

Our process of producing NIRSpec spectra for analysis was iterative. We first identified continuum, absorption due to ices, and gas-phase emission lines (Figure \ref{fig:systematics}). Using points free of known ice absorption, baselines were removed and lines fitted for each spaxel to produce images of several CO lines (Figure \ref{fig:co_imgs}). In Section \ref{apertures}, the images were used to choose apertures close to the central source that optimized the line to continuum ratios for the CO lines. The spectra obtained from summing over spaxels in these apertures were then fitted for continuum and ices again to produce a line-only spectrum (Figure \ref{fig:example_spaxels}). That spectrum was used to extract line fluxes (Table \ref{tab:co_lines_fluxes} in Appendix \ref{sect:flux_table}) for further analysis. These steps are explained in more detail below and in appendices. The MIRI/MRS spectra are used in Sections \ref{absorption_doppler} and \ref{implied_profile} to compare with our NIRSpec spectra and to characterize our line profiles (Figure \ref{fig:extinc_example}).

        \subsubsection{Spectral Line Identification}
        \label{cube_lines}    

For our sample of Class 0 protostars, we identify CO fundamental emission lines. The observed set are collectively called a forest of spectral lines, while band refers to a particular vibrational transition (e.g. $v=1-0$ or $v=2-1$).  
We focus on the CO line forest in bands of $v=1-0$ and $v=2-1$ vibrationally excited transitions above the ground state ($v=0, J=0$) up to approximately $J=40$ (with upper state energy per Boltzmann constant, $k_B$ above the ground state expressed in temperature units of about 3000 to 12000 K) detected throughout our sample (see spectra in Figure \ref{fig:systematics}). The spectral resolving power ($R \sim 1000$) is sufficiently high to separate the $^{12}$CO $v=1-0$ lines between about 4.3 and 5.2 \mic, but not to resolve the Doppler velocity structure or reveal self-absorption as seen by high-spectral resolution ground-based observations \citep[e.g.][]{Najita_2003, Herczeg_2011}.

To identify molecular line series, we use the HITRAN database \citep{HITRAN_2022} accessed in April 2023 to retrieve data tables for all relevant species and transitions between 4.3 \mic \ and 5.2 \mic\  of ${}^{12}$CO ($v=1-0$, $v=2-1$, and $v=3-2$) and ${}^{13}$CO ($v=1-0$). From the nomenclature of molecular spectra, $v=1-0$ CO lines between about 4.3 and 4.7 \mic \ are called the R branch (rotationally excited transitions of $J_u - J_l = +1$), and between 4.7 and 5.2 \mic \ the P branch ($J_u - J_l = -1$). The two branches meet at the center of the band as they approach $J_u=0$ (4.658 \mic) and increase in $J$ and upper state energies moving away from the center. 

Initially, we identify CO line series by overplotting bands of vibrationally excited lines over example spaxels and summed spectra to increase the signal-to-noise ratio (S/N). 
Then we use the software CARTA (Cube Analysis and Rendering Tool for Astronomy; \citealt{CARTA_2021}) to scan images and check if all detected lines of CO show the same or similar spatially resolved structures for each CO line detected. Identified lines are documented in Appendix \ref{sect:flux_table}, Table \ref{tab:co_lines_fluxes}. We also included atomic species using the NIST atomic database and H$_2$ lines as in \citet{Federman_2023} and \citet{Narang_2024} as they can overlap CO lines.  CO $v=2-1$ lines appear past $5.2$ \mic, and we did not confirm any detection of CO $v=3-2$ or ${}^{13}$CO $v=1-0$. ${}^{13}$CO $v=1-0$ lines can be used to estimate optical depth, so those lines are given special attention in Section \ref{implied_profile}.
    
        \subsubsection{Spectral Line Images from CO Forests}
        \label{forest_fitting}    
        Ideally, we would directly measure CO emission by fitting each spike in a given spectrum with a line profile. 
But R branch lines overlap at the spectral resolution of JWST/NIRSpec G395M, and spectroscopic analysis of line profiles is hampered by other spectral lines and ices present in each spaxel (see high S/N examples in the upper panels of Figure \ref{fig:systematics} in green). If line overlaps are ignored, line images, apertures, and analyses will include excess artificial continuum and extended emission. Narrow emission lines, such as H$_2$, H~I, and [Fe~II], overlap CO lines. Ordered by increasing wavelength, overlapping ice features include $^{12}$CO$_{2}$ (4.27 \mic), $^{13}$CO$_{2}$ (4.39 \mic), OCN$^{-}$ (4.60 \mic), $^{12}$CO (4.675 \mic), $^{13}$CO (4.779 \mic), and OCS (4.90 \mic). For detailed analyses of ice properties beyond the scope of this work, please see \citet{Brunken_2024}, \citet{Nazari_2024}, and \citet{Slavicinska_2024}, Tyagi et al. in prep. 

We summarize our procedure for extracting line images from IFU cubes. We first retrieve baselines for all spaxels with the goal of separating out continuum and ice features from our spectral cubes (see purple baselines in Figure \ref{fig:systematics}; explained in detail in Appendix \ref{sect:baselines}). We then simultaneously fit all lines for all spaxels using Gaussian profiles (detailed in Appendix \ref{sect:fit_all}) and identify any systematic effects from our fitting procedure as well as from the instrument itself (Appendix \ref{sect:systematics}). In general, any changes seen in line images as a function of quantum numbers ($v$, $J$) may be influenced by variations in S/N, temperature, column density, self-opacity, and extinction. Images of selected CO transitions are shown in Figure \ref{fig:co_imgs} for each source, including a low-$J$ $v=1-0$ transition, a high-$J$ $v=1-0$ transition, and a $v=2-1$ transition in each row.

        \subsubsection{Apertures and Extracted Line Fluxes}
        \label{apertures}    
        Apertures can capture CO in scattered light from protostellar disks or in protostellar outflows. 
With the exception of IRAS 16253-2429, CO emission is offset from the position of the central protostar (see Table \ref{tab:source_props} for mm/sub-mm continuum coordinates) because the central sources are obscured.
Comparing sources from our CO images (Figure \ref{fig:co_imgs}) against those from other tracers \citep{Federman_2023}, the positions of the CO emission sources do not completely match those of the continuum, ionized jet (e.g. [Fe~II]), or other molecules (e.g. H$_2$). 
Due to the previously mentioned complications, our centrally located apertures could contain a combination of CO emission in scattered light from a disk's surface and from the outflow cavity as well as emission directly from disk winds. 

We maximize emission towards the central sources of CO by choosing the aperture locations and sizes to encompass the emission across the outflow cavity and near the central dust continuum source. 
A single set of apertures is chosen for each target across all wavelengths (see Table \ref{tab:apertures}) to detect
the full range of rotational and vibrational CO transitions (see each row in Figure \ref{fig:co_imgs}).  
We sum the intensity 
within the set of magenta rectangular apertures shown in Figure \ref{fig:co_imgs}. The summed flux densities ($F_\lambda = \sum_{i} \Delta\Omega \ I_{\lambda,i}$, where $\Delta\Omega$ is the solid angle of a pixel) are converted from MJy~sr$^{-1}$ 
to erg~sec$^{-1}$~cm$^{-2}$~\micron$^{-1}$. We neglect the influence of beam dilution as our chosen apertures are all $>$0\farcs5 and are larger than the spatial resolution of JWST/NIRSpec. In Figure \ref{fig:systematics}, we display the extracted spectra from summing all spaxels within all apertures for each source before baseline removal. Table \ref{tab:apertures} lists the centers and dimensions of these apertures. 

\begin{deluxetable*}{cccccc}
    \tablewidth{0pt}
    \tablecaption{CO Aperture Properties Across Central Outflow Cavity}
    \tablehead{
        \colhead{CO Source} & \colhead{Center\tablenotemark{a}} & \colhead{Size ($l, w$) \tablenotemark{b}} & \colhead{NIRSpec Rel. Speed\tablenotemark{c}} & \colhead{MIRI Rel. Speed\tablenotemark{c}} & \colhead{$A_V$ \tablenotemark{d}} \\ 
        \colhead{(--)} & \colhead{(hh:mm:ss, dd:mm:ss)} & \colhead{(au)} & \colhead{(km sec$^{-1}$)} & \colhead{(km sec$^{-1}$)}  & \colhead{(--)}      }
    \startdata
        IRAS 16253-2429         & 16:28:21.62, -24:36:24.10     & $80, 80$   &       --           &              --            &  22.68$^{+0.45}_{-0.45}$  \\ 
        B335-W                  & 19:37:0.87, +07:34:09.39      & $110, 110$ &  \multirow{2}{*}{8$\pm$4.2} &  \multirow{2}{*}{24$\pm$18.5}     &   39$^{+2.7}_{-2.7}$ \\ 
        B335-E                  & 19:37:0.92, +07:34:09.39      & $110, 110$ &                     &                          &   36.86$^{+4.68}_{-4.79}$ \\ 
        HOPS 153-SE             & 05:37:57.04, -07:06:56.23     & $210, 210$ &  \multirow{2}{*}{1$\pm$1.9} &  \multirow{2}{*}{1$\pm$1.9}     &   31$^{+16.3}_{-7.2}$ \\ 
        HOPS 153-NW             & 05:37:56.99, -07:06:55.33     & $210, 210$ &                     &                         &  25.08$^{+0.71}_{-0.71}$  \\ 
        HOPS 370-S              & 05:35:27.64, -05:09:34.85     & $260, 260$ &  \multirow{2}{*}{2$\pm$2.5} &  \multirow{2}{*}{5$\pm$1.5}     &  7.6$^{+0.34}_{-0.36}$  \\ 
        HOPS 370-N              & 05:35:27.64, -05:09:34.05     & $260,260$  &                     &                         &  11.81$^{+0.30}_{-0.29}$  \\ 
        IRAS 20126+4104-SE      & 20:14:26.07, +41:13:32.34     & $760, 680$ &  \multirow{2}{*}{3$\pm$1.6} &  \multirow{2}{*}{3$\pm$1.7}     &  15$^{+1.3}_{-1.2}$  \\ 
        IRAS 20126+4104-NW      & 20:14:25.94, +41:13:33.54     & $760, 680$ &                     &                        &  9.42$^{+0.34}_{-0.34}$  \\ 
    \enddata
    \tablenotetext{a}{The central positions are measured in regions of elevated line-to-continuum emission ratios within an aperture. Compact continuum sources probing dust and measured with mm/sub-mm inteferometry (Table \ref{tab:source_props}) do not have locations that directly track CO emission. The relative calibration for sky coordinates is consistent with respect to MIRI \citep{Federman_2023}.} 
    \tablenotetext{b}{Rectangular apertures with longer side $l$ by shorter side $w$ (see Figure \ref{fig:co_imgs}). We converted to au using the distances from Table \ref{tab:source_props}. HOPS 153 and IRAS 20126+4104 had their apertures rotated by 50$^{\circ}$ and 35$^{\circ}$ counterclockwise about center, respectively.}
    \tablenotetext{c}{The mean velocity difference between line centroids from two apertures (e.g. top of Figure \ref{fig:extinc_example}). The uncertainties are from the sample standard deviation.}
    \tablenotetext{d}{The visual extinction values ($A_V$) are derived from fitting rotationally excited H$_2$ lines from NIRSpec and MIRI for each aperture (\citealt{Narang_2024}; see also \citealt{Neufeld_2024} and Salyk et al. 2024, Submitted). They are reported with approximate 1-$\sigma$ asymmetric uncertainties from propagating the measured uncertainties in fluxes.
    }
     \label{tab:apertures}
      \vspace{-0.3in}
\end{deluxetable*}

Before fitting the aperture spectra for spectral line measurements, we re-fitted a spline baseline to correct for minor deviations caused by ices using a reference baseline (from our algorithm in Appendix \ref{sect:baselines}, discussed in Appendix \ref{sect:systematics}, and shown in Figure \ref{fig:systematics}). 
After subtracting the splined baseline from the high S/N spectra, we optimized a fit to determine line fluxes for each spectral line identified (Section \ref{cube_lines}, Appendix \ref{sect:fit_all}) and for each protostar using software called \texttt{fityk} \citep{Wojdyr:ko5121}. For each line, we fitted the FWHM and peak intensity of a Gaussian and then repeated the procedure, simultaneously fitting neighboring lines to account for known line overlaps. We also calibrated line centers independently from the FWHM and peak intensity by permitting each Gaussian to be fitted individually and then in groups until the modeled lines collectively reached an approximately constant offset ($\lambda_0$) from the cube's default wavelength axis. If lines overlap but are separable (e.g. $v=2-1$ compared to $v=1-0$), then we average or interpolate our Gaussian fit parameters between lines of neighboring $J$. We repeated these steps until residuals reached a standard deviation similar to the RMS noise generated by default from the JWST pipeline (statistical $R^2 \sim$ 0.99). The numerically integrated flux ($F$) for each line is then found from a Gaussian with the measured fit parameters (width $\sigma$ from the FWHM, peak flux density $F_{\lambda}$, and 5 wavelength resolution elements to sum over $\lambda$) for each emission line (i) and set of wavelength resolution elements from a given line center (j) as:
\begin{equation}
    F = \sum_{i=0}^{152}  \sum_{j=-5}^{+5} F_{\lambda,ij} \times \exp{\bigg(\frac{-(\lambda_{ij} - \lambda_{0,i})^2}{2 \sigma^2}\bigg)}.\label{equ:flux_sum}
\end{equation}

\begin{figure*}
    \centering
        \includegraphics[width=0.8\linewidth, trim = 2in 0.5in 2in 0.55in,clip]{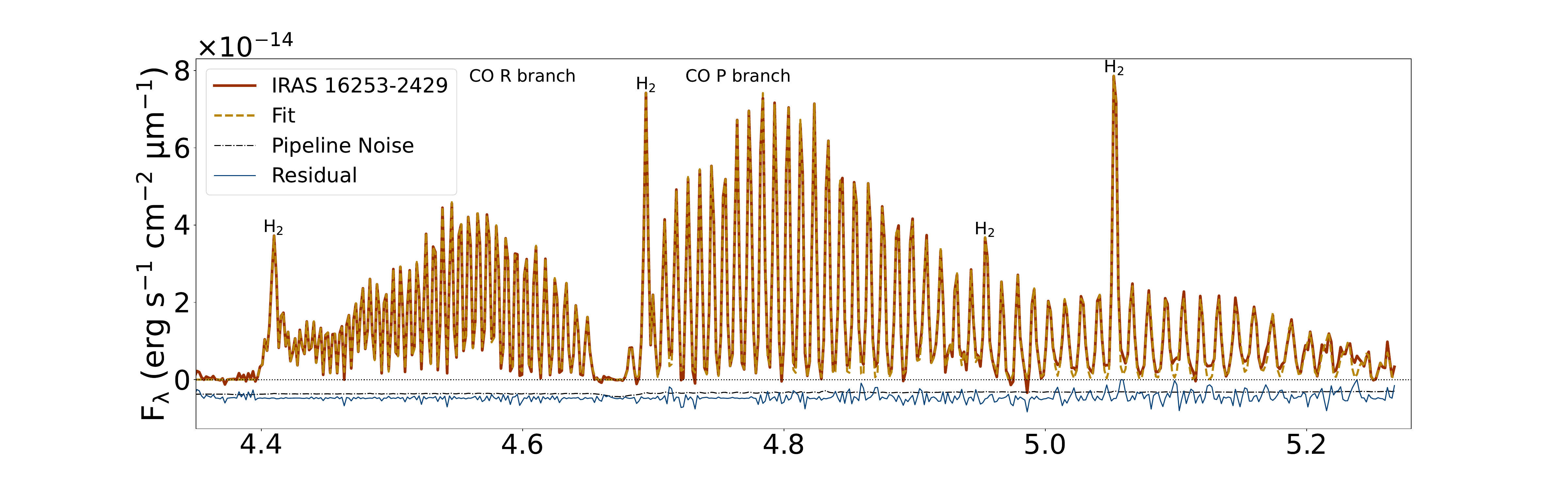}
        \includegraphics[width=0.8\linewidth, trim = 2.5in 0.5in 2in 0.55in,clip]{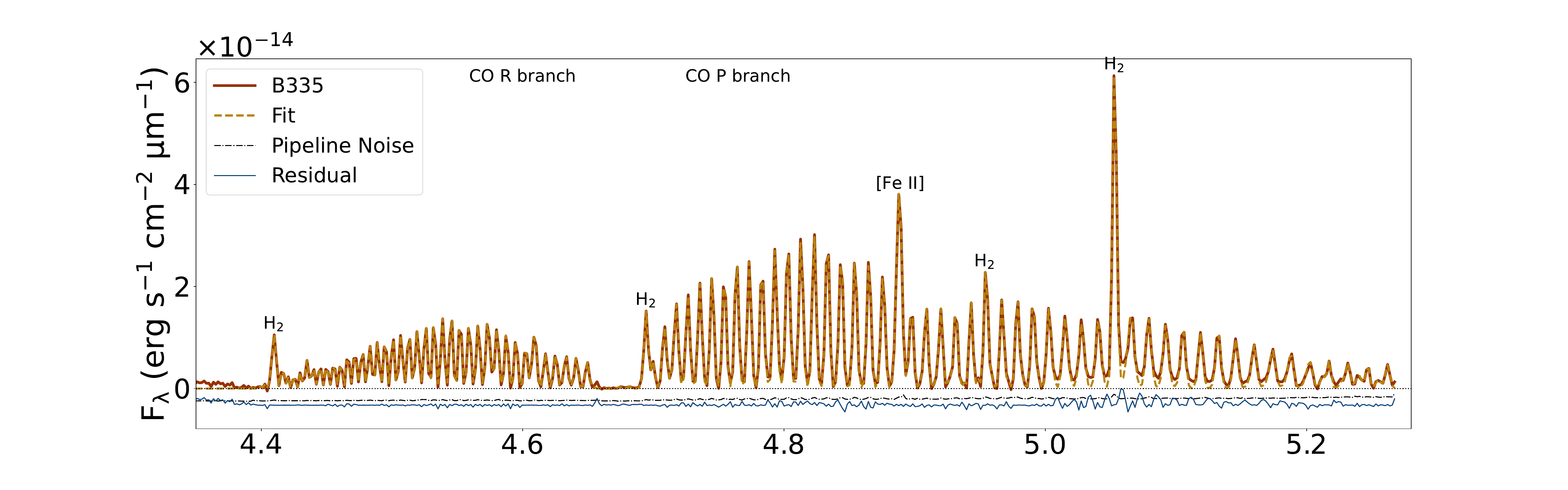}
        \includegraphics[width=0.8\linewidth, trim = 2in 0.5in 2in 0.55in,clip]{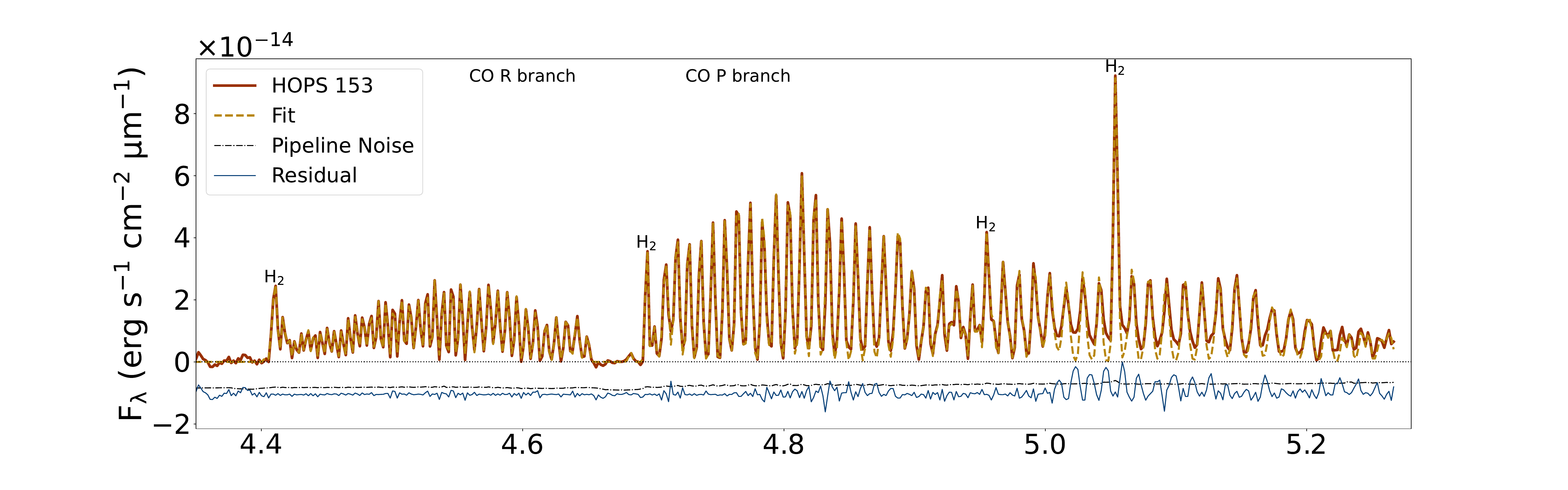}
        \includegraphics[width=0.8\linewidth, trim = 1.5in 0.5in 2in 0.55in,clip]{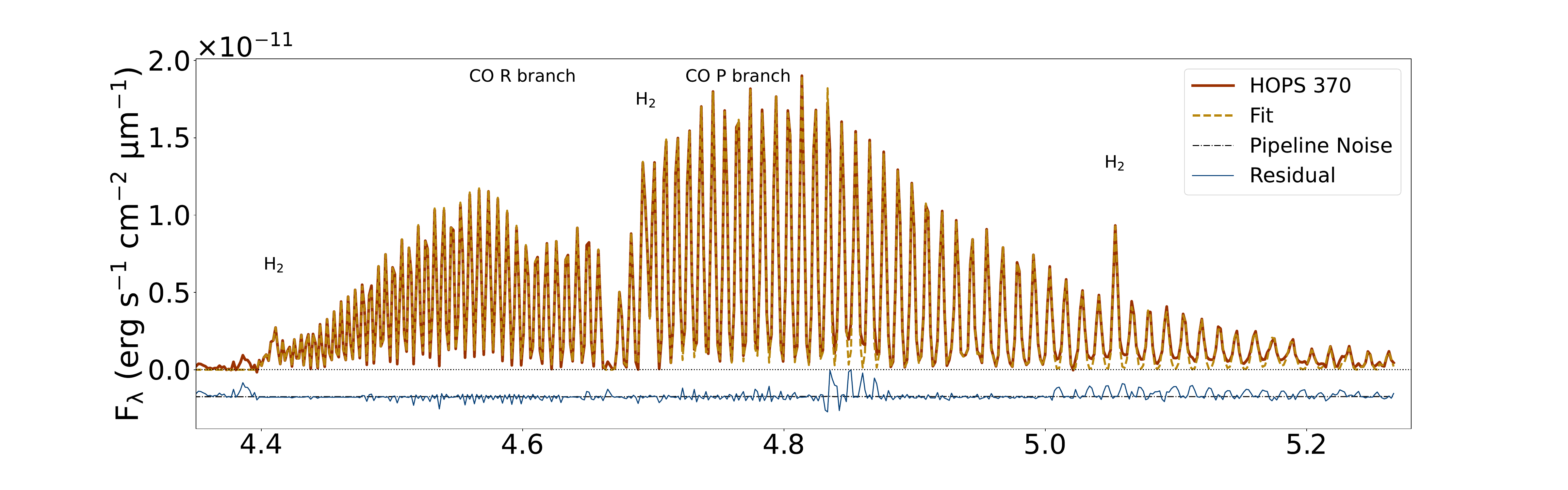}
        \includegraphics[width=0.8\linewidth, trim = 1.5in 0in 2in 0.55in,clip]{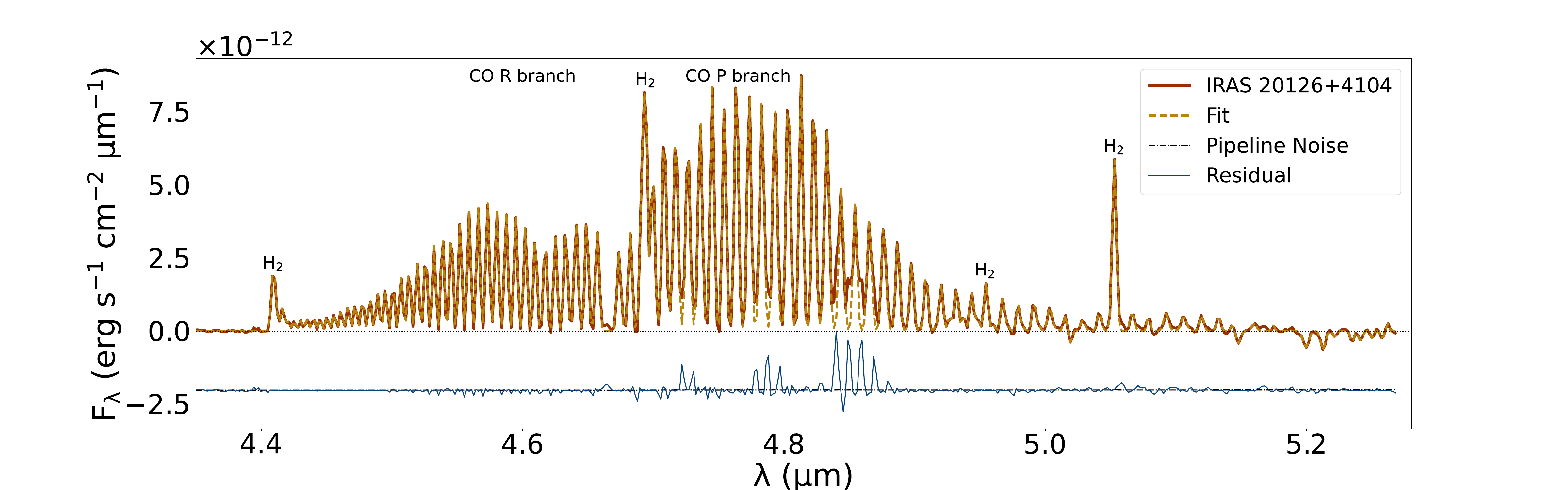}
    \caption{High S/N baseline-subtracted spectra (orange) for each protostar compared to our total fits (dashed yellow) for all spectral lines in Table \ref{tab:co_lines_fluxes}. 
    Selected bright non-CO species are labeled for reference. The CO $v=1-0$ R and P branches meet at 4.658 \mic\ ($J_u \sim 0$, E/$k_B \sim 3100$ K), where ice absorption causes an absence of emission. The two branches increase in $J_u$ away from center up to $J_u \sim 45$.
    Residuals (blue) with mean $\sim$ 0 and pipeline-derived noise (black dash-dot curve) are both offset by the residual's max (unscaled). Remaining residuals do not match known lines and may be unresolvable ${}^{12}$CO lines.}\label{fig:example_spaxels}
\end{figure*}

The full fit summed from all line profiles and the residuals leftover from fitting are shown in Figure \ref{fig:example_spaxels}. For complete results, including line flux measurements and their 1-$\sigma$ uncertainties for each protostellar source and for each CO $v=1-0$ and $v=2-1$ transition, see Table \ref{tab:co_lines_fluxes} in Appendix \ref{sect:flux_table}. For details about uncertainty and uncertainty propagation, see Appendix \ref{sect:uncert_propagation}. The shortest wavelength ($\sim 4.4\ \mu$m) R branch lines at $J>42$ can be seen in residuals and are not fitted because their line spacing was too fine to be resolved. Some residuals are prominent at the wings of CO lines, especially for the high-mass sources from 4.7 to 4.9 \mic. The effect may be due to spectrally unresolved ${}^{12}$CO $v=2-1$ lines causing deviations at the wings of the Gaussian profile \citep{Temmink_2024}, or unknown or uncommon ice features affecting our baselines (Section \ref{sect:systematics}). 

We quantified the systematic effects of our optimized fits by measuring the FWHM of each line (bottom panel of Figure \ref{fig:systematics}). Compared to the pre-launch spectral resolution of NIRSpec/395M\footnote{\label{ftnote:sp_res_table}For tabulated disperser and spectral resolution data, see https://jwst-docs.stsci.edu/jwst-near-infrared-spectrograph/nirspec-instrumentation/nirspec-dispersers-and-filters}, the broader lines (points below the median curve) at longer wavelengths mean the CO forest's P branch lines may be slightly resolved, and the R branch is at the same time narrower and blended. The $v=2-1$ lines are dim but are evident at wavelengths longer than 5.2 $\mu m$, so they may also be blended, therefore causing many of the points in Figure \ref{fig:systematics} that are above the G395M pre-launch profile at wavelengths longer than $4.8$ \mic.
        
    \subsection{MIRI/MRS Spectra}
    \label{jwst_miri}
    MIRI/MRS is an IFU similar to NIRSpec (Section \ref{jwst_nirspec}). It covers longer wavelengths from 4.9 to 27.9~\mic, has higher spectral resolution (from longer wavelengths to shorter wavelengths, $R \sim 1500 - 4000$, or from shorter to longer as $\Delta v \sim 75 - 200 \ \rm km \ {sec}^{-1} $)\footnote{\label{ftnote:sp_res_miri}For relevant references and tables, see https://jwst-docs.stsci.edu/jwst-mid-infrared-instrument/miri-observing-modes/miri-medium-resolution-spectroscopy\#gsc.tab=0}, but has lower spatial resolution (0\farcs27 at the shortest wavelength channel to 1\arcsec\ at longer wavelengths)\footnote{\label{ftnote:spatial_miri}For the FWHM as a function of wavelength, see https://jwst-docs.stsci.edu/jwst-mid-infrared-instrument/miri-performance/miri-point-spread-functions\#gsc.tab=0 or \citet{Rigby_2023}}. For all MIRI cubes, we used CRDS version 11.17.2. The pmaps are 1100 for B335, HOPS 370, and IRAS 20126+4104, while 1105 was used for IRAS 16253-2429 and HOPS 153.

\begin{figure*}
    \centering
        \includegraphics[width=\linewidth, trim = 0in 0in 0in 0in,clip]{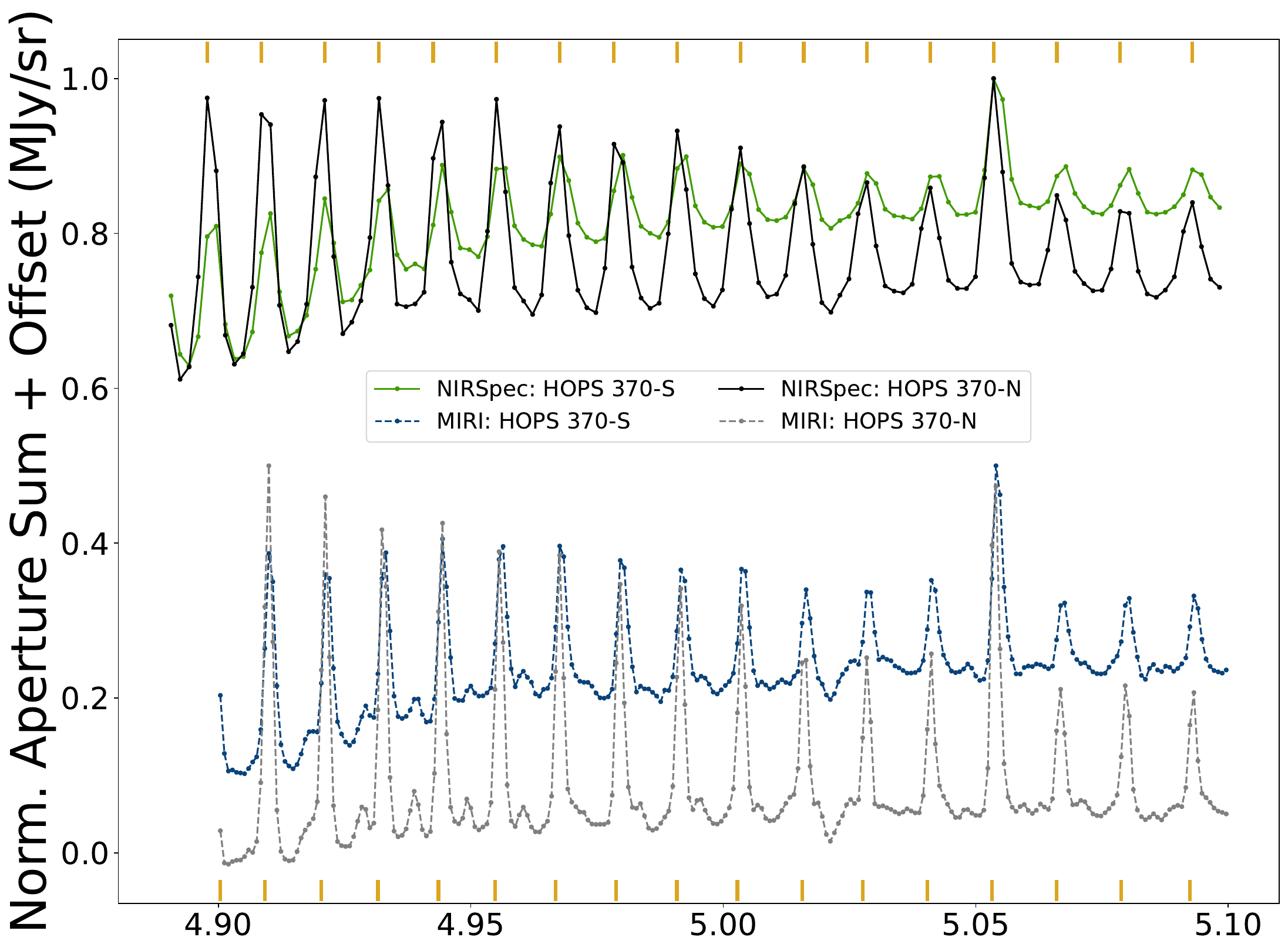}
        \includegraphics[width=\linewidth, trim = 0in 0in 0in 0in,clip]{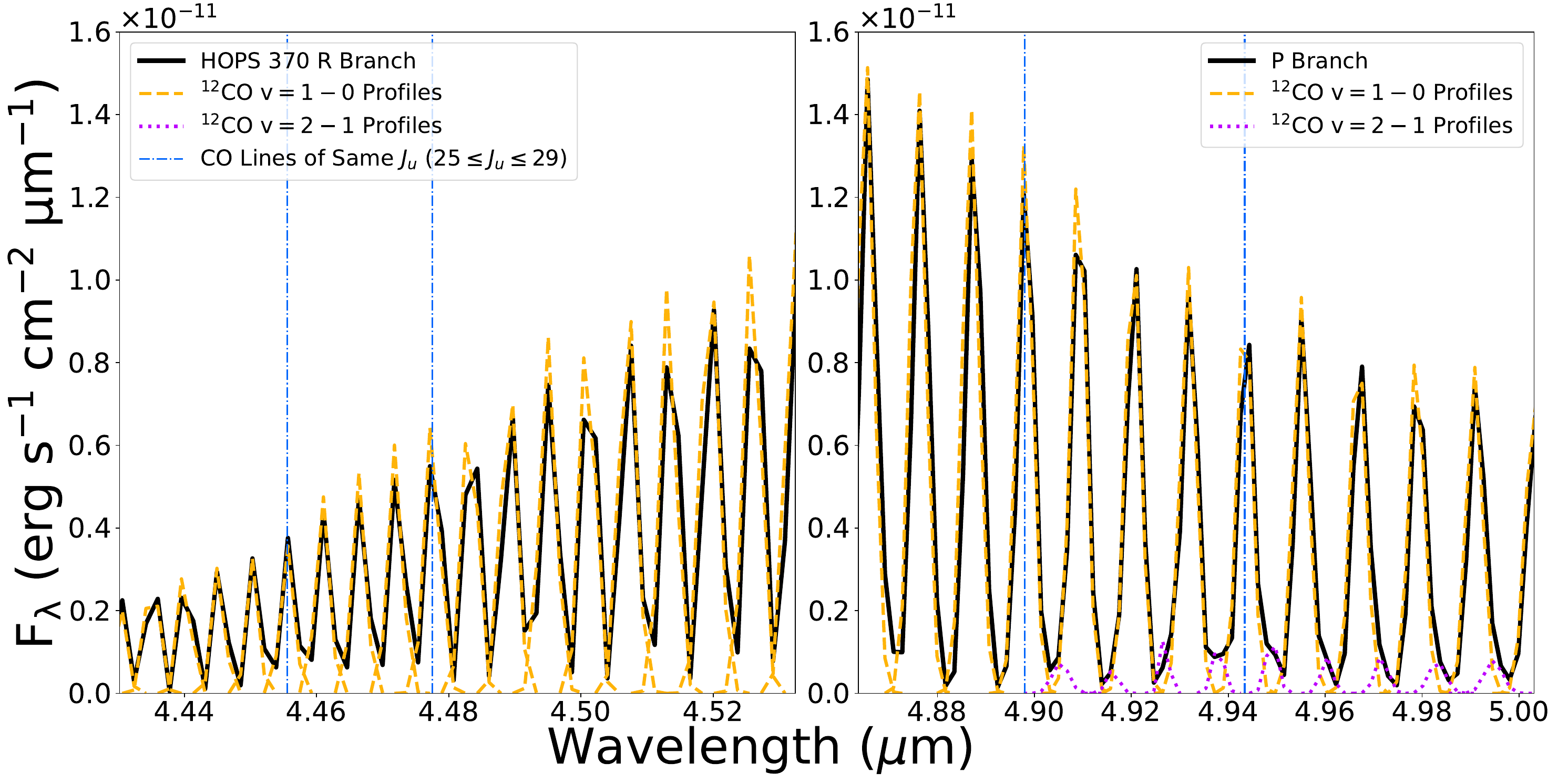}
    \caption{Complexities of CO spectra illustrated by HOPS 370. \textit{\underline{Top}}: Comparing the summed and normalized raw spectra from apertures across the outflow cavity (Table \ref{tab:apertures}, Figure \ref{fig:co_imgs}). For reference, the H$_2$ S(8) line is at 5.053 $\mu$m. The solid green and black lines are from NIRSpec, while the dashed blue and gray lines are from MIRI. 
    \textit{\underline{Bottom}}: A zoomed in, baseline-subtracted spectrum (solid black lines) to show asymmetry between the CO R (left) and P (right) branches. The dashed yellow lines shown the fitted $v=1-0$ line profiles and the dotted purple lines show the measured $v=2-1$ line profiles. For reference, lines from common upper states are marked within the blue dash-dot vertical lines.}\label{fig:extinc_example}
\end{figure*}

MIRI/MRS spectra are used to assess velocity information (Section \ref{absorption_doppler}), check our line profiles for indications of isotopologues (Section \ref{implied_profile}), and measure rotationally excited H$_2$ emission lines to estimate extinction (Section \ref{co_ext}). In the top panel of Figure \ref{fig:extinc_example}, we show example raw spectra (the highest S/N, HOPS 370) extracted with the same apertures as our NIRSpec spectra (see Table \ref{tab:apertures} and Figure \ref{fig:co_imgs}). The spectra are from NIRSpec (solid green and black lines) and MIRI (dashed blue and gray lines), and they are normalized and offset for comparison. For reference, the vertical solid yellow lines mark ${}^{12}$CO $v=1-0$ lines, the H$_2$ S(8) line is at 5.053 $\mu$m, and the absorption line at approximately 5.025 $\mu$m is possibly due to H$_2$O. The spectral region shown includes the lines with our best residuals (best separated CO $v=2-1$ lines) and the overlapping wavelengths between NIRSpec and MIRI (approximately 4.90 to 5.10 $\mu$m). The bottom panel shows CO observations in black, CO $v=1-0$ fits in dashed yellow, and $v=2-1$ fits in dotted purple. Lines with a common upper state between the R (bottom left) and P (bottom right) branches are highlighted with the vertical dashed blue lines where the P branch was best constrained (see Figure \ref{fig:example_spaxels}).

        \subsubsection{CO Velocity Information Across the Outflow Cavity}
        \label{absorption_doppler}

If CO gas emission is part of an outflow,  we would expect a difference in the velocity of the gas between opposite sides across the outflow cavity.
To precisely determine relative velocities, we measure the centroids of all available CO $v=1-0$ emission lines (one is excluded near the H$_2$ S(8) line at 5.053 $\mu$m) in the wavelength range shown in Figure \ref{fig:extinc_example} when two apertures are available. The relative velocities are found as  ${|\Delta v_{rel}|}/{c} = {|\Delta \lambda|}/{\lambda_0}$. Here, $\Delta v_{rel}$ is the relative velocity derived from the difference in line centroids between two apertures ($\Delta \lambda = \lambda_{aper,1} - \lambda_{aper,2}$). The centroids themselves are found from the flux-weighted average of the closest 5 points to $\lambda_0$, the rest wavelength of a given CO line. The mean and sample standard deviation of the absolute value of the relative velocity is reported in Table \ref{tab:apertures} for all our sources except IRAS 16253-2429.

In general, there is no strong evidence for a velocity shift for any of our sources. Though the B335 CO apertures may have a relative motion of $\sim$10 km sec$^{-1}$, we are limited by spectral resolution and S/N. The abundance of lines from the CO forest and other molecule-rich data using NIRSpec and MIRI may help to measure sources with more extreme Doppler shifts (e.g. those in absorption further from the protostar in \citealt{Federman_2023}). We leave more extended molecular sources and absorption signatures to follow-up work and focus on our higher sensitivity NIRSpec observations of CO emission, using MIRI/MRS spectra to clarify the nature of the underlying line profiles as needed.

\section{Analysis and Results}
    \subsection{Optical Depth}    \label{implied_profile}
Physical properties (e.g. temperature or total gas mass) can be directly extracted from gas populations if ${}^{12}$CO emission line fluxes are optically thin \citep[e.g. ][]{Blake_2004, Brittain_2009}.
%
Detecting isotopologues in emission reveal if gas-phase ${}^{12}$CO emission is optically thin or thick. 
In Class I and II disks, $^{13}$CO is often detected, and $^{12}$CO is generally found to be at least partially optically thick \citep[e.g.][]{Herczeg_2011,Brown_2012, Brown_2013}.

Within our sample, we did not discern any ${}^{13}$CO $v=1-0$ lines. 
Most ${}^{13}$CO lines in NIRSpec overlap ${}^{12}$CO lines, especially  $v=2-1$ lines. 
We also investigated the MIRI data where ${}^{12}$CO $v=1-0$ and $v=2-1$ lines could be separated, but ${}^{13}$CO $v=1-0$ was still undetected, limited by lower sensitivity (top of Figure \ref{fig:extinc_example}).
With no clear detection of ${}^{13}$CO $v=1-0$, the problem reduces to determining an upper limit from the noise by fitting the higher sensitivity NIRSpec line fluxes.

To find upper limits for ${}^{13}$CO $v=1-0$ lines,  
we assume the noise 
includes the rotationally excited lines of the ${}^{13}$CO $v=1-0$ forest.
If true, ${}^{13}$CO $v=1-0$ lines have
FWHM set by that of the ${}^{12}$CO $v=1-0$ with a matching $J$ (using the bottom panel of Figure \ref{fig:systematics}). For the isotopologue's peak intensity, we find the average noise power for each $J$ transition ($\sigma_{13}$) by adding in quadrature our two independent sources of uncertainty over the number of points ($N_p$) within $\pm1$ FWHM from the ${}^{13}$CO line's center: 
\begin{equation}
    \sigma_{13}(J) = \sqrt{\frac{ \sum_{i}^{N_p} (\sigma_{r,i} - \overline{\sigma_{r}})^2 + \sum_{j}^{N_p} (\sigma_{pl,j} - \overline{\sigma_{pl}})^2} {N_p-1} }
\end{equation}
where $\sigma_{r}$ is the residual from fitting, and $\sigma_{pl}$ is the pipeline-derived noise. $\sigma_{r}$ measures point to point scatter and our systematic choices, and $\sigma_{pl}$ measures the instrument's known sensitivity and limits on S/N.

\begin{figure*}
    \centering
    \includegraphics[width=\linewidth, trim = 0in 0in 0in 0in,clip]{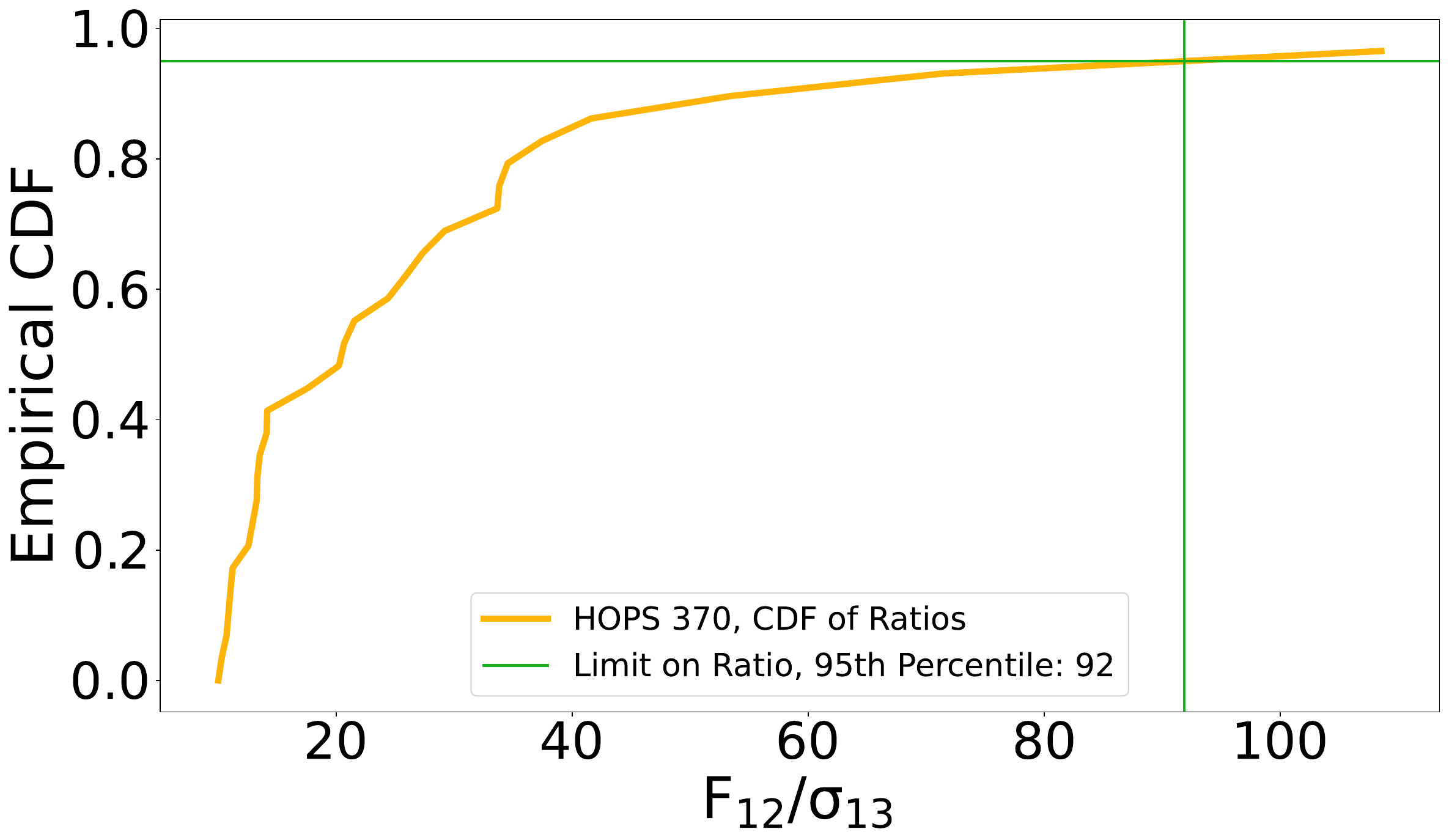}
    \caption{An example ${}^{13}$CO isotopologue ratio given the residuals from HOPS 370. 
    The EDF (yellow curve) shows a consistently chosen percentile (green) that informs our lower limit. 
    }\label{fig:upperlimit_analysis}
\end{figure*}

For each $J$, 
we integrate the noise power ($\sigma_{13}(J)$) in the same manner as the ${}^{12}$CO lines (Section \ref{forest_fitting}, Equation \ref{equ:flux_sum}). 
In Figure \ref{fig:upperlimit_analysis}, we plot the empirical distribution function (EDF) of the ratios of $^{12}$CO line fluxes to $^{13}$CO integrated noise power ($F_{12}(J) / \sigma_{13}(J)$). 
We included $\sim 30-40$ lines in each branch ($\sim 70$ lines total) with $2<J_u<40$. We only show HOPS 370 to demonstrate the procedure.
We excluded from analysis stretches of the spectrum where the fits were poor, due to ice features at 4.7 to 4.8 \mic\ and near 4.85 \mic. The edges of the spectrum ($<4.6$ \mic\ and $>5$ \mic) are also excluded as the gas-phase $^{12}$CO lines become weaker and more sensitive to our baseline (see systematic effects in Appendix \ref{sect:systematics}).
In table \ref{tab:co_rot_vib_props}, we report the ${}^{12}$CO/${}^{13}$CO flux ratios at the 95th percentile for consistent comparison. 

\begin{deluxetable*}{ccccccccccc}
    \centering
    \tablewidth{0pt}
    \tablecaption{Ro-vibrational CO Gas Properties and Total Warm Gas Mass}
    \tablehead{
        \colhead{CO Source} & \colhead{$\frac{F_{12}}{\sigma_{{13}}}$}  & \colhead{$T_{1-0, 1}$} & \colhead{$T_{1-0, 2}$} & \colhead{$T_{2-1}$} & \colhead{$N^{tot}_{1-0, 1}$} & \colhead{$N^{tot}_{1-0,2}$} & \colhead{$N^{tot}_{2-1}$} 
        & \colhead{$\overline{T_{vib}}$} & \colhead{$N_{CO,tot}$} & \colhead{$M_{gas,NIR}$} \\
        \colhead{(--)} & \colhead{(--)} &  \colhead{(10$^{3}$ K)} &  \colhead{(10$^{3}$ K)} & \colhead{(10$^{3}$ K)}  &    \colhead{(--)}  &    \colhead{(--)}  &    \colhead{(--)} &         \colhead{(10$^{3}$ K)} & \colhead{(--)} & \colhead{(g)} \\
        \colhead{(1)} & \colhead{(2)} &  \colhead{(3)} &  \colhead{(4)} & \colhead{(5)}  &    \colhead{(6)}  &    \colhead{(7)}  &    \colhead{(8)} &         \colhead{(9)} & \colhead{(10)} & \colhead{(11)}
        }
    \startdata
        IRAS 16253-2429 & $>65$ & 1.03$\pm 0.033$ & $73 \pm 138$ & $301 \pm 13.5$ & $1.6 \times 10^{41}$ & $6.4 \times 10^{40}$ & $2.8 \times 10^{39}$ & 1.2 $\pm$ 0.49 & $2.5 \times 10^{41}$  & $7.0 \times {10}^{21}$ \\
        B335 & $>71$ & 1.17$\pm 0.111$ & 5.13$\pm 0.763$ & -- & $1.5 \times 10^{41}$ & $2.6 \times 10^{40}$ & $1.7 \times 10^{40}$ & 2.0 $\pm$ 0.86 & $3.7 \times 10^{41}$  & $1.0 \times {10}^{22}$ \\
        HOPS 153 & $>105$ & 1.08$\pm 0.06$ & 8.78$\pm 4.697$ & -- & $1.1 \times 10^{42}$ & $1.8 \times 10^{41}$ & $5.6 \times 10^{40}$& 1.5 $\pm$ 1.1 & $2.1 \times 10^{42}$  & $6.0 \times {10}^{22}$ \\
        HOPS 370 & $>92$ & 1.15$\pm 0.022$ & 2.55$\pm 0.121$ & 3.68$\pm 0.0$ & $1.2 \times 10^{44}$ & $2.3 \times 10^{43}$ & $1.6 \times 10^{42}$ & 0.9 $\pm$ 0.23 & $1.7 \times 10^{44}$  &  $4.8 \times {10}^{24}$ \\
        IRAS 20126+4104 & $>106$ & 0.616$\pm 0.0425$ & 1.59$\pm 0.386$ & 3.45$\pm 0.0$ & $7.6 \times 10^{45}$ & $1.1 \times 10^{44}$ & $9.3 \times 10^{42}$ & 0.99 $\pm$ 0.25 & $1.1 \times 10^{46}$  & $3.1 \times {10}^{26}$ \\
    \enddata
    \tablecomments{Gas population properties for each source calculated directly by using the line fluxes in Table \ref{tab:co_lines_fluxes} as described in Sections \ref{implied_profile}, \ref{co_rotation}, \ref{co_vib}, and \ref{co_mass}. (\underline{Column 2}) The intensity ratio of the ${}^{12}$CO and ${}^{13}$CO isotopologues using the unmodified distribution of noise drawn from the 95th percentile with respect to the distribution of line intensity ratios. (\underline{Columns 3 to 5}) The rotational temperature components ($T_{1-0,1}$) are found from the slope of linear fits in Figure \ref{fig:co_ext_rot}. The uncertainties are from the standard error, the inverse of the covariance matrix given the propagated uncertainties for each measured point. For IRAS 16253-2429, the rotational temperatures for the higher excited states are nearly horizontal fits, which may imply a non-thermal deviation from our model. For B335 and HOPS 153, their $v=2-1$ lines have lower SNR, so their $v=2-1$ rotational temperatures are not reported. (\underline{Columns 6 to 8}) The total numbers of molecules are the intercepts of the fits for all measured $J_u$ states for each set of vibrational transitions. 
    We only report uncertainties in the temperatures as the number of molecules is an extrapolation dependent on systematic effects (optical depth, extinction).  
    (\underline{Column 9}) Vibrational temperatures ($\overline{T_{vib}}$) are found from the mean ratio of rotational lines at the same $J_u$ from adjacent vibrational states using Equation \ref{equ:tvib}. Uncertainties in the $\overline{T_{vib}}$ come from the sample variance of values among all measurements. (\underline{Column 10 to 11}) Total numbers of CO molecules ($N_{CO,tot}$) and NIR gas masses ($M_{gas,NIR}$) are estimated by applying a Boltzmann factor to the total number of molecules in $v=1-0$. The gas masses assume an average fractional abundance of CO to H$_2$ of $\frac{1}{6000}$. $N_{CO,tot}$ and $M_{gas,NIR}$ are lower limits since colder gas may go undetected if optically thicker than we assumed.
     }\label{tab:co_rot_vib_props}
      \vspace{-0.3in}
\end{deluxetable*}


For optical depth, we compare to the $^{12}$C/$^{13}$C abundance ratio of 67 in molecular clouds around Orion \citep{Langer_1993}. Note the local ISM conditions may vary for some of our sources, but the standard abundance ratio of the ISM is generally $\sim$60 \citep{Jacob_2020}.
The protostars all have flux ratios exceeding the standard ISM conditions, so they may be modestly optically thick only in the strongest lines or at low-$J$ that we cannot probe. 
The higher-mass sources, with higher S/N, consistently show elevated values of approximately 100 or higher. 
Enhanced ratios are not unprecedented around Class II protoplanetary disks and diffuse nebulae and are often attributed to UV photochemistry \citep{Lambert_1994, Federman_2003, Goto_2003, Smith_2015}  or differences in sublimation temperatures for colder gas \citep{L_Smith_2021}.

Restricted to analyzing noise, our estimated CO isotopologue flux ratios show forests of $^{12}$CO emission that are overall inconsistent with being optically thick. 
Thus, we proceed to analyzing extinction and the rovibrationally excited CO gas populations assuming ${}^{12}$CO is optically thin, though we warn that this assumption may not be applicable for the brightest or lowest $J$ lines. In practice, we minimize use of low-$J_u$ lines (e.g. $<10$), especially as they are also affected by the OCN$^{-}$ and ${}^{12}$CO ice features.

    \subsection{Extinction and P/R Asymmetry}
    \label{co_ext}     
We correct CO fluxes for extinction to estimate physical properties of CO gas within our apertures, such as temperature and column density \citep[e.g.][]{Goldsmith_1999}.
We use the extinction model by K. Pontoppidan ("KP5"), which is implemented with \texttt{OpTool} \citep{Dominik_2021} and applicable to deeply embedded protostars \citep{Pontoppidan_2024}, to derive the total extinction cross section ($\kappa_{ext}$). $\kappa_{ext}$ is the sum of contributions due to dust scattering ($\kappa_{sca}$) and absorption ($\kappa_{abs}$).

The most direct method to estimate extinction would use the fact that we have the full spectrum of the CO fundamental with both the R and P branches. In the absence of spectral-line opacity or radiative pumping, transitions from a common upper level at different wavelengths provide a measure of reddening  (e.g. \citealt{Miller_1968}), which can be translated into extinction using a model of opacities. Applied to our spectra, this method would produce nonphysical results, such as individual CO line luminosities ($J_u$ as high as 20) being predicted to exceed the protostar's bolometric luminosity (HOPS 370 in particular). Figure \ref{fig:example_spaxels} shows that the R branch is always substantially weaker than the P branch for our sources. Infrared radiation in the vibrational band (either continuum or line emission from hotter CO) will transfer the $v = 0$ populations to the $v = 1$ levels while producing a strong P/R asymmetry \citep{GonzalezAlfonso_2002, Lacy_2013}. The asymmetry may be caused by favoring absorption in the R branch and emission in the P branch, which relies on having a radiation temperature that differs from the kinetic temperature \citep{Lacy_2013}.

P Cygni profiles are often seen in R branch lines in velocity-resolved observations (for past observations of young stars see \citealt{Evans_1991}; \citealt{Rettig_2005}; \citealt{Barentine_2012}, and for recent, similar JWST/NIRSpec observations, but of AGNs, see \citealt{Buiten_2023}; \citealt{PereiraSantaella_2024}; \citealt{GarciaBernete_2024}; \citealt{GonzalezAlfonso_2024}). Our observations lack the spectral resolution needed for confirmation. We see hints in the top of Figure \ref{fig:extinc_example} that, relative to HOPS 370-S, HOPS 370-N appears to have a systematic sub-spectral pixel offset to redder wavelengths in NIRSpec, which is not seen in the MIRI spectra. The spectra indicate possible blueshifted absorption (e.g. around 4.95 to 5.00 $\mu$m), and the systematic offset we observed in NIRSpec could therefore result from blending an unresolved P Cygni profile as found with similar spectral resolutions to ours in \citet{GonzalezAlfonso_2002} or \citet{Lacy_2013}. Blended P Cygni profiles may also explain why the R branch is narrower than expected (bottom of Figure \ref{fig:systematics}).

Instead, we use the purely rotationally excited H$_2$ emission lines for extinction correction because they are shown to produce consistent \Av\ when compared to extinction derived from ice features and H$_2$ \citep{Salyk_2024} and, like CO, they do not show a significant velocity shift between our apertures (e.g. Figure \ref{fig:extinc_example}). We use the method by \citet{Narang_2024} based upon molecular rotation diagrams (see also \citealt{Neufeld_2024}). As with our extracted aperture fluxes (Section \ref{apertures}), we neglect the influence of beam dilution or convolving the spatial and spectral PSFs for each line because the chosen apertures are larger than the largest MIRI beam size. We included S(1), S(2), S(3), S(4), S(7), S(8), S(11), S(12), S(13), and S(14) for all sources except HOPS 153-SE, for which we have only upper limits to the S(3) line likely due to high extinction. The visual extinctions ($A_V$) are reported in Table \ref{tab:apertures} for each aperture. The values are also given with asymmetric uncertainties (approximately 1-$\sigma$ when propagated through the fitting procedure). IRAS 16253-2429 has H$_2$-derived extinction similar to that derived from OH and CO$_2$ emission lines (Watson et al. 2024 in prep.). For sources with two apertures, $A_V$ agrees within a factor of 2, which is less for optical depth at the wavelengths of CO $v=1-0$ lines (e.g. a factor of $<$1.5 at 4.7 \mic), so we use the mean $A_V$ for correcting line fluxes.
We do not include the systematic uncertainties from the extinction law into later calculations.  

    \subsection{CO Population Diagrams}
    \label{co_rotation}    

The population diagram is a tool to analyze the temperature and column density of extinction-corrected molecular line emission (e.g. Appendix A in \citealt{Turner_1991}, \citealt{Blake_1995}, and \citealt{Goldsmith_1999}; or observational examples in \citealt{Manoj_2013} and \citealt{Green_2013}).   
Assuming the lines are optically thin, population diagrams are constructed from sets of rotationally excited states characterized by a single excitation temperature $T_{ex}$, also called the rotational temperature $T_{rot}$ (as distinct from kinetic gas temperature, $T_K$). 
In general, derived values of temperatures and total column densities can also include deviations from local equilibrium thermodynamics, or LTE \citep{Goldsmith_1999, Yildiz_2015}. 

Per Section \ref{implied_profile} and Section \ref{co_ext}, we proceed by assuming that our chosen set of ${}^{12}$CO lines are optically thin based on our isotopologue flux ratios (Section \ref{implied_profile}, Table \ref{tab:co_rot_vib_props}). We excluded low-$J$ transitions ($J_u < 10$ for $v=1-0$ and $J_u < 13$ for $v=2-1$) due to 
proximity to ice features and only use the CO P branch lines, which are observed with higher spectral resolution (bottom of Figure \ref{fig:systematics}) and therefore better separated from blending with $v=2-1$ and absorption due to possible P Cygni profiles unlike the R branch (Section \ref{co_ext}, Figure \ref{fig:extinc_example}).

\begin{figure*}
    \centering
    \begin{minipage}{0.49\textwidth}
        \includegraphics[width=\textwidth, trim = 0in 0.1in 0in 0.1in,clip]{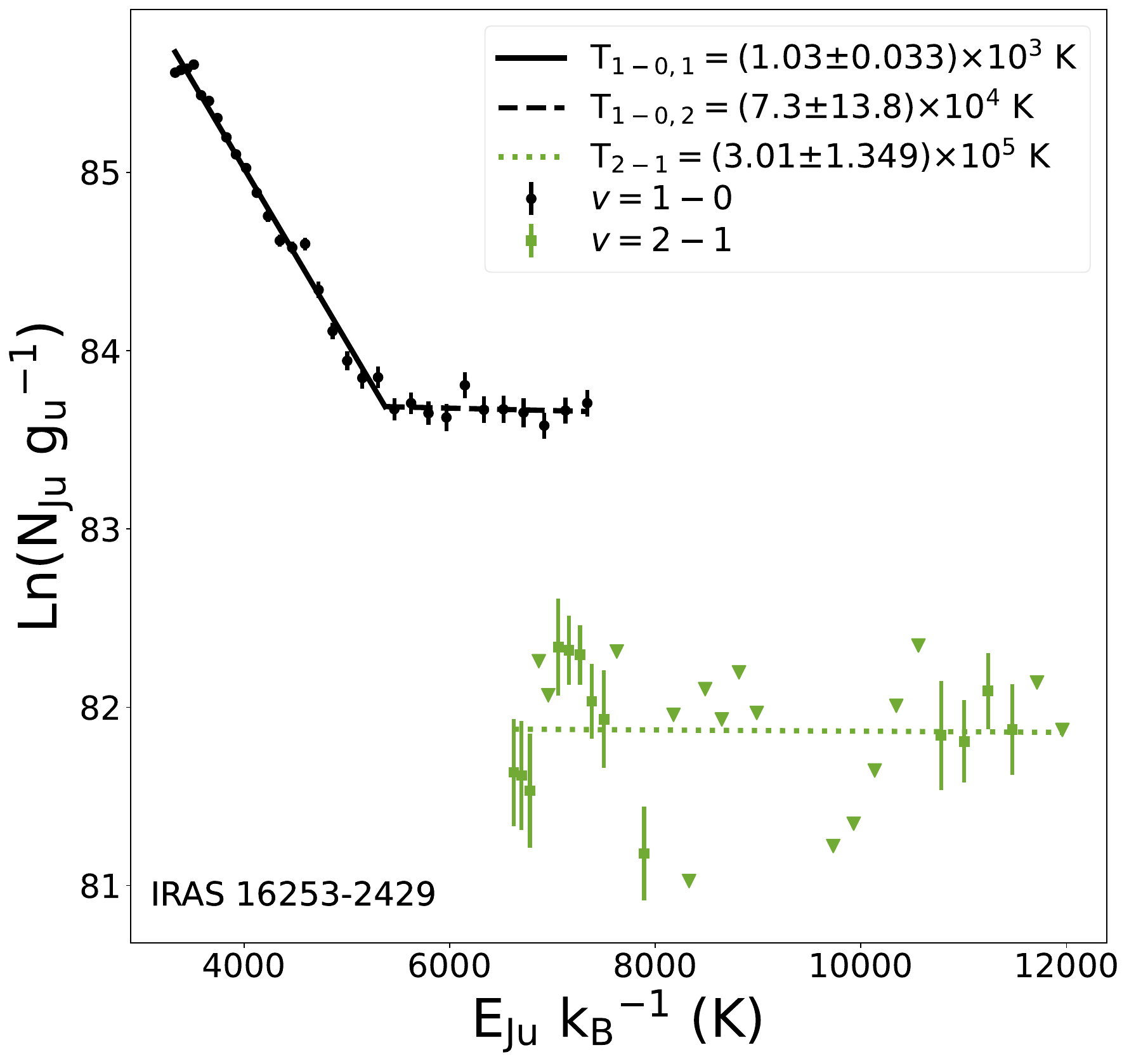}
    \end{minipage}
    \begin{minipage}{0.49\textwidth}
        \includegraphics[width=\textwidth, trim = 0.1in 0.1in 0in 0.1in,clip]{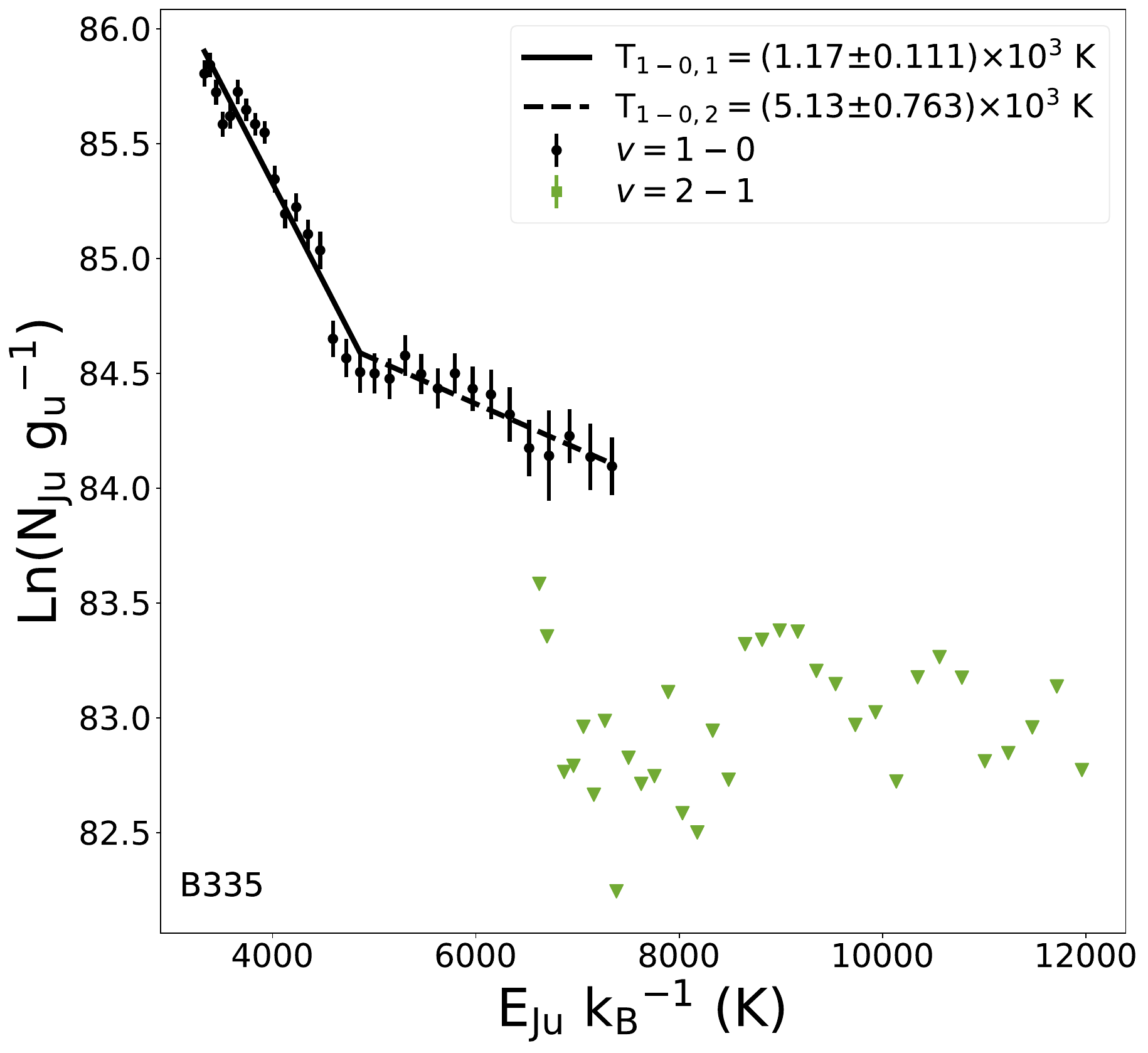}
    \end{minipage}
    \caption{
    CO population diagrams with propagated 1-$\sigma$ uncertainties from extinction-corrected fluxes. See lower left of each panel for source name. The estimated values listed in the legend and in Table \ref{tab:co_rot_vib_props} assume a set of optically thin CO transitions based on Section \ref{implied_profile} ($J_u>10$) only from the P branch (based on Section \ref{absorption_doppler}). The $v=1-0$ data (black points) are fit with two temperature components (cooler is solid, hotter is dashed). When possible, data from $v=2-1$ (green squares) is fit with only one line (dotted). Note the error bars do not show the systematic uncertainty from extinction correction, and 3-$\sigma$ upper limits to $v=2-1$ lines are shown by downward pointing triangles. Rotational temperatures with large uncertainties may imply a deviation from kinetic temperature.}\label{fig:co_ext_rot}
\end{figure*}

To compute $\frac{{N_{J_{u}}^{tot}}}{g_{u}}$, the total number of molecules in upper state $J_u$ per degeneracy of that state, 
we get the line luminosity from dereddened P branch line fluxes by using distances ($d_{star}$) from Table \ref{tab:source_props}, assuming the flux is output isotropically. We find ${N_{J_{u}}^{tot}}$, the number of molecules in $J_{u}$, by turning the line luminosity into a number of molecules using $E_{J_u}$, Einstein A-coefficients retrieved from HITRAN \citep{HITRAN_2022}, and line wavelengths noted in Table \ref{tab:co_lines_fluxes}:
\begin{align}
    \frac{{N_{J_{u}}^{tot}}}{g_{u}} = \frac{1}{hc \ g_u(J_u)} \ F_P(J_u) \frac{\lambda_P(J_u) \ e^{{\tau_{ext,avg}(\lambda_P)}}}{A_{ij,P}} \ (4 \pi {d_{star}}^2), \label{co_units}
\end{align}
where $F_P$ is the integrated CO line flux for a given P branch transition.
Using Section \ref{co_ext}, we de-extinct the fluxes for the wavelength $\lambda_P$ of each line, using the average optical depth
\begin{equation}
    \tau_{ext,avg}(\lambda_P) = \frac{\overline{A_V}}{2.5\ \rm{log_{10}}(e) \ \kappa_{ext}(\lambda = V)} \kappa_{ext}(\lambda_P),
\end{equation}
where $\kappa_{ext}$ is the total opacity from the KP5 law and $\overline{A_V}$ is the visual extinction for each source from Table \ref{tab:apertures} (taken as the mean between apertures when two are used).
Population diagrams are then plotted on a semilog plot: 
\begin{equation}
    \centering
    \ln{\left( \frac{{N_{J_{u}}^{tot}}}{g_{u}} \right)} = \ln{\left( \frac{N_{v}^{tot}}{Z_J(T_{rot})} \right)}- \frac{E_{J_{u}}}{T_{rot}},
    \label{eq:co_rot_line}
\end{equation}
where $N_{v}^{tot}$ is the total number of molecules in a vibrational state of CO gas ($v$), $E_{J_{u}}$ is the energy level identified by $J_{u}$ divided by the Boltzmann constant $k_B$ to have units of temperature for convenience, $T_{rot}$ is a representative temperature of rotationally excited CO gas, and $Z_J$ is the partition function for rotational CO transitions in equipartition (taken from Appendix A3 in \citealt{Evans_1991}, assumed with temperature equal to $T_{rot}$ (i.e. $Z_J(T_{rot}) \approx \frac{k_B T_{rot}}{hcB}$). 
$Z_J$ uses $h$ (Planck's constant), $c$ (speed of light), and $B$ (rotational constant of a given molecule, 1.9225 ${\rm cm}^{-1}$ for CO, according to \citealt{Nolt_1987, Rank_1965}). 
A semilog plot fit by a straight line correlation then shows a chosen set of CO line transitions that occur at approximately constant $T_{rot}$. We also find the total number of molecules in vibrational level $v$ from the line's y-intercept. 
We construct population diagrams where the total number of molecules for each $J_{u}$ is plotted as a function of its level energy (Figure \ref{fig:co_ext_rot}). 
The uncertainties plotted for ${N_{J_{u}}^{tot}}$ due to noise are propagated through adding the uncertainties in the line fluxes from the P branch 
(see Appendix \ref{sect:uncert_propagation}). All CO $v=2-1$ lines with low SNR are plotted as 3-$\sigma$ upper limits (upside down triangles), and including or excluding these points does not significantly affect the results that follow.
We fit straight lines for each component with the form of Equation \ref{eq:co_rot_line}, and for $v=1-0$, we fit a piecewise function consisting of two lines by moving the breakpoint between them until the residuals are minimized. 
For $v=1-0$, the optimal break points tend to be near the P20 (4.85 \mic) but can vary. The $v=1-0$ rotational temperatures are then broken into two components, $T_{1-0,1}$ and $T_{1-0,2}$. The corresponding number of CO molecules from each component of the fit are  $N^{tot}_{1-0,1}$ and $N^{tot}_{1-0,2}$. For $v=2-1$ we only fit one linear component, and due to lower SNR, for B335 and HOPS 153 we are only able to estimate $N^{tot}_{2-1}$ from upper limits. We summarize the rotational temperatures and number of molecules over all $J_{u}$ in CO, for each $v$, and for each source in Table \ref{tab:co_rot_vib_props}. For the rotational temperatures, we show uncertainties from fitting for each representative gas population that we defined. 

\begin{figure*}
    \centering
    \begin{minipage}{0.49\textwidth}
        \includegraphics[width=\textwidth, trim = 0in 0.1in 0in 0.1in,clip]{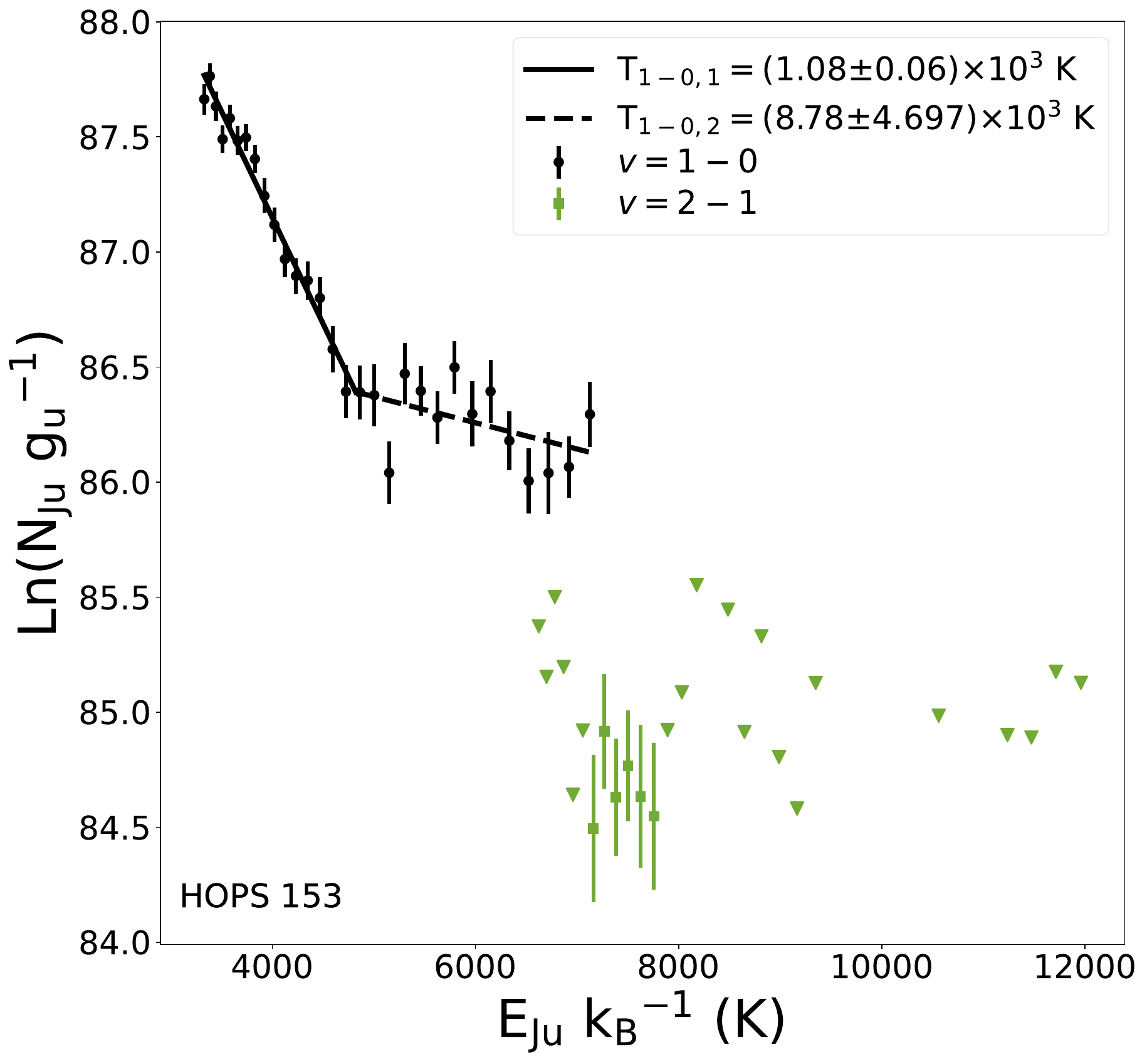}
    \end{minipage}
    \begin{minipage}{0.49\textwidth}
        \includegraphics[width=\textwidth, trim = 0in 0.1in 0in 0.1in,clip]{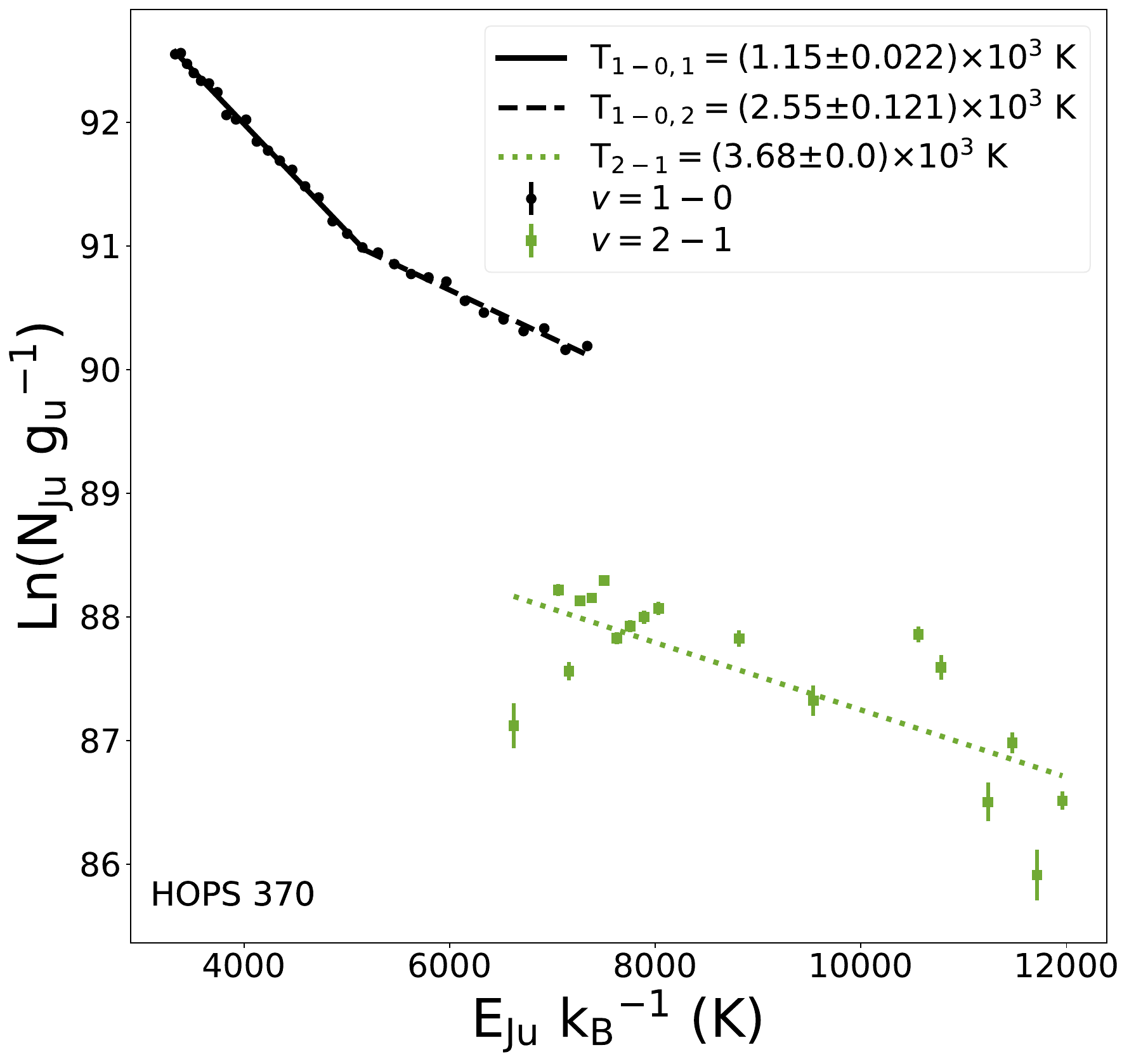}
    \end{minipage}
    \begin{minipage}{0.49\textwidth}
        \includegraphics[width=\textwidth]{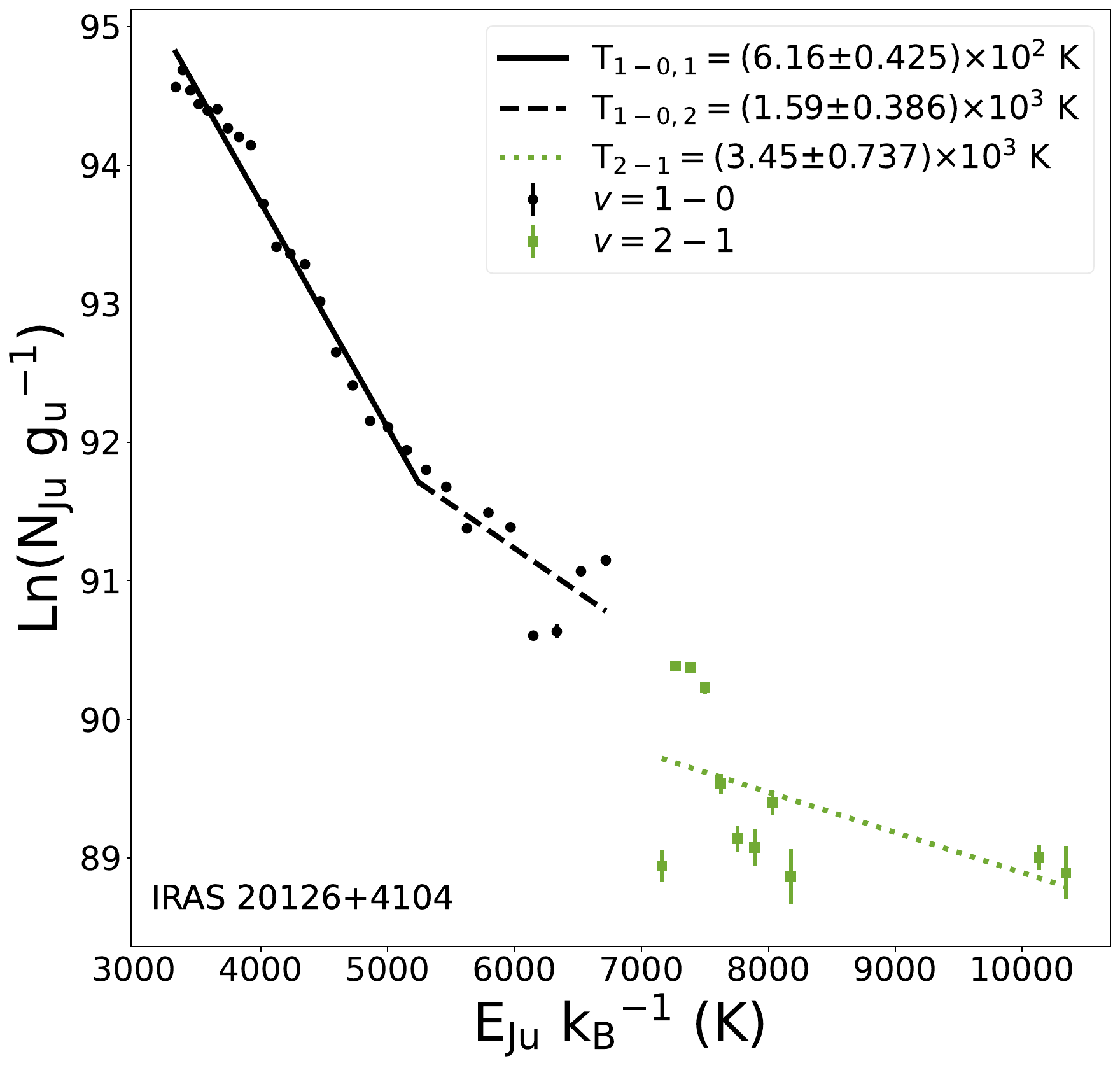}
    \end{minipage}
    \textbf{Figure \ref{fig:co_ext_rot}.} \textit{continued from last page}
\end{figure*}

More lines can be fit with arbitrarily chosen cutoff points for different thermal populations, but these populations may not be representative of $T_K$ \citep[e.g.][]{Neufeld_2012, Green_2013, Manoj_2013}. The $v=1-0$ populations do not show points deviating from their respective straight line fits, which is consistent with our assumption of optically thin emission.
Meanwhile, the scatter in $v=2-1$ does deviate from linear fits, which could imply a non-thermal population, but these measurements are inherently noisier.
Systematic effects from optical depth, extinction, and modeling of the population diagram probably dominate the uncertainty in the number of molecules from fitting the line's intercept, so we do not report its uncertainty.

    \subsection{CO Vibrational Transitions and Bulk Gas Properties}
    \label{co_vib}     With two series of vibrational states between levels $v=1$ and $v=2$, we can estimate $T_{vib}$, the vibrational excitation temperature. 
An average $\overline{T_{vib}}$ of the CO gas population is found from the ratio of $N_v$ particles between two neighboring vibrationally excited states $v_u$ and $v_l$ of the same $J_u$:
\begin{equation}
      T_{vib} = \frac{ -\frac{\Delta E_v(J_u)}{k_B} }      {\ln{\bigg(\frac{N^{tot}_{v_l}(J_u)}{N^{tot}_{v_u}(J_u)}\bigg)}} \\
     \label{equ:tvib}
\end{equation}
where $\Delta E_v(J_u) = E_{v_u}(J_u) - E_{v_l}(J_u)$ is the energy difference between two consecutive vibrational bands for each pair of lines at a given $J_u$. Note that the two states are from the same $J_u$, so the ratio of degeneracies, $g_u$, is dropped. 
Considering that $\overline{T_{vib}}$ agrees within uncertainties with $T_{1-0,1}$ (Table \ref{tab:co_rot_vib_props}), we may derive the total number of molecules for the majority of CO gas.

$N_{CO,tot}$, the total number of molecules in the CO population, comes from the $N_{v}^{tot}$ measured in the population diagram for each source and each component of the vibrationally excited CO population (see Section \ref{co_rotation}, Table \ref{tab:co_rot_vib_props}, and Figure \ref{fig:co_ext_rot}).
We assume a Boltzmann distribution and calculate $N_{CO,tot}$ for the two $v=1-0$ components from our rotation diagrams:
\begin{equation}
    \frac{N_{v=1-0}^{tot}}{N_{CO,tot}}
    = \frac{1}{Z_{v}} \exp{\bigg(\frac{-\overline{T_{vib}}}{E_{v=1-0} / k_B}\bigg)},
\end{equation}
$E_{v=1-0}$ is approximately the $J_u=0$ via the mean energy of the R0 and P1 lines (noticing CO gas has no P0 transition) and is 3087 K when dividing out $k_B$. $Z_{v}$ is the partition function for CO vibrational states:
\begin{equation}
    Z_{v} \approx { [ 1 - \exp({\frac{-hc \nu_i}{k_B \overline{T_{vib}}}}) ] }^{-d_1}.
\end{equation}
In the vibrational partition function, $\nu_i$ is the band frequency for each upper state transition, about 2143.27 $\rm cm^{-1}$ (Table 9 from \citealt{Evans_1991}), 
and $d_i$, the degeneracy for a vibrational state $v$, is 1 since CO is a diatomic molecule with no spin degeneracy. Our $Z_{v} \times Z_{J}$ agrees to within 2\% of what is found from HITRAN's partition function at the same temperature (Table 7 in \citealt{Li_2015}), although theirs combines the rotational and vibrational partition functions assuming $T_{rot} = T_{vib}$, which is not correct in general for the ISM.

We display vibrational temperatures and total number of CO molecules in Table \ref{tab:co_rot_vib_props}. $\overline{T_{vib}}$ values are found from the mean value using all pairs of lines from the two neighboring vibrationally excited bands we can measure. 
The $N_{CO,tot}$ values are lower limits if some gas-phase CO emission lines are optically thick.
We do not propagate uncertainties in $N_{CO,tot}$ since the bounds set by $T_{vib}$ and by our systematic effects (e.g. baseline fitting; see Appendix \ref{sect:systematics}) can vary greatly.

    \subsection{Inferred Total Gas Masses}
    \label{co_mass}         We compute lower limits to the gas mass for each source. Given the total number of CO gas molecules, we infer the total gas mass, $M_{gas, NIR}$, probed by the NIR CO populations in Figure \ref{fig:co_ext_rot}. We use an averaged abundance of $\frac{1}{6000}$ 
    for the fractional amount of CO relative to molecular hydrogen gas present in dense molecular clouds (\citealt{Lacy_1994} and Table 5 in \citealt{Lacy_2017}). The value is from studying CO rovibrational absorption seen through neutral molecular gas but does not apply the same extinction law we use. At a mean molecular weight per hydrogen atom of $\mu_{H_2} = 2.809$ (e.g. \citealt{Evans_2022} or see Appendix A in \citealt{Kauffmann_2008}): 
    \begin{equation}
        M_{gas, NIR} \geq 6000 N_{CO,tot} \times {\mu_{H_2}} {m_H},\label{eq:tot_gas_mass} 
    \end{equation}
    where ${m_H} = 1.67 \times 10^{-24}$ g is the mass of a hydrogen atom. 
    All we need is the total number of observed CO molecules ($N_{CO,tot}$), which we found in Section \ref{co_vib}. 
    See Table \ref{tab:co_rot_vib_props} for the inferred masses. Our gas masses for the low L$_{bol}$ sources agree within a factor of 2 to 3 with that of IRAS 15398-3359, another Class 0 source with an L$_{bol}$ of approximately 1.5 \Lsun\ \citep{Salyk_2024}.
    
    The masses inferred from the CO emission are, strictly speaking, lower limits for two reasons. One is that the lines may be optically thicker than assumed, which we have addressed earlier (Section \ref{implied_profile}). The other is that we detect only light scattered in our direction (Section \ref{apertures}). If the dust grains were like those in the diffuse ISM, the fraction scattered would be small compared to that absorbed. However, the large icy grains in protostellar envelopes have much higher ratios of scattering to absorption. For the KP5 dust model we have adopted, both $\kappa_{\rm abs}$ and $\kappa_{\rm sca}$ over the range of the CO line emission, but outside the CO ice feature around 4.67 \micron, has an average ratio of $\kappa_{\rm sca}/\kappa_{\rm abs} = 0.82$. In contrast, the average ratio, calculated the same way for the diffuse ISM model by \citet{Hensley_2022} is 0.034. While the actual mass estimate depends on unknown geometry, we expect that the correction factor for incomplete scattering is small.  

\section{Discussion} 
    \label{discussion}  
In the following sections, we discuss our CO sources and their possible physical origins.  
The apertures in this work extend beyond the dust disk radii measured by sub-mm/mm continuum emission (Table \ref{tab:source_props}, Figure \ref{fig:co_imgs}), so multiple temperature components measured from our population diagrams could in principle arise from different regions around a YSO, including scattered light from disks, CO gas entrained in disk winds, and the surrounding outflow cavity \citep[e.g.][]{vanKempen_2010A, vanKempen_2010B, Goicoechea_2012, Herczeg_2011, Herczeg_2012, Dionatos_2013, Green_2013, Karska_2013, Karska_2014, Manoj_2013, Yildiz_2013, Matuszak_2015, Yang_2018}. 
Our analyses also show the observed CO line profiles and the asymmetry between the P and R branches could be due to a mixture of emission and blueshifted absorption (Figure \ref{fig:extinc_example}, Section \ref{co_ext}).

\begin{figure*}
        \centering
        \begin{minipage}{0.49\textwidth}
            \includegraphics[width=\linewidth, trim = 0in 0in 0in 0in,clip]{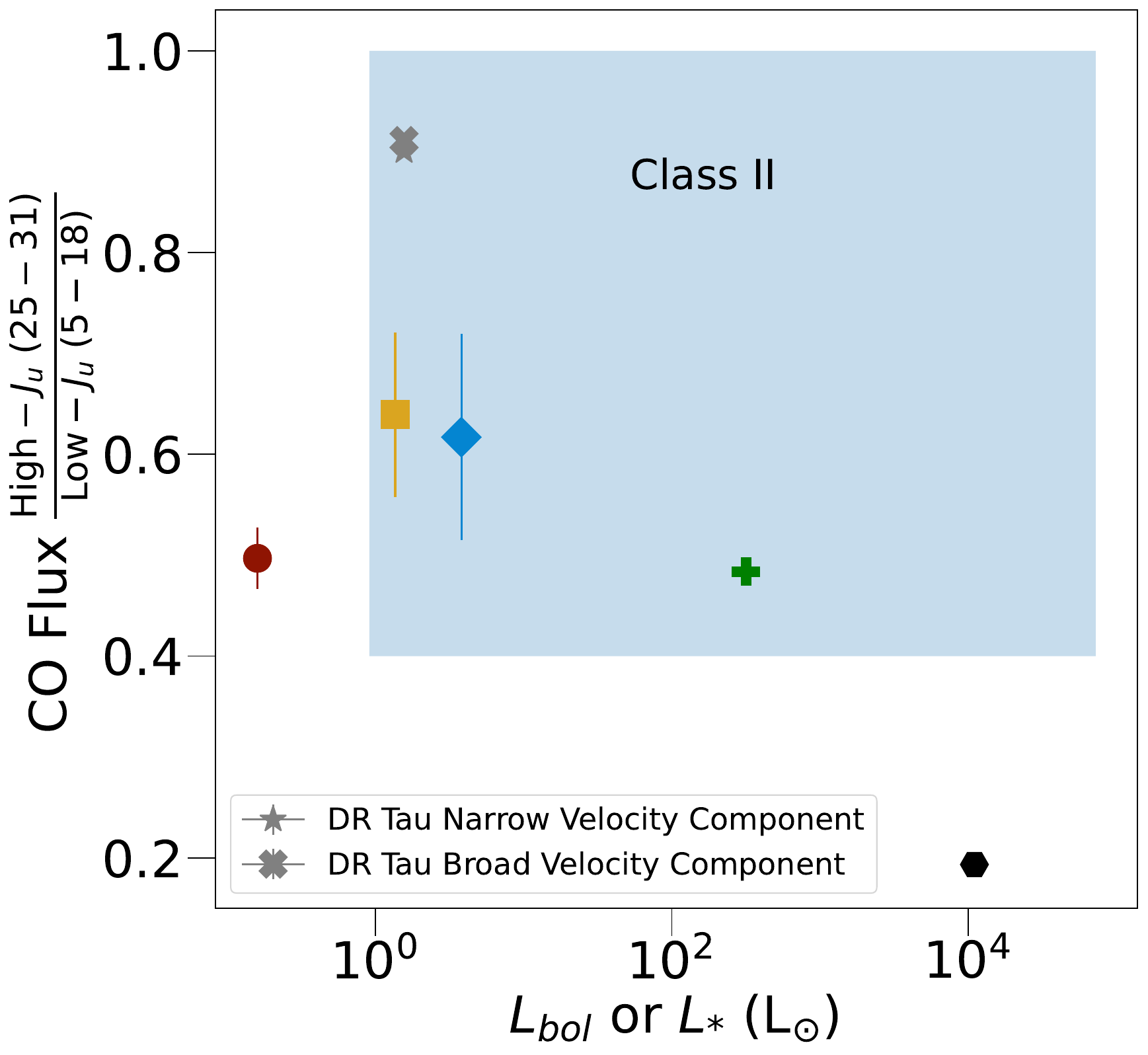}
        \end{minipage}
        \begin{minipage}{0.49\textwidth}
            \includegraphics[width=\linewidth, trim = 0in 0in 0in 0in,clip]{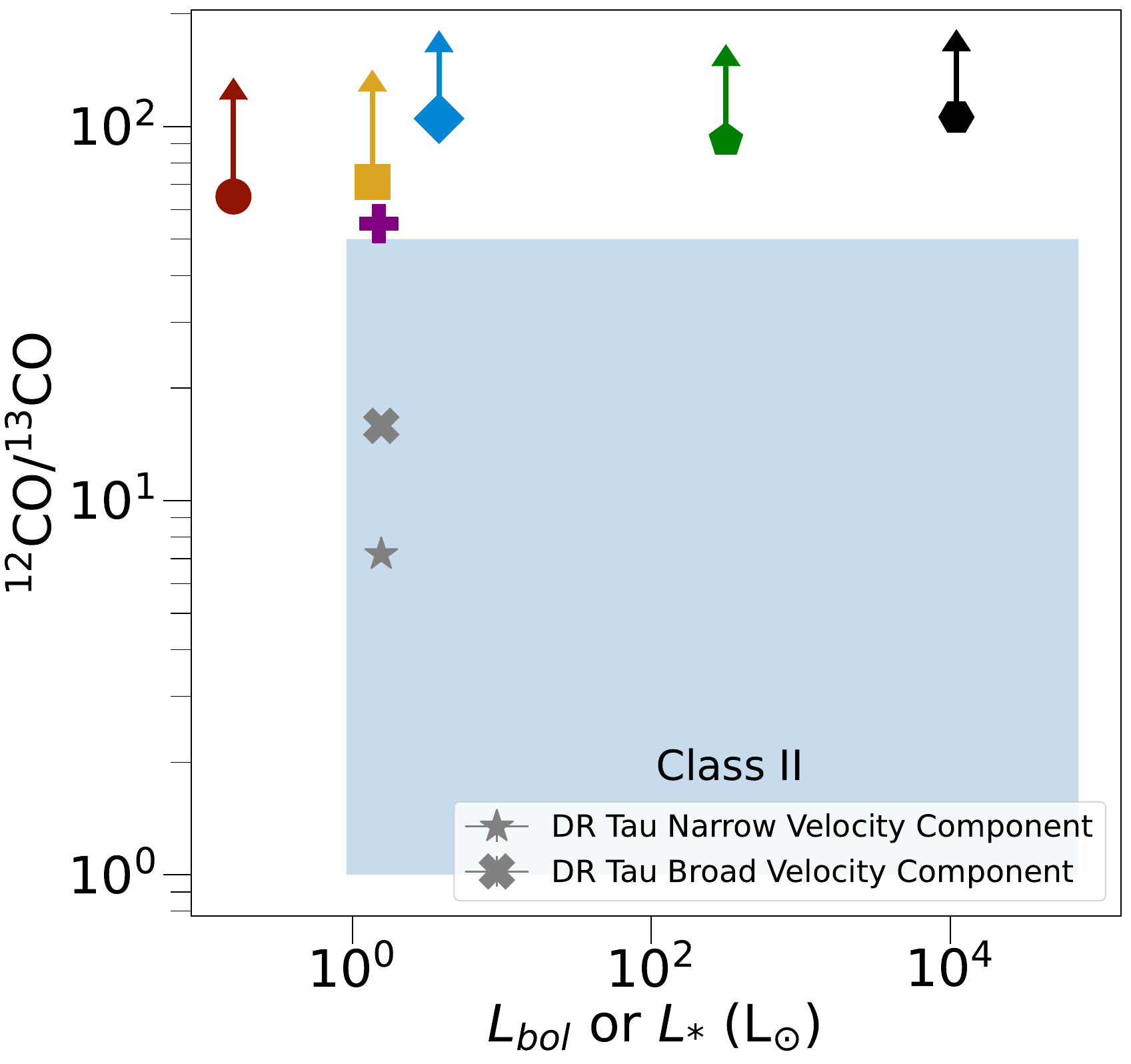}
        \end{minipage}
        \includegraphics[width=0.65\linewidth, trim = 0in 0in 0in 0in,clip]{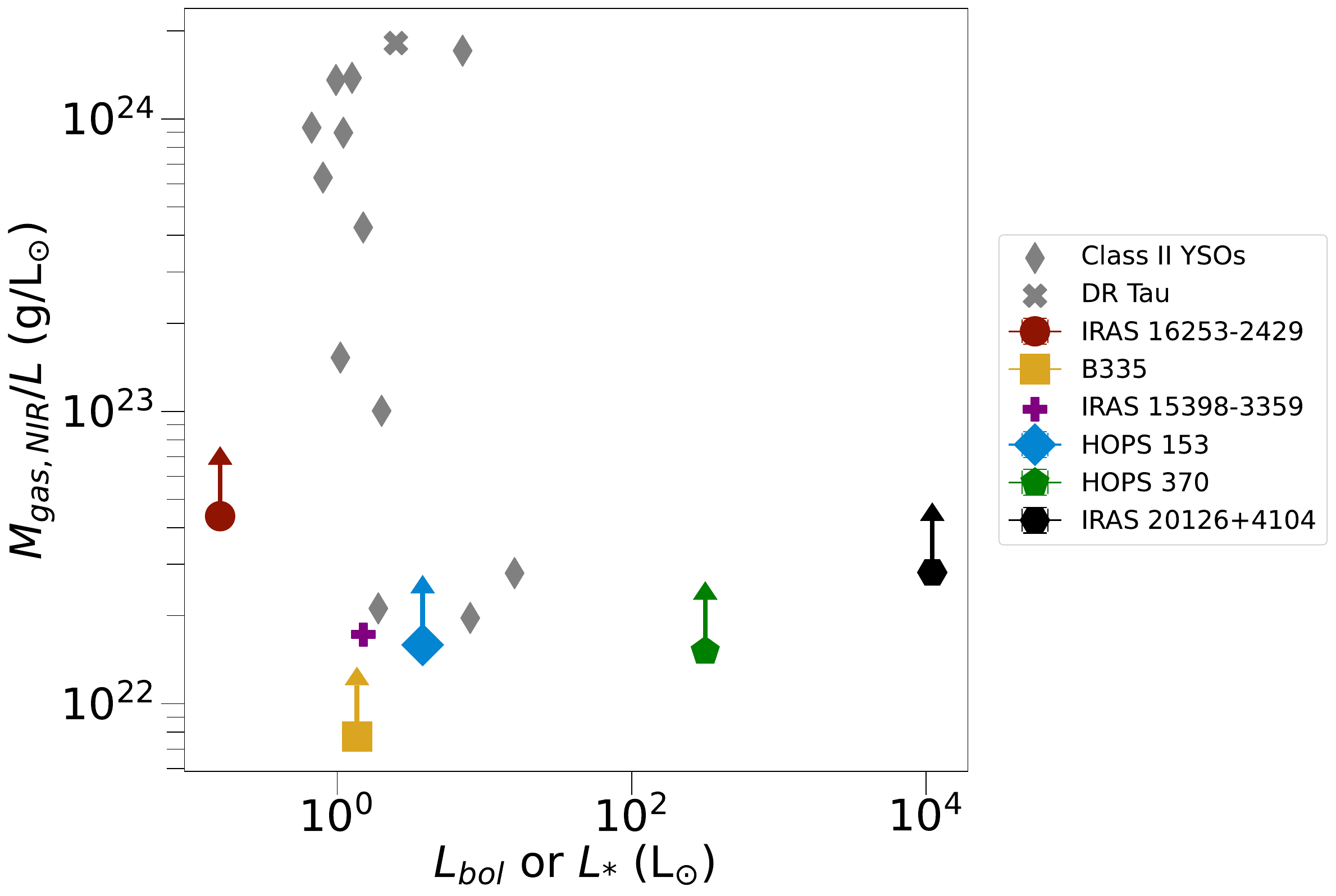}
        \caption{Context of Class 0 with respect to Class II protostars (Appendix E in \citealt{Banzatti_2022} for upper panels and Table 10 of \citealt{Salyk_2011} for the bottom panel). The x-axis of each plot shows the protostellar luminosity, which is either $L_{bol}$ for Class 0 or $L_*$ for Class II. In the upper panels, the large light blue square represents the general region of the diagram occupied by Class II YSOs, and the x or star highlights DR Tau, an example T Tauri star. The purple plus marks model values for a Class 0 source from \citet{Salyk_2024}. \textit{\underline{Upper Left}}: Plot of the ratio of high-$J_u$ to low-$J_u$ CO fluxes with propagated uncertainties (using Table \ref{tab:co_lines_fluxes}), where our Class 0 sample tends to lower values. \textit{\underline{Upper Right}}: The CO isotopologue ratio for Class 0 sources (lower limits from Table \ref{tab:co_rot_vib_props}) are above Class II protostars. \textit{\underline{Bottom}}: Gas masses probed by CO (Table \ref{tab:co_rot_vib_props}) per protostellar luminosity  
        when plotted as a function of protostellar luminosity (Table \ref{tab:source_props}). The masses for our Class 0 YSOs are lower limits because of systematic effects, so these values would move together on the plot given any systematic changes (e.g. extinction law).
        }\label{fig:mgas_lbol_corr}
    \end{figure*}

To contextualize these possibilities, in Figure \ref{fig:mgas_lbol_corr}, we compare our measurements (the colored points) to that of Class II sources (the blue box and gray markers). 
The direct comparisons included: the unextincted flux ratio of high-$J_u$ to low-$J_u$ lines that indicates the temperature distribution of spectra (upper left panel), the isotopologue ratio that indicates the optical depth throughout spectra (upper right panel), and the NIR total gas masses per luminosity (bottom panel). The first two comparisons use results from \citet{Banzatti_2022}, while the third panel uses the column densities and emitting areas modeled in \citet{Salyk_2011}. DR Tau is highlighted because it is a Class II source with relatively high accretion rate, has yielded past high-resolution spectra \citep{Banzatti_2022}, and JWST observations \citep{Temmink_2024}. We also include the modeled isotopologue ratio and total warm gas mass for another Class 0 source, IRAS 15398-3359, \citep{Salyk_2024} from the JWST program CORINOS \citep{Yang_2022}. Class I sources are difficult to incorporate in Figure \ref{fig:mgas_lbol_corr} because the VLT-CRIRES surveys with modeled column densities and temperatures had fewer comparable $^{12}$CO $v=1-0$ lines \citep{Herczeg_2011, Brown_2013}. 
To summarize, compared to Class II sources, our isotope ratios are higher, $T_{rot}$ similar or lower, and gas masses consistent, given that they are lower limits. 


\subsection{CO Gas Temperatures and Excitation Processes}    \label{gas_temp_pumping}
    We have shown two temperature components for $v=1-0$ ($T_{1-0,1}$ and $T_{1-0,2}$), one component for $v=2-1$ ($T_{2-1}$), and the mean vibrational temperature ($\overline{T_{vib}}$). There are no general correlations based on the uncertainties and small sample, so we review average values from Table \ref{tab:co_rot_vib_props}. 
    The lower mass sources have higher median rotational temperatures compared to that of the higher mass sources ($T_{1-0,1} \sim 1100$~K and $T_{2-1} \sim 10^4$~K  compared to $T_{1-0,1} \sim 850$~K and $T_{2-1} \sim 3550$~K), though $T_{2-1}$ values have greater uncertainty and cannot be well-measured for B335 and HOPS 153. For the medians of $T_{1-0,2}$, our lowest mass source IRAS 16253-2429 has the highest rotational temperature (nearly flat in its population diagram), while our highest-mass source IRAS 20126+4104 has the lowest rotational temperature of $\sim$1600~K. The median $T_{1-0,2}$ for the other sources remain between these two values. The median $\overline{T_{vib}}$ among all sources is $\sim1000$~K and is within uncertainties of $T_{1-0,1}$ for all our sources.

    The excitation temperature is not necessarily equal to 
    the kinetic temperature of a collisionally excited gas, which would require modeling the curvature of rotation diagrams \citep{Neufeld_2012, Brown_2013, Yang_2018}. 
    For $v=1-0$, 
    since values of $\overline{T_{vib}}$ are similar to that of the $T_{1-0,1}$ component, there may be collisional excitation at $J_u \lesssim 30$ (Figure \ref{fig:co_ext_rot}). 
    Furthermore, our CO spectra tend to be brighter toward low-$J_u$ compared to the same averaged set of lines from Class II sources neglecting extinction correction (Figure \ref{fig:mgas_lbol_corr}).

    The rotational levels in $v=2$ appear to be populated at higher T$_{rot}$, though we warn the fits are noisy.
    To explain a higher $T_{rot}$ in $v=2$, we could consider a secondary source of excitation like radiation by IR or UV pumping. This is further evidenced by the second component of $v=1-0$ and the P/R asymmetry in our spectra (Section \ref{co_ext}). 
    For example, UV photons may excite electronic states and produce a higher $T_{rot}$ in the hotter $v=1-0$ and $v=2-1$ components. UV would also have to contend with extinction and scattering, which is even higher at shorter wavelengths, but some CO is likely near the top surface of a protostellar disk (e.g. \citealt{Banzatti_2022}) and directly exposed to UV from the accretion shock onto the star. Alternatively, some CO could be exposed to UV in dense shocks along the jet \citep[e.g.][]{Neufeld_2024}. In past work, $\overline{T_{vib}}$ of approximately 3000 to 5000~K are known to indicate UV excitation \citep[e.g.][]{Blake_2004, Brittain_2009, Bast_2011, Bruderer_2012, Brown_2013, Thi_2013}, so our $\overline{T_{vib}}$ of 1000~K may be too low to be reproduced by only UV.

\subsection{Isotope-Selective Radiative Processing} 
    Our targets all lack $^{13}$CO detections (see Section \ref{implied_profile}, the flux ratios in Table \ref{tab:co_rot_vib_props}). 
    Deviations far above the standard ISM abundance ratio are rarely found in past studies similar to ours (see the upper right of Figure \ref{fig:mgas_lbol_corr}). 
    There is only some evidence seen for massive, evolved sources with variable ratios of $^{12}$CO to $^{13}$CO \citep{Smith_2021}. The process may be connected to UV radiation sources either deep in accretion columns of the protostellar disk or in shocks via outflows \citep[e.g. Table 5, Figures 9 and 13 in ][]{Visser_2009}. Enhancing the flux ratio of ${}^{12}$CO to ${}^{13}$CO at high temperatures may mostly occur by preferentially dissociating ${}^{13}$CO relative to $^{12}$CO via UV radiation \citep{Morris_1983, Glassgold_1985, Mamon_1988, vanDishoeck_1988, Visser_2009, Saberi_2019} in the inner regions of protoplanetary disks and diffuse nebulae \citep[e.g.][]{Scoville_1979, Bally_1982, Langer_1984, Warin_1996}.    
    Our Class 0 sources show evidence for radiative pumping (Section \ref{gas_temp_pumping}) by UV or IR photons and may provide a novel opportunity to study such interactions in dense conditions. 
     

\subsection{Origins of Protostellar CO Emission}
    Warm gas masses for Class 0 sources are on the order of $10^{22}$ to $10^{23}$ g for the lower mass sources (IRAS 16253-2429, IRAS 15398-3359, B335, and HOPS 153) and on the order of 5 to 300 $\times 10^{24}$ g for the higher mass sources (HOPS 370 and IRAS 20126+4104).
    The bottom panel of Figure \ref{fig:mgas_lbol_corr} shows these warm gas masses per protostellar luminosity since both quantities scale with distance. All our sources are consistently near the lower bound of what is seen from Class II disks. The gas masses for our Class 0 sources are strict lower limits due to systematic effects (Section \ref{co_mass}), so the points may collectively be at higher values on the plot. 

    We explore collisional excitation to explain the similarities observed between Class 0 and Class II sources, which primarily depends on our first rotational temperature component $T_{1-0,1}$ (Section \ref{gas_temp_pumping}).     
    Approximate densities in astrophysical conditions come from modeling CO collisional rate coefficients ($\gamma_{ij}$)  
    with a collider element. 
    Einstein coefficients (A$_{ij}$) for our vibrational bands equal $20 - 50 \ \rm {sec}^{-1}$ (from HITRAN), which we can use with $\gamma_{ij}$ to compute a critical number density on the order of $A_{ij} \ {(\gamma_{ij})}^{-1}$. 
    For CO $v=1-0$ $\gamma_{ij}$ is in the range of $10^{-11}$ to $10^{-10}\ \rm {cm}^{3} \ {sec}^{-1}$ for all $J_u$ we use and for colliders of para-H$_2$ and ortho-H$_2$ (Stancil, P. C. priv. comm.). 
    %
    %
    A typical critical number density among all $\gamma_{ij}$ results in ${10}^{12} \ \rm {cm}^{-3}$. 
    If CO gas were primarily in outflows \citep[e.g.][]{Tabone_2020}, shocked and ionized gas is modeled with Hydrogen number densities on the order of at most ${10}^{4} \ \rm {cm}^{-3}$ \citep[e.g.][]{Rubinstein_2023, Narang_2024, Neufeld_2024}.      
    While the CO may be in dense winds from the disk rather than around ionized gas from the jet, even the lowest critical densities derived from rate coefficients are orders of magnitude too high for vibrational excitation only in outflows, especially for IRAS 16253-2429 (our lowest mass and least active protostar; see \citealt{Narang_2024}, Watson et al. 2024 in prep.). 

    CO $v=1-0$ lines can appear in absorption through colder winds as in Class I sources (\citealt{Thi_2010}, \citealt{Herczeg_2011}, and Appendix B of \citealt{Harsono_2023}). Past work also used line profiles to distinguish whether emission arises from the disk surface due to Keplerian rotation or due to low-velocity ($<1 \ \rm km \ {sec}^{-1}$) winds \citep[e.g.][]{Bast_2011, Brown_2013}. 
    Outflows including slow disk winds that pile up gas and cause absorption are a possible explanation for the asymmetry between the P and R branch measured using CO (Section \ref{co_ext}), and perhaps the large total gas masses that HOPS 370 and IRAS 20126+4104 (Table \ref{tab:co_rot_vib_props}) have compared to Class II sources. 
    According to the relative velocities we measure between our CO apertures (Table \ref{tab:apertures}), we see no significant evidence for velocity shifts, but our limits do not rule out slow winds.
    Any velocity shifts may be weak because our apertures are near the central protostar where extinction is higher \citep[e.g.][]{Narang_2024} and because the fastest outflows around the jet are perpendicular to our line of sight. 
    \citet{Banzatti_2022} also found intermediate- to high-mass Class II sources often have weak or absent inner disk winds reflected in their line profiles. 

    The innermost surfaces of the protostellar disk is the most plausible origin of the $v=1-0$ CO emission sources at low-$J_u$, especially considering our gas masses are lower limits. Yet the brightest emitting regions are spatially resolved from the central protostar and central source of dust continuum (Figure \ref{fig:co_imgs}). Our program's initial NIRSpec data release showed regions of Br-$\alpha$ that spatially overlapped continuum emission and CO emission, possibly indicating scattered light from the inner disk \citep{Federman_2023}. Accordingly, our images would be explained by enhancing disk emission via forward-scattering light through dust in the winds \citep{Pontoppidan_2002} or resonant scattering with CO gas in the outflow cavity \citep{Lacy_2013}.





\section{Conclusion} 
    This work provides the first in depth analysis using JWST on the full CO fundamental bands from a sample of Class 0 protostars with orders of magnitude different bolometric luminosities. The data covers a forest of rotational states in multiple vibrational bands, which cannot be completely observed with only ground-based observations. 
For visual extinction across the outflow cavity of each source, we estimate $A_V$ of 10 to 40 mag (Table \ref{tab:apertures}) by using H$_2$ lines as in \citet{Narang_2024} and apply then to de-extincting CO (see also Salyk et al. 2024 Submitted). Assuming CO is optically thin in the P branch and using the extinction estimates, we created population diagrams and characterized the ro-vibrational CO gas temperatures and number of molecules (Table \ref{tab:co_rot_vib_props}). The main findings are as follows:
\begin{itemize}
    \item{From Section \ref{jwst_nirspec}, we use NIRSpec to image CO emission, which is often obscured toward the central protostar. 
    }

    \item From Section \ref{jwst_miri}, we use higher spectral resolution MIRI data to measure the average Doppler velocity of CO lines between apertures across outflow cavities, and we find no relative velocity shifts down to $15$~km~sec$^{-1}$.

    \item From Section \ref{implied_profile}, we did not detect ${}^{13}$CO lines. Based on the limits set by the noise, the $^{12}$CO appears to be optically thin outside ice features (approximately $J > 10$), with better constraints set for the higher mass protostars HOPS 370 and IRAS 20126+4104 (i.e. isotopologue flux ratios $>100$). The limits set by the non-detection of ${}^{13}$CO from our high-mass sources may indicate selective UV photodissociation whether close to the protostar due to accretion or near dense shocks along the jet.

    \item From Section \ref{co_ext}, there may be a weak blueshifted absorption component in our CO line profiles that is seen with MIRI and is blended out when viewed with NIRSpec. Such signatures may explain the asymmetry observed between the P and R branches for all our sources, which is difficult to reproduce using extinctions derived from CO. 

    \item From Section \ref{co_rotation}, the rotational temperatures of $600$~K to $\sim 10^4$~K among all the different sources and vibrational bands do not correlate with bolometric luminosity. 
    
    \item From Section \ref{co_vib}, the average vibrational temperature ($\overline{T_{vib}})$ is $\sim1000$~K, does not vary with respect to bolometric luminosity, and is similar to our lowest rotational temperatures from $v=1-0$ regardless of extinction, which indicates collisional excitation of the rotational levels at $J_u \lesssim 30$. The presence of a higher $T_{rot}$ in the $v=2$ levels may indicate non-LTE effects, such as UV or IR pumping in a smaller number of CO molecules (cf.\ columns 6 and 7 of Table \ref{tab:co_rot_vib_props}). 
    
    \item From Section \ref{co_mass}, the lower limits to gas masses inferred from the majority of our $v=1-0$ populations range from $\sim 7 \times 10^{21}$~g for our lowest mass protostar to $\sim 3 \times 10^{26}$~g \ for our highest mass protostar (Table \ref{tab:co_rot_vib_props}). These gas masses probed by CO around our Class~0 sources, when compared with protostellar luminosity (Figure \ref{fig:mgas_lbol_corr}), appear consistent with that of Class II sources.
 
\end{itemize}

Our analyses show the range of properties for CO gas around Class 0 protostars. 
One source, IRAS 16253-2429 has relatively low outflow and accretion rates  (\citealt{Narang_2024}, Watson et al. in prep.), so the CO gas may be found purely on the inner disk's surface in the region of terrestrial planet formation \citep[like for Class II sources, e.g. ][]{Najita_2003, Salyk_2009}. Our other sources (B335, HOPS 153, HOPS 370, and IRAS 20126+4104) likely present a mixture of CO gas in disk winds and light from the inner disk scattered off dust in the outflow cavity walls.

\section{Acknowledgments} 
    We thank the anonymous referee for very thoughtfully reviewing our work. Phillip C. Stancil graciously provided rate coefficients for CO-H$_2$ collisional rate coefficients in advance of publication.  Colette Salyk shared relevant results about IRAS 15398-3359. Steven Federman, Daniel Harsono, John Lacy, and Rachel L. Smith gave fruitful discussion that helped to interpret CO line profiles and isotopologues at lower spectral resolution. A.E.R thanks Alice C. Quillen for guidance, David A. Neufeld and Mayra Osorio for rechecking our line luminosities and extinction laws, and Andrea S. J. Lin for discussing data and table presentation. 

This work is based on observations made with the NASA/ESA/CSA James Webb Space Telescope. 
The data in this paper were obtained from the Mikulski Archive for Space Telescopes (MAST) at the Space Telescope Science Institute (STScI), which is operated by the Association of Universities for Research in Astronomy, Inc., under NASA contract NAS 5-03127 for JWST. These observations are associated with GO program \#1802.
All the JWST data used in this paper can be found in MAST: \dataset[10.17909/3kky-t040]{http://dx.doi.org/10.17909/3kky-t040}. 

Support for AER, DMW, RG, SF, STM, WF, JG, and JJT in program \#1802 was provided by NASA through a grant from the STScI, which is operated by the Association of Universities for Research in Astronomy, Inc., under NASA contract NAS 5-03127. NJE thanks the University of Texas at Austin for research support. 
 P.N. acknowledges support by the NWO grant 618.000.001, Danish National Research Foundation (Grant agreement no.: DNRF150) and ESO. 
ACG acknowledges from PRIN-MUR 2022 20228JPA3A “The path to star and planet formation in the JWST era (PATH)” and by INAF-GoG 2022 “NIR-dark Accretion Outbursts in Massive Young stellar objects (NAOMY)” and Large Grant INAF 2022 “YSOs Outflows, Disks and Accretion: towards a global framework for the evolution of planet forming systems (YODA)”.
EFvD and WRMR is supported by EU A-ERC grant 101019751 MOLDISK.Fellowship Program.

This research used the VizieR catalogue access tool, CDS, Strasbourg, France (DOI: 10.26093/cds/vizier) originally published in 2000, A\&AS 143, 23. The work also made use of the NIST atomic database and the HITRAN molecular database.

\software{ SpectralCube and wcsaxes from \texttt{Astropy} \citep{astropy:2013, astropy:2018, astropy:2022}; \texttt{CARTA} image and spectral viewer \citep{CARTA_2021}; \texttt{pybaselines} to fit baselines \citep{pybaselines}; \texttt{SciPy} for filters and smoothing options \citep{2020SciPy-NMeth}; \texttt{PyTorch} \citep{PyTorch} and \texttt{torchimize} \citep{torchimize} to improve runtime; \texttt{NumPy}'s \citep{harris2020array} einsum heuristic by Mark Wiebe; \texttt{fityk} for optimal fits to individual spectral lines \citep{Wojdyr:ko5121}.
}


\appendix

\section{Spectral Fitting Methods and Creating Line Images}
    \label{sect:fitting_methods}
    \subsection{Baselines Including Ices and Continuum}\label{sect:baselines}

    To image CO, we must break each spaxel in a spectral cube from the IFU down into broader features and narrower lines. 
    As supported by \citet{Federman_2023}, there are 6 spectral features we noticed to produce such a curve in the CO forest: broad ice features, continuum, CO purely appearing in emission, CO purely in absorption, and CO in a mixture of the two cases (generally in absorption at higher $J$ and in emission at lower $J$), and other molecules or ions in emission (e.g. H$_2$ and [Fe~II]). A baseline of broader features that strictly follows the bottom of peaks, or is locally determined by averaging noise nearby a given line, can be unrealistically sharp and fails in line forests.
    Rather than modeling each contribution to continuous parts of the spectrum, we employ a baseline-fitting library, \texttt{pybaselines}, which implements methods for similarly line-rich Raman spectroscopy \citep{pybaselines}. Our baseline criteria are to extract ice features (e.g. OCN$^{-}$ or OCS as presented in \citealt{Brunken_2024} and \citealt{Nazari_2024}) and to maintain a smooth enveloping curve with minimal sharp edges or corners. In the case of pure noise, we simply fit a smoothed cubic polynomial through the center of the noise.
    
    CO in absorption and emission are distinguished by initially checking for extrema within two spectral resolution elements (about $\pm 0.003$ \mic) of each CO line's central wavelength. A local maximum at the line's central wavelength is larger than its neighborhood and is assigned +1, while a local minimum would be smaller and then given a -1. If neither, then a value of 0 is given.  
    For a given spaxel, the baseline method to use is determined by the mode of these assigned values.
    The Boolean assignments are not reliable indicators for individual lines due to noise or potential Doppler shifts, but they form trends in aggregate.
    
    Using \texttt{pybaselines}, we apply the "Joint Baseline Correction and Denoising" method when CO is in emission and their "Penalized Spline Asymmetric Least Squares" in cases of CO in absorption. Fundamentally, these methods treat solving for baselines as a least squares problem. The different methods were chosen to tend to the bottom of emission lines and to the top of absorption lines, though other methods also work. In either method, we use two parameters to be robust. One, for smoothing, and the other, a morphological parameter to place baselines between signal and noise.
    
    Ice features can cause broad but smooth deviations, so we reinforce smoothness around ices by breaking each spaxel into spectral sub-regions and tailoring the two baseline fitting parameters for each sub-region. 
    Points for sub-regions are also chosen to be far from the brightest emission lines (e.g. from [Fe~II] or H$_2$) with as few points as possible to minimize influence on measurements (i.e. one or two per ice feature). Imposing sub-regions can also affect whether maps show CO in emission or absorption, but we did not apply an algorithm to decide sub-region points in this work. One may compute a spectrum's derivative to find changes in curvature due to ices, but line forests create beats or harmonics in derivatives, which complicates determining concavity. For now, we take precautions by smoothing around our chosen points after fitting a baseline. 
    
    An example automated baseline is shown by the purple dashed curve in Figure \ref{fig:example_spaxels}. Sub-region markers are shown by purple diamonds. Continuum is also shown by the pink dotted curve and found by fitting a smooth polynomial to baselines. The wavelength is set to shorter wavelengths than shown in Figure \ref{fig:example_spaxels} to demonstrate the utility of our baselines for initial guesses even around more complicated ice features and is modeled in other work \citep{Nazari_2024}. We contrast our algorithmic baselines with adjustments needed for an ideal splined baseline. Splines and polynomials can smoothly follow data \citep[see other work using NIRSpec data and fitting CO lines, such as][]{Boersma_2023, Buiten_2023, Sturm_2023, PereiraSantaella_2024, GarciaBernete_2024, GonzalezAlfonso_2024} but can require many more points or be unstable given all variations in ice and continuum throughout our sample and IFU cubes (see blue curve and x's in Figure \ref{fig:systematics} compared to the pink curve and diamonds).
    
    With the above methods, to summarize the baseline-fitting in each spaxel for a given spectral cube, we:
    \begin{enumerate}
         \item[1)] Select 1--2 points to mark spectral-sub-regions around each ice feature.
         \item[2a)] If the spectrum consists of a median S/N $<$10 or has mostly CO in absorption, we solve for the baseline by a weighted least squares.
         \item[2b)] Else, if the sub-region mostly shows CO in emission, as a precaution, we need to apply a low-pass (top-hat) filter to trace the spectrum's curvature. Then we apply a least squares baseline that also accounts for fitting between signal and noise.
         \item[3)] As a final precaution, we apply a smoothing filter (Savitzky-Golay from \texttt{SciPy}, which is applicable for data with constant spacing) around control points marking sub-regions to guarantee smoothness. 
    \end{enumerate} 
    
    We find that the main drawback of our precautions is that repeated smoothing by the top-hat or Savitzky-Golay filter can underestimate the effect of ice features (for an application related to ices, see \citealt{Nazari_2024}). Therefore, for our optimized fits to line profiles in Section \ref{forest_fitting} and upper panels of Figure \ref{fig:systematics}, the automated baseline is a default first guess that works well throughout our sample. Then adjusting that baseline with a splined version works when presenting final results and when the automated baseline deviates from the data for a known reason (i.e. around the bottoms of ice features).

\subsection{Rapidly and Simultaneously Fitting Spectral Lines for Images}\label{sect:fit_all}

    For a single spaxel, we found that fits to one spectrum including a polynomial continuum, ices and a parameter for line widths can take up to 10 minutes. In CO forests, we must also fit all emission line profiles simultaneously to account for blending between CO lines (e.g. the R branch CO lines in Figure \ref{fig:example_spaxels}). To make all such line images, each cube is $\sim 90 \times 90$ spaxels for a total runtime of $\sim60$ days. 
    
    At each spaxel across a spectral cube, a baseline-subtracted spectrum consists of narrow, unresolved spectral line profiles, which are approximately Gaussian. Line centers ($\lambda_0$) are noted in the Appendix \ref{sect:flux_table}, Table \ref{tab:co_lines_fluxes}. For efficiency, line widths are set by the spectral resolving power $\frac{\lambda}{\Delta\lambda}$, where $\Delta\lambda$ is according to the G395M grating on NIRSpec (bottom panel of Figure \ref{fig:systematics}). Using the nominal G395M spectral resolution speeds up fitting by pre-calculating the exponential part of a Gaussian:
    \begin{equation}
        f(\lambda, \lambda_0, \Delta\lambda) = \exp{\frac{-{(\lambda - \lambda_0)}^2}{2 \sigma^2}}, 
    \end{equation}
    where $\sigma = \frac{1}{8\ln{2}} \Delta\lambda$. 
    Summing over many lines, the objective function to fit becomes:
    \begin{equation}
        C(a) = I_{obs} - \sum_{i=0}^{i=N}a_i \ f(\lambda, \lambda_{0,i}, \Delta\lambda_i), 
        \label{min_func}
    \end{equation}
    where $I_{obs}$ is the observed baseline-subtracted spectrum and $a_i$ is the amplitude of the i$^{th}$ spectral line. $N=152$ is the total number of spectral lines we documented (Appendix \ref{sect:flux_table}).
    
    Before fitting, the wavelength axis for each spaxel varies spatially and does not match the rest wavelengths for CO lines. We added a constant offset $<0.0002$ \mic \ to $\lambda$ to better match the rest wavelengths, which should only fail for lines having Doppler shifts much larger than the spectral resolution. The parameter left to explore is $a_i$, so our fit is rendered linear and offers ease for statistical uncertainties and goodness of fit pending systematic effects (i.e. fixing the spectral resolution to the pre-launch profile; using a uniform wavelength offset throughout a cube). 
    
    We apply a gradient descent algorithm to find $a_i$, sum all Gaussian profiles, and minimize the residual spectrum (baseline-subtracted spectrum minus predicted). For optimal $\sim$40 msec runtimes per spaxel ($\sim$3 min per cube), we use a library called \texttt{torchimize}, which implements \texttt{PyTorch} to maximize CPU usage (\citealt{torchimize}), though there are alternatives to \texttt{PyTorch}. 
    \texttt{PyTorch} also offers a necessary built-in method from \texttt{Python}'s \texttt{NumPy} library, einsum, which is based on Einstein summation notation and uses heuristics to quickly perform matrix multiplication. 
    
    When checking spatial features in images and line forests in spectra, some CO lines overlap with bright ionic or molecular lines. In such cases, CO line fluxes should continuously increase or decrease from transition to transition (e.g. flux at CO $v=1-0$ P16 should be an average value between the neighboring P15 and P17). Excess line flux above the averaged CO line is redistributed to the overlapping line. This may fail if neighboring lines are dim or buried within an ice feature, but those lines are not significantly detected relative to our baseline and are identifiable outliers. 
    For more on bright lines and images of [Fe~II], H~I, and H$_2$, see \citet{Federman_2023} and \citet{Narang_2024}.
    

\subsection{Systematic Effects with JWST/NIRSpec's G395M}\label{sect:systematics}
    For the line widths needed to create line images, we used the resolving power defined by the pre-launch profile. The optimized fit produces a different curve of FWHMs that peaks around 4.4 \mic \ and tapers below the pre-launch profile past 5 \mic\ (bottom panel of Figure \ref{fig:systematics}). The distribution of widths reflects how the R branch lines tend to be narrower than expected, while the P branch lines are wider. For creating images, we found the lowest residuals by multiplying the pre-launch profile by a factor of 1.15. The systematic differences between the expected and observed CO line profiles may be from an apparent broadening of CO $v=1-0$ P branch lines with spectrally unresolved CO $v=2-1$ lines, or under-resolving R branch lines blended with an absorbing component. The baseline could also be poorly estimated in parts of the spectrum near ice features, but we estimate this would cause a less than 10\% constant offset on our fluxes (similar to or less than the propagated uncertainties), and the relative differences between the R and P branches would remain. The spatial domain or aperture chosen can influence the S/N, raising spectral resolution more than expected, but the effect tends to influence spectrographs with slits rather than those with IFUs.

    The wavelength solution from the pipeline is offset by a different but approximately constant amount for each protostar's spectral cube in both the individual spaxels from the line image procedure (Appendix \ref{sect:fit_all}) and the high S/N spectral fits (Section \ref{apertures}). No unique wavelength offset works throughout a given cube, but each individual spectrum we extract has an offset that works well, possibly averaging instrumental and physical effects depending on source properties.     
    For example, the wavelength offset may change due to overlaps of G395M's sampling rate, blended CO $v=2-1$ lines, and the source's Doppler shift. Our high S/N fit is well-controlled for our analyses by excluding points from a given spectral line that were greater than a channel width away, so the CO forest maintains an offset that is nearly constant with respect to NIRSpec's spectral resolution. There may still be an interference-like effect in our residuals in the unresolved wings of CO $v=1-0$ P branch lines (see the residuals in Figure \ref{fig:example_spaxels}) caused by CO $v=2-1$ (see Section \ref{absorption_doppler} and the top panel of Figure \ref{absorption_doppler}). 
    Meanwhile, the line image procedure could be sensitive to outflowing material.     
    We can guarantee the relative distribution of brightness in the images because our baselines match the centers of broad ice features and the continuum (upper panels of Figure \ref{fig:systematics}), but the absolute value of the fluxes in a given line image will be difficult to precisely constrain.

\section{Line Flux Measurements}
    \label{sect:flux_table}
    Here, we present line fluxes for our sample. Wavelengths are taken from HITRAN \citep{HITRAN_2022}, and line fluxes with 1-$\sigma$ uncertainties are measured using the apertures centered on each central source labeled in Figure \ref{fig:co_imgs}. Fluxes for lines without a measurement are filled in with a dash. 

Tentative, unresolved combinations of H$_2$O ($v=010-000$) lines are also identified in absorption for IRAS 20126+4104 (and potentially HOPS 370) around 4.4093, 4.4163, 4.6942, 4.9540, 5.0528 \mic \ (bottom panel of Figure \ref{fig:example_spaxels}), but they require re-inspection with MIRI. When mapping lines in CARTA, a line tracing jets and outflows is identified between the CO $v=1-0$ R7 and R6 lines for B335 (and potentially IRAS 16253-2429), but our apertures are not ideal for measuring this line. The best match is the He~I at 4.6066 \mic\ line, which is what we report in our tabulated line list.

\startlongtable
 \begin{deluxetable*}{ccccccc} 
        \centering
        \tablewidth{0pt}
        \tabletypesize{\scriptsize}
        \tablecaption{Emission Line Fluxes for Central CO Apertures}
        \tablehead{
            %
            %
            %
            %
            %
            %
            %
            \colhead{$\lambda_0$} & \colhead{Species}  & \colhead{IRAS 16253-2429} & \colhead{B335} & \colhead{HOPS 153}  &  \colhead{HOPS 370} &  \colhead{IRAS 20126+4104} \\ 
            \colhead{--} & \colhead{--} & \colhead{$\times {10}^{-17}$} & \colhead{$\times {10}^{-17}$}   &  \colhead{$\times {10}^{-17}$}    &  \colhead{$\times {10}^{-14}$}    &  \colhead{$\times {10}^{-15}$} \\    
            \colhead{\mic} & \colhead{...}  &    \colhead{erg $\rm sec^{-1} \ {cm}^{-2}$} & \colhead{erg $\rm sec^{-1} \ {cm}^{-2}$}  &    \colhead{erg $\rm sec^{-1} \ {cm}^{-2}$}  &  \colhead{erg $\rm sec^{-1} \ {cm}^{-2}$}   &  \colhead{erg $\rm sec^{-1} \ {cm}^{-2}$}   
            }
        \startdata
4.3984 & CO $v = 1 - 0$  R 42 & 1.053$\pm$0.332 & 0.142$\pm$0.193 & -- & 0.134$\pm$0.005 & -- \\
4.4027 & CO $v = 1 - 0$  R 41 & 3.490$\pm$0.352 & 0.168$\pm$0.180 & 0.257$\pm$0.395 & 0.326$\pm$0.007 & -- \\
4.4070 & CO $v = 1 - 0$  R 40 & 3.416$\pm$0.354 & 0.426$\pm$0.241 & 0.537$\pm$0.470 & 0.329$\pm$0.007 & -- \\
4.4098 & H$_2$ $v=0-0$ S~10 & 13.068$\pm$0.524 & 3.578$\pm$0.359 & 10.311$\pm$1.044 & 0.712$\pm$0.011 & 8.180$\pm$0.061 \\
4.4115 & CO $v = 1 - 0$  R 39 & 3.505$\pm$0.379 & 0.364$\pm$0.195 & 0.898$\pm$0.584 & 0.410$\pm$0.008 & 0.648$\pm$0.039 \\
4.4160 & CO $v = 1 - 0$  R 38 & 3.782$\pm$0.368 & 0.368$\pm$0.164 & 1.097$\pm$0.426 & 0.439$\pm$0.008 & 0.885$\pm$0.039 \\
4.4166 & H$_2$ $v=1-1$ S~11 & 4.035$\pm$0.517 & 0.817$\pm$0.284 & 3.884$\pm$0.963 & -- & 2.295$\pm$0.069 \\
4.4207 & CO $v = 1 - 0$  R 37 & 3.670$\pm$0.357 & 0.455$\pm$0.178 & 2.286$\pm$0.711 & 0.501$\pm$0.008 & 0.822$\pm$0.036 \\
4.4254 & CO $v = 1 - 0$  R 36 & 3.444$\pm$0.316 & 0.682$\pm$0.198 & 2.040$\pm$0.625 & 0.442$\pm$0.007 & 1.137$\pm$0.046 \\
4.4302 & CO $v = 1 - 0$  R 35 & 4.493$\pm$0.360 & 0.735$\pm$0.191 & 2.920$\pm$0.740 & 0.501$\pm$0.007 & 1.108$\pm$0.042 \\
4.4348 & [Fe~II] & -- & 0.727$\pm$0.361 & 0.932$\pm$1.105 & -- & -- \\
4.4351 & CO $v = 1 - 0$  R 34 & 5.465$\pm$0.417 & 1.151$\pm$0.285 & 2.930$\pm$0.739 & 0.733$\pm$0.008 & 1.343$\pm$0.045 \\
4.4401 & CO $v = 1 - 0$  R 33 & 5.072$\pm$0.372 & 1.224$\pm$0.260 & 3.158$\pm$0.692 & 0.702$\pm$0.007 & 1.336$\pm$0.041 \\
4.4452 & CO $v = 1 - 0$  R 32 & 4.387$\pm$0.301 & 1.415$\pm$0.269 & 3.043$\pm$0.638 & 0.681$\pm$0.007 & 1.227$\pm$0.035 \\
4.4503 & CO $v = 1 - 0$  R 31 & 4.564$\pm$0.314 & 1.329$\pm$0.237 & 3.650$\pm$0.752 & 0.692$\pm$0.007 & 1.491$\pm$0.040 \\
4.4556 & CO $v = 1 - 0$  R 30 & 4.470$\pm$0.292 & 1.619$\pm$0.290 & 3.511$\pm$0.820 & 0.872$\pm$0.008 & 1.875$\pm$0.044 \\
4.4609 & CO $v = 1 - 0$  R 29 & 4.863$\pm$0.305 & 1.462$\pm$0.237 & 3.035$\pm$0.656 & 1.181$\pm$0.010 & 2.199$\pm$0.044 \\
4.4664 & CO $v = 1 - 0$  R 28 & 6.009$\pm$0.331 & 2.019$\pm$0.278 & 4.289$\pm$0.682 & 1.295$\pm$0.010 & 2.688$\pm$0.044 \\
4.4719 & CO $v = 1 - 0$  R 27 & 7.311$\pm$0.383 & 2.363$\pm$0.298 & 4.988$\pm$0.750 & 1.534$\pm$0.011 & 3.000$\pm$0.047 \\
4.4775 & CO $v = 1 - 0$  R 26 & 9.284$\pm$0.424 & 2.378$\pm$0.298 & 5.765$\pm$0.925 & 1.839$\pm$0.011 & 3.352$\pm$0.046 \\
4.4832 & CO $v = 1 - 0$  R 25 & 9.247$\pm$0.450 & 2.915$\pm$0.343 & 6.069$\pm$0.934 & 1.926$\pm$0.011 & 3.623$\pm$0.047 \\
4.4891 & CO $v = 1 - 0$  R 24 & 8.806$\pm$0.366 & 3.081$\pm$0.313 & 6.654$\pm$0.819 & 2.199$\pm$0.013 & 4.287$\pm$0.051 \\
4.4950 & CO $v = 1 - 0$  R 23 & 7.932$\pm$0.301 & 3.007$\pm$0.288 & 6.627$\pm$0.798 & 2.463$\pm$0.013 & 4.708$\pm$0.049 \\
4.5010 & CO $v = 1 - 0$  R 22 & 9.453$\pm$0.384 & 3.160$\pm$0.290 & 5.993$\pm$0.719 & 2.474$\pm$0.012 & 4.807$\pm$0.044 \\
4.5071 & CO $v = 1 - 0$  R 21 & 10.170$\pm$0.382 & 3.687$\pm$0.347 & 7.217$\pm$0.875 & 2.857$\pm$0.014 & 6.370$\pm$0.055 \\
4.5132 & CO $v = 1 - 0$  R 20 & 10.138$\pm$0.402 & 3.626$\pm$0.319 & 7.326$\pm$0.936 & 2.822$\pm$0.013 & 6.954$\pm$0.051 \\
4.5195 & CO $v = 1 - 0$  R 19 & 11.198$\pm$0.410 & 3.967$\pm$0.361 & 7.905$\pm$0.975 & 3.281$\pm$0.014 & 7.992$\pm$0.056 \\
4.5259 & CO $v = 1 - 0$  R 18 & 13.505$\pm$0.440 & 4.302$\pm$0.370 & 8.657$\pm$0.906 & 3.255$\pm$0.013 & 8.636$\pm$0.052 \\
4.5324 & CO $v = 1 - 0$  R 17 & 13.121$\pm$0.347 & 4.279$\pm$0.328 & 8.976$\pm$0.846 & 3.629$\pm$0.015 & 9.774$\pm$0.058 \\
4.5389 & CO $v = 1 - 0$  R 16 & 15.423$\pm$0.432 & 4.732$\pm$0.374 & 7.992$\pm$0.906 & 3.762$\pm$0.015 & 10.491$\pm$0.053 \\
4.5456 & CO $v = 1 - 0$  R 15 & 16.270$\pm$0.440 & 4.809$\pm$0.341 & 8.696$\pm$0.818 & 3.698$\pm$0.013 & 10.995$\pm$0.052 \\
4.5524 & CO $v = 1 - 0$  R 14 & 16.091$\pm$0.405 & 4.426$\pm$0.325 & 8.775$\pm$0.848 & 4.002$\pm$0.014 & 12.778$\pm$0.062 \\
4.5592 & CO $v = 1 - 0$  R 13 & 17.348$\pm$0.469 & 4.407$\pm$0.359 & 8.399$\pm$0.902 & 4.147$\pm$0.015 & 13.984$\pm$0.065 \\
4.5662 & CO $v = 1 - 0$  R 12 & 18.159$\pm$0.496 & 4.382$\pm$0.364 & 9.215$\pm$0.967 & 4.216$\pm$0.014 & 14.572$\pm$0.065 \\
4.5732 & CO $v = 1 - 0$  R 11 & 18.496$\pm$0.497 & 4.445$\pm$0.353 & -- & 4.285$\pm$0.015 & 15.509$\pm$0.066 \\
4.5755 & H$_2$ $v=1-0$ O~9 & -- & 0.951$\pm$0.285 & 9.657$\pm$1.019 & -- & -- \\
4.5804 & CO $v = 1 - 0$  R 10 & 14.776$\pm$0.432 & 4.231$\pm$0.365 & 9.025$\pm$0.922 & 3.989$\pm$0.013 & 14.648$\pm$0.063 \\
4.5876 & CO $v = 1 - 0$  R 9 & 14.039$\pm$0.406 & 3.724$\pm$0.331 & 8.821$\pm$0.870 & 3.728$\pm$0.013 & 14.310$\pm$0.060 \\
4.5950 & CO $v = 1 - 0$  R 8 & 12.562$\pm$0.352 & 3.341$\pm$0.312 & 7.598$\pm$0.772 & 3.357$\pm$0.011 & 13.795$\pm$0.057 \\
4.6024 & CO $v = 1 - 0$  R 7 & 11.485$\pm$0.357 & 2.775$\pm$0.303 & 6.116$\pm$0.710 & 3.081$\pm$0.011 & 12.510$\pm$0.054 \\
4.6066 & He~I & -- & 2.061$\pm$0.436 & -- & -- & -- \\
4.6100 & CO $v = 1 - 0$  R 6 & 12.676$\pm$0.417 & 2.633$\pm$0.340 & 5.895$\pm$0.737 & 3.030$\pm$0.010 & 11.070$\pm$0.048 \\
4.6177 & CO $v = 1 - 0$  R 5 & 11.511$\pm$0.443 & 2.426$\pm$0.338 & 4.891$\pm$0.747 & 2.941$\pm$0.010 & 10.316$\pm$0.044 \\
4.6254 & CO $v = 1 - 0$  R 4 & 10.060$\pm$0.415 & 2.485$\pm$0.338 & 5.076$\pm$0.739 & 2.971$\pm$0.011 & 11.517$\pm$0.050 \\
4.6333 & CO $v = 1 - 0$  R 3 & 9.457$\pm$0.440 & 2.484$\pm$0.379 & 5.957$\pm$0.916 & 3.173$\pm$0.012 & 12.470$\pm$0.053 \\
4.6374 & [Fe~II] & 0.351$\pm$0.366 & 0.064$\pm$0.183 & -- & -- & -- \\
4.6412 & CO $v = 1 - 0$  R 2 & 7.102$\pm$0.436 & 2.165$\pm$0.341 & 6.287$\pm$0.971 & 3.303$\pm$0.013 & 13.874$\pm$0.056 \\
4.6493 & CO $v = 1 - 0$  R 1 & 6.177$\pm$0.465 & 1.885$\pm$0.331 & 2.962$\pm$0.767 & 3.476$\pm$0.013 & 13.348$\pm$0.056 \\
4.6538 & H~I Pf $\beta$ & 0.660$\pm$0.295 & 0.124$\pm$0.232 & -- & -- & -- \\
4.6575 & CO $v = 1 - 0$  R 0 & -- & -- & -- & 2.837$\pm$0.013 & 12.563$\pm$0.060 \\
4.6742 & CO $v = 1 - 0$  P 1 & -- & 0.115$\pm$0.264 & -- & 1.879$\pm$0.010 & 9.904$\pm$0.052 \\
4.6826 & CO $v = 1 - 0$  P 2 & 3.900$\pm$0.326 & 0.126$\pm$0.237 & 1.055$\pm$0.528 & 3.173$\pm$0.012 & 12.075$\pm$0.054 \\
4.6912 & CO $v = 1 - 0$  P 3 & 5.483$\pm$0.460 & 0.244$\pm$0.232 & -- & 3.821$\pm$0.014 & 17.213$\pm$0.068 \\
4.6946 & H$_2$  $v=0-0$ S~9 & 23.831$\pm$0.503 & 5.511$\pm$0.397 & 11.252$\pm$0.712 & 3.577$\pm$0.016 & 29.095$\pm$0.073 \\
4.6999 & CO $v = 1 - 0$  P 4 & 7.256$\pm$0.430 & 1.932$\pm$0.369 & 3.774$\pm$0.784 & 5.206$\pm$0.017 & 21.315$\pm$0.073 \\
4.7088 & CO $v = 1 - 0$  P 5 & 16.580$\pm$0.641 & 5.099$\pm$0.442 & 16.157$\pm$1.392 & 6.052$\pm$0.018 & 28.883$\pm$0.087 \\
4.7157 & CO $v = 2 - 1$  R 0 & -- & 0.493$\pm$0.260 & 1.384$\pm$0.630 & -- & 1.174$\pm$0.038 \\
4.7177 & CO $v = 1 - 0$  P 6 & 18.258$\pm$0.548 & 6.670$\pm$0.482 & 17.201$\pm$1.230 & 6.413$\pm$0.020 & 28.300$\pm$0.086 \\
4.7267 & CO $v = 1 - 0$  P 7 & 18.691$\pm$0.536 & 7.045$\pm$0.472 & 16.012$\pm$1.147 & 6.589$\pm$0.021 & 26.210$\pm$0.086 \\
4.7359 & CO $v = 1 - 0$  P 8 & 19.510$\pm$0.540 & 7.718$\pm$0.479 & 15.378$\pm$1.138 & 6.910$\pm$0.021 & 29.214$\pm$0.091 \\
4.7451 & CO $v = 1 - 0$  P 9 & 21.888$\pm$0.569 & 8.228$\pm$0.486 & 15.877$\pm$1.110 & 6.940$\pm$0.020 & 30.755$\pm$0.086 \\
4.7545 & CO $v = 1 - 0$  P 10 & 22.630$\pm$0.579 & 8.599$\pm$0.495 & 17.607$\pm$1.176 & 6.734$\pm$0.020 & 27.873$\pm$0.084 \\
4.7589 & CO $v = 2 - 1$  P 4 & 0.494$\pm$0.282 & -- & 1.011$\pm$0.662 & -- & -- \\
4.7640 & CO $v = 1 - 0$  P 11 & 25.131$\pm$0.575 & 9.786$\pm$0.524 & 21.303$\pm$1.187 & 7.426$\pm$0.021 & 34.434$\pm$0.092 \\
4.7678 & CO $v = 2 - 1$  P 5 & 0.318$\pm$0.256 & -- & 1.879$\pm$0.668 & -- & -- \\
4.7736 & CO $v = 1 - 0$  P 12 & 27.501$\pm$0.607 & 9.471$\pm$0.520 & 20.329$\pm$1.298 & 7.382$\pm$0.022 & 32.251$\pm$0.092 \\
4.7769 & CO $v = 2 - 1$  P 6 & 0.676$\pm$0.238 & -- & -- & -- & -- \\
4.7833 & CO $v = 1 - 0$  P 13 & 30.354$\pm$0.607 & 8.901$\pm$0.477 & 19.016$\pm$1.130 & 7.376$\pm$0.022 & 31.470$\pm$0.094 \\
4.7861 & CO $v = 2 - 1$  P 7 & 2.076$\pm$0.480 & 0.808$\pm$0.357 & 2.028$\pm$0.802 & -- & -- \\
4.7931 & CO $v = 1 - 0$  P 14 & 27.372$\pm$0.600 & 9.899$\pm$0.548 & 22.329$\pm$1.332 & 7.404$\pm$0.022 & 32.104$\pm$0.096 \\
4.7954 & CO $v = 2 - 1$  P 8 & 0.636$\pm$0.420 & 0.914$\pm$0.489 & -- & -- & -- \\
4.8031 & CO $v = 1 - 0$  P 15 & 28.294$\pm$0.624 & 11.773$\pm$0.615 & 21.639$\pm$1.358 & 7.717$\pm$0.022 & 34.573$\pm$0.097 \\
4.8048 & CO $v = 2 - 1$  P 9 & 0.893$\pm$0.481 & 0.307$\pm$0.484 & 2.622$\pm$1.691 & -- & -- \\
4.8131 & CO $v = 1 - 0$  P 16 & 27.261$\pm$0.620 & 11.573$\pm$0.582 & 23.282$\pm$1.381 & 7.612$\pm$0.022 & 31.887$\pm$0.093 \\
4.8143 & CO $v = 2 - 1$  P 10 & -- & -- & 1.931$\pm$1.288 & -- & -- \\
4.8233 & CO $v = 1 - 0$  P 17 & 25.862$\pm$0.609 & 11.513$\pm$0.578 & 22.432$\pm$1.400 & 6.680$\pm$0.020 & 31.626$\pm$0.093 \\
4.8240 & CO $v = 2 - 1$  P 11 & 0.830$\pm$0.568 & -- & 1.933$\pm$1.172 & 0.756$\pm$0.015 & -- \\
4.8336 & CO $v = 1 - 0$  P 18 & 24.775$\pm$0.615 & 11.732$\pm$0.581 & 20.157$\pm$1.536 & 6.779$\pm$0.022 & 31.341$\pm$0.111 \\
4.8337 & CO $v = 2 - 1$  P 12 & 0.890$\pm$0.559 & -- & 1.554$\pm$1.179 & 0.749$\pm$0.017 & -- \\
4.8436 & CO $v = 2 - 1$  P 13 & 1.153$\pm$0.509 & 0.612$\pm$0.469 & 1.128$\pm$1.063 & 0.161$\pm$0.017 & -- \\
4.8440 & CO $v = 1 - 0$  P 19 & 24.104$\pm$0.720 & 10.089$\pm$0.592 & 18.701$\pm$1.398 & 7.093$\pm$0.024 & 21.529$\pm$0.097 \\
4.8536 & CO $v = 2 - 1$  P 14 & 1.187$\pm$0.357 & 0.927$\pm$0.536 & 1.446$\pm$1.109 & 0.077$\pm$0.014 & -- \\
4.8546 & CO $v = 1 - 0$  P 20 & 21.943$\pm$0.664 & 9.077$\pm$0.578 & 16.843$\pm$1.327 & 6.208$\pm$0.021 & 16.478$\pm$0.082 \\
4.8637 & CO $v = 2 - 1$  P 15 & 1.243$\pm$0.380 & 0.496$\pm$0.553 & 1.206$\pm$0.961 & -- & -- \\
4.8652 & CO $v = 1 - 0$  P 21 & 20.116$\pm$0.635 & 9.791$\pm$0.602 & 16.373$\pm$1.272 & 6.017$\pm$0.021 & 16.345$\pm$0.093 \\
4.8739 & CO $v = 2 - 1$  P 16 & 1.211$\pm$0.386 & -- & 1.836$\pm$1.434 & -- & -- \\
4.8760 & CO $v = 1 - 0$  P 22 & 18.222$\pm$0.611 & 9.069$\pm$0.581 & 16.707$\pm$1.393 & 5.760$\pm$0.021 & 15.756$\pm$0.091 \\
4.8843 & CO $v = 2 - 1$  P 17 & 1.284$\pm$0.454 & 0.293$\pm$0.350 & 1.683$\pm$1.035 & -- & -- \\
4.8869 & CO $v = 1 - 0$  P 23 & 18.218$\pm$0.629 & 8.802$\pm$0.713 & 16.100$\pm$1.467 & 5.551$\pm$0.020 & 12.509$\pm$0.081 \\
4.8891 & [Fe~II] & -- & 11.337$\pm$0.675 & 6.668$\pm$2.151 & -- & -- \\
4.8948 & CO $v = 2 - 1$  P 18 & 0.934$\pm$0.456 & 0.523$\pm$0.311 & 1.049$\pm$0.616 & -- & -- \\
4.8980 & CO $v = 1 - 0$  P 24 & 19.304$\pm$0.693 & 6.224$\pm$0.498 & 13.382$\pm$1.335 & 5.028$\pm$0.019 & 8.984$\pm$0.072 \\
4.9054 & CO $v = 2 - 1$  P 19 & 3.165$\pm$0.862 & 0.550$\pm$0.421 & 1.613$\pm$0.803 & 0.302$\pm$0.015 & -- \\
4.9091 & CO $v = 1 - 0$  P 25 & 15.448$\pm$0.702 & 5.938$\pm$0.493 & 11.535$\pm$1.334 & 4.752$\pm$0.018 & 7.313$\pm$0.070 \\
4.9161 & CO $v = 2 - 1$  P 20 & 3.258$\pm$0.633 & 0.420$\pm$0.332 & 2.752$\pm$0.884 & 0.164$\pm$0.012 & 0.363$\pm$0.042 \\
4.9204 & CO $v = 1 - 0$  P 26 & 12.673$\pm$0.590 & 5.791$\pm$0.519 & 11.904$\pm$1.392 & 4.048$\pm$0.018 & 5.839$\pm$0.069 \\
4.9269 & CO $v = 2 - 1$  P 21 & 3.313$\pm$0.551 & 0.717$\pm$0.442 & 4.383$\pm$1.092 & 0.302$\pm$0.009 & 1.595$\pm$0.061 \\
4.9318 & CO $v = 1 - 0$  P 27 & 11.070$\pm$0.589 & 5.952$\pm$0.525 & 12.136$\pm$1.647 & 3.771$\pm$0.019 & 5.752$\pm$0.072 \\
4.9379 & CO $v = 2 - 1$  P 22 & 2.655$\pm$0.560 & 0.236$\pm$0.259 & 3.429$\pm$0.875 & 0.321$\pm$0.011 & 1.644$\pm$0.059 \\
4.9434 & CO $v = 1 - 0$  P 28 & 10.355$\pm$0.616 & 6.002$\pm$0.530 & 8.927$\pm$1.221 & 3.470$\pm$0.018 & 5.013$\pm$0.074 \\
4.9490 & CO $v = 2 - 1$  P 23 & 2.495$\pm$0.682 & 0.620$\pm$0.422 & 4.087$\pm$0.983 & 0.384$\pm$0.012 & 1.474$\pm$0.062 \\
4.9541 & H$_2$ $v=1-1$ S~9 & 5.375$\pm$0.609 & 2.516$\pm$0.526 & 5.536$\pm$1.165 & -- & 1.665$\pm$0.061 \\
4.9550 & CO $v = 1 - 0$  P 29 & 10.699$\pm$0.657 & 6.844$\pm$0.605 & 14.140$\pm$1.899 & 3.421$\pm$0.018 & 4.473$\pm$0.073 \\
4.9603 & CO $v = 2 - 1$  P 24 & 1.805$\pm$0.659 & 0.656$\pm$0.364 & 3.711$\pm$1.151 & 0.249$\pm$0.012 & 0.762$\pm$0.056 \\
4.9668 & CO $v = 1 - 0$  P 30 & 9.199$\pm$0.574 & 6.507$\pm$0.575 & 13.502$\pm$1.448 & 3.199$\pm$0.018 & 4.057$\pm$0.070 \\
4.9716 & CO $v = 2 - 1$  P 25 & -- & 0.515$\pm$0.454 & 3.530$\pm$1.127 & 0.284$\pm$0.014 & 0.532$\pm$0.050 \\
4.9788 & CO $v = 1 - 0$  P 31 & 9.759$\pm$0.590 & 6.279$\pm$0.552 & 12.343$\pm$1.403 & 3.025$\pm$0.018 & 3.082$\pm$0.064 \\
4.9831 & CO $v = 2 - 1$  P 26 & 1.302$\pm$0.344 & 1.185$\pm$0.537 & 2.519$\pm$0.926 & 0.315$\pm$0.017 & 0.514$\pm$0.068 \\
4.9908 & CO $v = 1 - 0$  P 32 & 9.461$\pm$0.621 & 6.890$\pm$0.607 & 15.755$\pm$1.796 & 3.013$\pm$0.018 & 3.534$\pm$0.068 \\
4.9947 & CO $v = 2 - 1$  P 27 & 1.371$\pm$0.457 & 0.669$\pm$0.346 & 2.835$\pm$1.204 & 0.348$\pm$0.018 & 0.731$\pm$0.065 \\
5.0031 & CO $v = 1 - 0$  P 33 & 9.465$\pm$0.735 & 6.613$\pm$0.634 & 13.193$\pm$1.858 & 2.979$\pm$0.020 & 3.257$\pm$0.077 \\
5.0065 & CO $v = 2 - 1$  P 28 & 1.388$\pm$0.542 & 0.524$\pm$0.365 & 3.615$\pm$2.324 & -- & 0.442$\pm$0.088 \\
5.0154 & CO $v = 1 - 0$  P 34 & 11.600$\pm$0.864 & 6.607$\pm$0.709 & 14.885$\pm$2.046 & 2.604$\pm$0.019 & 1.521$\pm$0.048 \\
5.0184 & CO $v = 2 - 1$  P 29 & 0.323$\pm$0.299 & 0.660$\pm$0.646 & -- & -- & -- \\
5.0279 & CO $v = 1 - 0$  P 35 & 10.335$\pm$0.774 & 6.205$\pm$0.741 & 12.296$\pm$1.579 & 2.417$\pm$0.019 & 1.601$\pm$0.081 \\
5.0304 & CO $v = 2 - 1$  P 30 & 1.448$\pm$0.745 & 0.605$\pm$0.519 & 2.093$\pm$2.663 & -- & -- \\
5.0405 & CO $v = 1 - 0$  P 36 & 10.573$\pm$0.798 & 5.475$\pm$0.671 & 10.536$\pm$1.498 & 2.332$\pm$0.021 & 2.519$\pm$0.073 \\
5.0425 & CO $v = 2 - 1$  P 31 & 1.129$\pm$0.686 & 1.592$\pm$0.805 & 1.914$\pm$1.390 & -- & -- \\
5.0531 & H$_2$ $v=0-0$ S~8 & 24.219$\pm$0.680 & 19.225$\pm$0.795 & 22.564$\pm$1.720 & 1.609$\pm$0.017 & 19.682$\pm$0.091 \\
5.0532 & CO $v = 1 - 0$  P 37 & 10.589$\pm$0.861 & 5.411$\pm$1.066 & 11.133$\pm$1.990 & 2.162$\pm$0.022 & 2.783$\pm$0.109 \\
5.0548 & CO $v = 2 - 1$  P 32 & 1.513$\pm$0.913 & 1.201$\pm$0.996 & 4.108$\pm$1.779 & 0.310$\pm$0.021 & -- \\
5.0623 & [Fe~II] & 1.149$\pm$0.848 & 1.813$\pm$0.640 & 2.391$\pm$1.196 & -- & -- \\
5.0661 & CO $v = 1 - 0$  P 38 & 10.043$\pm$0.727 & 6.026$\pm$0.703 & 11.670$\pm$1.558 & 2.252$\pm$0.022 & 2.524$\pm$0.077 \\
5.0673 & CO $v = 2 - 1$  P 33 & 1.171$\pm$0.765 & 1.637$\pm$0.945 & 2.149$\pm$1.191 & -- & -- \\
5.0792 & CO $v = 1 - 0$  P 39 & 11.116$\pm$0.821 & 5.599$\pm$0.807 & 14.921$\pm$2.108 & 1.924$\pm$0.019 & 1.468$\pm$0.052 \\
5.0798 & CO $v = 2 - 1$  P 34 & -- & 1.585$\pm$0.989 & 1.200$\pm$1.159 & -- & -- \\
5.0924 & CO $v = 1 - 0$  P 40 & 11.798$\pm$0.882 & 5.485$\pm$0.689 & 12.995$\pm$1.922 & 2.017$\pm$0.021 & 2.789$\pm$0.097 \\
5.0925 & CO $v = 2 - 1$  P 35 & -- & 1.020$\pm$0.970 & 1.467$\pm$2.264 & -- & -- \\
5.1054 & CO $v = 2 - 1$  P 36 & -- & 1.009$\pm$0.924 & -- & 0.204$\pm$0.025 & -- \\
5.1057 & CO $v = 1 - 0$  P 41 & 12.416$\pm$0.957 & 4.705$\pm$0.605 & 15.375$\pm$2.003 & 1.629$\pm$0.019 & 2.564$\pm$0.103 \\
5.1184 & CO $v = 2 - 1$  P 37 & 0.306$\pm$0.494 & 0.881$\pm$0.788 & -- & -- & -- \\
5.1191 & CO $v = 1 - 0$  P 42 & 8.851$\pm$0.664 & 4.046$\pm$0.595 & 12.522$\pm$1.781 & 1.601$\pm$0.019 & 2.176$\pm$0.083 \\
5.1315 & CO $v = 2 - 1$  P 38 & 0.552$\pm$0.504 & 1.165$\pm$0.777 & -- & -- & -- \\
5.1328 & CO $v = 1 - 0$  P 43 & 10.324$\pm$0.818 & 4.236$\pm$0.665 & 16.501$\pm$2.159 & 1.682$\pm$0.023 & 1.846$\pm$0.077 \\
5.1448 & CO $v = 2 - 1$  P 39 & 1.124$\pm$0.569 & 0.619$\pm$0.672 & -- & -- & 0.643$\pm$0.058 \\
5.1465 & CO $v = 1 - 0$  P 44 & 12.212$\pm$1.179 & 4.037$\pm$0.714 & 17.380$\pm$2.258 & 1.299$\pm$0.020 & -- \\
5.1582 & CO $v = 2 - 1$  P 40 & 1.407$\pm$0.913 & 1.194$\pm$1.013 & -- & -- & 0.587$\pm$0.113 \\
5.1605 & CO $v = 1 - 0$  P 45 & 10.446$\pm$1.004 & 3.268$\pm$0.665 & 15.189$\pm$2.297 & 1.465$\pm$0.023 & 0.729$\pm$0.068 \\
5.1718 & CO $v = 2 - 1$  P 41 & 2.621$\pm$1.096 & 1.243$\pm$1.153 & 1.480$\pm$2.183 & 0.379$\pm$0.024 & -- \\
5.1745 & CO $v = 1 - 0$  P 46 & 7.621$\pm$0.760 & 3.251$\pm$0.719 & 11.130$\pm$2.577 & 1.174$\pm$0.025 & 0.425$\pm$0.045 \\
5.1855 & CO $v = 2 - 1$  P 42 & 3.629$\pm$1.113 & 0.902$\pm$1.159 & -- & 0.295$\pm$0.029 & -- \\
5.1887 & CO $v = 1 - 0$  P 47 & 7.211$\pm$1.010 & 3.042$\pm$0.741 & 10.732$\pm$2.251 & 1.078$\pm$0.025 & -- \\
5.1994 & CO $v = 2 - 1$  P 43 & 3.561$\pm$0.820 & 0.778$\pm$0.771 & -- & -- & -- \\
5.2031 & CO $v = 1 - 0$  P 48 & 4.646$\pm$0.721 & 2.481$\pm$0.797 & 10.562$\pm$2.597 & 0.711$\pm$0.021 & -- \\
5.2134 & CO $v = 2 - 1$  P 44 & 4.788$\pm$1.023 & 1.185$\pm$0.687 & 3.810$\pm$1.304 & 0.102$\pm$0.016 & -- \\
5.2177 & CO $v = 1 - 0$  P 49 & 3.989$\pm$0.664 & 1.908$\pm$0.574 & 3.734$\pm$1.343 & 0.626$\pm$0.016 & -- \\
5.2276 & CO $v = 2 - 1$  P 45 & 3.916$\pm$0.994 & 0.926$\pm$0.924 & 3.302$\pm$1.491 & 0.167$\pm$0.014 & -- \\
5.2323 & CO $v = 1 - 0$  P 50 & 3.933$\pm$0.749 & 1.794$\pm$0.616 & 3.894$\pm$1.676 & 0.589$\pm$0.017 & -- \\
5.2420 & CO $v = 2 - 1$  P 46 & 2.353$\pm$0.937 & 1.481$\pm$0.998 & 4.849$\pm$1.872 & 0.058$\pm$0.012 & -- \\
5.2472 & CO $v = 1 - 0$  P 51 & 2.428$\pm$0.526 & 1.666$\pm$0.594 & 3.285$\pm$1.273 & 0.549$\pm$0.018 & -- \\
5.2565 & CO $v = 2 - 1$  P 47 & 1.872$\pm$0.713 & 0.851$\pm$0.769 & 3.844$\pm$2.092 & 0.107$\pm$0.008 & -- \\
5.2622 & CO $v = 1 - 0$  P 52 & 2.799$\pm$0.711 & 2.070$\pm$0.742 & 5.428$\pm$2.283 & 0.568$\pm$0.020 & 0.807$\pm$0.065 \\
        \enddata
    \tablecomments{The emission line species detected throughout the CO forest, their associated fluxes (not extinction corrected), and 1-$\sigma$ uncertainties for each source. Note that any flux calibration uncertainties are not included. The apertures used for each source are shown in Figure \ref{fig:co_imgs}. Any undetected lines are filled with a -- to show they have not been measured.}
\end{deluxetable*}\label{tab:co_lines_fluxes}



\section{Uncertainties and Propagation}
    \label{sect:uncert_propagation} 
    For Table \ref{tab:co_lines_fluxes} in Appendix \ref{sect:flux_table}, the uncertainty in the total line flux is computed similarly to the line profiles
\begin{equation}
    \Delta(F_{line}) = \sum_{i,j} \Delta(F_{\lambda, i}) \times \exp{\frac{-(\lambda_j - \lambda_0)^2}{2 \sigma^2}},
\end{equation}
where $\Delta(F_{line})$ is the total uncertainty and $\Delta(F_{\lambda, i})$ is the measured uncertainty from the default error cube generated from the JWST pipeline. This cube of noise is also called the error array and indexed with keyword 'ERR' in a cube's header. We opted for the pipeline values because they are independent to our residuals from fitting the spectra and are more stable near a given line, which is seen in the pipeline-derived noise plotted in Figure \ref{fig:example_spaxels}.


Uncertainties in the number of molecules per degeneracy ($\frac{{N_{J_{u}}^{tot}}}{g_{u}}$) can be approximated from P branch fluxes from standard propagation:
\begin{align}
    \Delta \bigg(\frac{{N_{J_{u}}^{tot}}} {g_{u}} \bigg) \approx
    \frac{1}{hc \ g_u(J_u)} \ \frac{\lambda_P(J_u) \ e^{({\tau_{dust,\lambda_P}})}}{A_{ij,P}} \Bigg( \Big[\frac{\Delta\big({F_P(J_u)}\big)}{F_P(J_u)} \Big]^2 \Bigg)^{\frac{1}{2}} \times (4 \pi {d_{star}}^2).
\end{align}
Here we neglect adding systematic effects, such as an additional term for $\Delta \tau_{dust}$, distance, and other constants determined in labs. Similarly, standard propagation for the natural log of this ratio is: 
\begin{equation}
    \Delta \Bigg( 
    \ln{\bigg( \frac{{N_{J_{u}}^{tot}}} {g_{u}} \bigg)}
    \Bigg) = \Delta \bigg(\frac{{N_{J_{u}}^{tot}}} {g_{u}} \bigg) \times \frac{1}{\frac{{N_{J_{u}}^{tot}}} {g_{u}}}
\end{equation}

\bibliography{biblio}{}
\bibliographystyle{aasjournal}

\end{document}